\documentclass[11pt,a4paper]{article}
\usepackage{jcappub}

\pdfoutput=1

\usepackage{graphicx}
\usepackage{epsfig}
\usepackage{rotate}
\usepackage{amsmath}
\usepackage{amssymb}
\usepackage{amsfonts}
\usepackage{enumerate}
\usepackage{afterpage}
\usepackage{placeins}
\usepackage{caption}
\usepackage{xcolor}
\usepackage[T1]{fontenc}
\usepackage{booktabs}
\usepackage{braket}
\usepackage{subfig}
\usepackage{todonotes}
\usepackage[utf8]{inputenc}
\hypersetup{
 citecolor=blue,filecolor=blue,linkcolor=blue,urlcolor=blue}
\usepackage{array}
\newcolumntype{P}[1]{>{\centering\arraybackslash}p{#1}}
\newcolumntype{M}[1]{>{\centering\arraybackslash}m{#1}}

\allowdisplaybreaks[1]

\def\ra{\rightarrow}

\newcommand{\bx}{{\mathbf x}}

\newcommand{\bn}{{\mathbf n}}
\newcommand{\bv}{{\mathbf v}}
\newcommand{\bk}{{\mathbf k}}

\newcommand{\br}{{\mathbf r}}

\newcommand{\pa}{{\parallel}}

\newcommand{\HH}{{\cal H}}
\newcommand{\cd}{\cdot}
\newcommand{\al}{\alpha}

\newcommand{\de}{\delta}
\newcommand{\De}{\Delta}

\newcommand{\ga}{\gamma}

\newcommand{\ka}{\kappa}

\newcommand{\La}{\Lambda}
\newcommand{\la}{\lambda}
\newcommand{\Om}{\Omega}

\newcommand{\si}{\sigma}

\newcommand{\be}{\begin{equation}}
\newcommand{\ee}{\end{equation}}
\newcommand{\bea}{\begin{eqnarray}}
\newcommand{\eea}{\end{eqnarray}}
\newcommand{\bean}{\begin{eqnarray*}}
\newcommand{\eean}{\end{eqnarray*}}
\newcommand{\beal}{\begin{align}}
\newcommand{\enal}{\end{align}}
\newcommand{\pd}{\partial}
\newcommand{\dd}{\text{d}}

\newcommand{\kcos}{\nu}
\newcommand{\nuP}{\nu}

\usepackage{ntheorem}

\newtheorem*{theorem-non}{Theorem}

\newenvironment{proof}[1][Proof]{\begin{trivlist}
\item[\hskip \labelsep {\bfseries #1}]}{\end{trivlist}}
\newcommand{\qed}{\nobreak \ifvmode \relax \else
      \ifdim\lastskip<1.5em \hskip-\lastskip
      \hskip1.5em plus0em minus0.5em \fi \nobreak
      \vrule height0.75em width0.5em depth0.25em\fi}

\definecolor{revisioncolor}{RGB}{22, 158, 230}

\begin{document}
\title{ The full-sky relativistic correlation function and power spectrum of galaxy number counts:\\ I. Theoretical aspects}
\author{Vittorio Tansella, Camille Bonvin,  Ruth Durrer, Basundhara Ghosh and Elena Sellentin}
\affiliation{D\'epartement de Physique Th\'eorique and Center for Astroparticle Physics, Universit\'e de Gen\`eve, 24 Quai Ansermet, CH--1211 Gen\`eve 4, Switzerland}
\emailAdd{vittorio.tansella@unige.ch}
\emailAdd{camille.bonvin@unige.ch}
\emailAdd{ruth.durrer@unige.ch}
\emailAdd{basundhara.ghosh@unige.ch}
\emailAdd{elena.sellentin@unige.ch}
\vspace{1 em}
\date{\today}

\abstract{We derive an exact expression for the correlation function in redshift shells including all the relativistic contributions. This expression, which does not rely on the distant-observer or flat-sky approximation, is valid at all scales and includes both local relativistic corrections and integrated contributions, like gravitational lensing. We present two methods to calculate this correlation function, one which makes use of the angular power spectrum $C_\ell(z_1,z_2)$ and a second method which evades the costly calculations of the angular power spectra. The correlation function is then used to define the power spectrum as its Fourier transform. In this work theoretical aspects of this procedure are presented, together with quantitative examples. In particular, we show that gravitational lensing modifies the multipoles of the correlation function and of the power spectrum by a few percent at redshift $z=1$ and by up to 30\% and more at $z=2$. We also point out that large-scale relativistic effects and wide-angle corrections generate contributions of the same order of magnitude and have consequently to be treated in conjunction. These corrections are particularly important at small redshift, $z=0.1$, where they can reach 10\%. This means in particular that a flat-sky treatment of relativistic effects, using for example the power spectrum, is not consistent.}

\maketitle

\section{Introduction}

Upcoming redshift surveys of the distribution of galaxies~\cite{Laureijs:2011gra,Amendola:2016saw,Abate:2012za,Abell:2009aa,Aghamousa:2016zmz,Carilli:2004nx} are going to probe the large-scale structure of the universe at high redshift and for wide patches of the sky with unprecedented precision. To exploit the information delivered by these surveys in an optimal way, it is crucial to have  reliable theoretical predictions of the signal. Redshift surveys generally associate two quantities to each galaxy they detect: the direction from which photons are received, $\bn$, and the redshift $z$. It has therefore been argued in the past~\cite{Heavens:1994iq,Heavens:1997gq,Szalay:1997cc,Szapudi:2004gh,Raccanelli:2010hk,Bonvin:2011bg}, that galaxy correlation functions are truly functions of two redshifts and an angle. The angular-redshift power spectrum is then 
given by $C_\ell(z_1,z_2)$. This quantity has been introduced in~\cite{Bonvin:2011bg,Challinor:2011bk}, where it has also been shown that due to relativistic projection effects, the linear power spectrum is not simply given by density fluctuations and redshift-space distortions, but it acquires several additional terms from lensing, ordinary and integrated Sachs Wolfe terms, gravitational redshift, Doppler terms, and Shapiro time delay. These projection effects had been previously identified in~\cite{Yoo:2009au,Yoo:2010ni}.
 
Subsequently, linear Boltzmann codes like {\sc camb}~\cite{Lewis:1999bs} and {\sc class}~\cite{Blas:2011rf} have been generalized to calculate this galaxy count angular power spectrum~\cite{DiDio:2013bqa,DiDio:2013sea}. To determine the $C_\ell(z_1,z_2)$ observationally, one correlates the number of galaxies in a redshift  bin around $z_1$  and in a small solid angle around direction $\bn_1$ with those in a redshift  bin around $z_2$  and in a small solid angle around direction $\bn_2$.  Due to statistical isotropy, the resulting correlation
function only depends on the angle $\theta$ between $\bn_1$ and $\bn_2$, $\cos\theta=\bn_1\cd\bn_2$ and is related to the angular power spectrum  in the well known way,
\be\label{e:xiCl}
\xi(\theta,z_1,z_2)= \frac{1}{4\pi} \sum_\ell (2\ell+1)C_\ell(z_1,z_2)L_\ell(\cos\theta) \,,
\ee
where $L_\ell$ denotes the Legendre polynomial of degree $\ell$.

Before the introduction of the $C_\ell(z_1,z_2)$'s,  cosmologists have mainly concentrated on determining the correlation function and the power spectrum in Fourier space. In comoving  gauge, on sub-horizon scales
 the latter is given by~\cite{Kaiser1987}
\bea\label{e:poldk}
P_g(k, \nuP, \bar{z}) &=& D_1^2(\bar{z})\left[b(\bar z) +f(\bar z)(\hat\bk\cdot \bn)^2\right]^2P_m(k) \\
&=&D_1^2(\bar z)\left[b^2+\frac{2bf}{3}+\frac{f^2}{5}+\left(\frac{4bf}{3}+\frac{4f^2}{7} \right)L_2(\kcos)
+\frac{8f^2}{35} L_4(\kcos)\right]P_m(k)\, .\nonumber
\eea
Here $\bar z$ is the mean redshift or the survey, $P_m(k)$ is the matter density power spectrum today,  $D_1(\bar z)$ is the growth factor normalized to $D_1(0)=1$, $b(\bar z)$ is the galaxy bias and
\be\label{e:growth}
f(\bar z) = -\frac{D_1'}{D_1}(1+\bar z)=\frac{d\ln D_1}{d\ln(a)} \,,
\ee
is the growth rate, where the prime denotes the derivative with respect to the redshift $\bar z$.
The direction cosine $\kcos$ is the cosine of the angle between $\bk$ and the observation direction $\bn$ (in the literature this direction cosine is often denoted as $\mu$ but here we reserve $\mu$ for the corresponding angle in real space and in order to avoid confusion we denote it by $\kcos$ in Fourier space). 

Equation~\eqref{e:poldk} has an interesting property: projecting out the monopole, quadrupole and hexadecapole in $\kcos$, one can directly measure the bias $b$ and the growth rate $f$. This has been exploited in previous observations and has led to the best determinations of $f$ so far (see~\cite{Contreras:2013bol,Oka:2013cba,Achitouv:2016mbn,Adams:2017val,Alam:2016hwk,Satpathy:2016tct} and refs. therein). It is clear that the form \eqref{e:poldk} of the power spectrum can only be valid if the bins are not too far apart in the sky. Eq.~\eqref{e:poldk} indeed implicitly assumes that the galaxies are observed in {\it one} single direction $\bn$ so that a 'flat-sky approximation' with a well defined angle $\kcos$ is a reasonably good approximation. 

An  observable alternative to the power spectrum, which is routinely used in galaxy surveys is the correlation function $\xi(r,\mu,\bar {z})$, where $r$ denotes the separation between the galaxies, $\mu$ is the orientation of the pair with respect to the direction of observation $\bn$ and $\bar z$ is the mean redshift of the survey. The correlation function is observed in terms of $z_1, z_2$ and $\theta$. To express it in terms of $r,\mu$ and $\bar {z}$, the redshifts $z_1$ and $z_2$ have to be converted into comoving distances and a direction cosine $\mu$ has to be defined.

Neglecting spatial curvature we can use the cosine law to express $r$ in terms of the comoving distances to $z_1$ and $z_2$,
\be\label{e:rofztheta}
r(z_1,z_2,\theta) = \sqrt{\chi(z_1)^2 + \chi(z_2)^2-2\chi(z_1)\chi(z_2)\cos\theta} \,,
\ee
where 
\be
\chi(z) = \frac{1}{H_0}\int_0^z\frac{dz}{\sqrt{\Om_m(1+z)^3 + \Om_Xg_X(z)}} \,.
\ee
Here $\Om_m$ is the matter density parameter and $\Om_Xg_X(z)$ is the dark energy density in units of the critical density today; $g_X$ is normalized to $g_X(0)=1$. Hence the correlation function $\xi(r,\mu,\bar {z})$, as well as the power spectrum, are not directly observable: they both require the use of a fiducial cosmology to calculate $r$ and $\chi(z)$. If the redshift is small, $z\ll 1$, we can write $\chi(z) \simeq z/H_0$,  and the dependence on $H_0$ is taken into account by measuring cosmological distances in units of Mpc/$h$, where Mpc denotes a megaparsec ($\simeq 3.1\times 10^6$ light years) and $h=H_0/100$\,km/s/Mpc. However, in present and upcoming catalogues which go out to $z=2$ and more, this is no longer sufficient and $r$ depends in a non-trivial way on the dark matter and dark energy density, on the dark energy equation of state and on curvature (which is set to zero in this work for simplicity).
Fortunately this dependence can be accounted for by introducing correction parameters, which allow for deviations from the fiducial cosmology, see e.g.~\cite{Xu:2012fw}. In the flat-sky approximation, the standard correlation function takes the simple form~\cite{Hamilton:1997zq} 
\be
\label{e:corrflat}
\xi^{\rm st}(r,\mu,\bar {z})=D_1^2(\bar z)\left[\left(b^2+\frac{2bf}{3}+\frac{f^2}{5}\right)c_0(r)-\left(\frac{4bf}{3}+\frac{4f^2}{7} \right)c_2(r)L_2(\mu)
+\frac{8f^2}{35} L_4(\mu)c_4(r)\right]\,,
\ee
with 
\be
c_\ell(r)=\frac{1}{2\pi^2}\int dk \, k^2P_m(k)j_\ell(rk)\, .
\ee
Note that the terms containing the growth factor $f$ come from the Jacobian transforming real space positions $\bx$ into redshifts
\footnote{We point out that the original derivation of redshift-space distortion from~\cite{Kaiser1987} contains a contribution proportional to $\bn\cdot\mathbf{v}=v_r$. This term does contribute to the monopole and quadrupole and it consequently modifies~\eqref{e:corrflat}. It is however neglected in most redshift-space distortion analysis and therefore we do not consider it as 'standard' and we do not include it in~\eqref{e:corrflat}. We include it however in the relativistic corrections, along with the other Doppler corrections, which are of the same order of magnitude (see Eq.~\eqref{e:Dd1}). Note that, as discussed in more detail in Section~\ref{sec:corr_mult}, this specific contribution has been studied in detail in~\cite{Szalay:1997cc,Szapudi:2004gh,Papai:2008bd,Raccanelli:2010hk} and its impact on the correlation function was found to be important at small redshift and large separation.}.

In Appendix~\ref{app:theorem} we derive the general relation between the $c_\ell(r)$ and the corresponding pre-factors of the Legendre polynomials in the power spectrum.

Expressions~\eqref{e:poldk} and~\eqref{e:corrflat} are currently used to analyse redshift surveys~\footnote{Note that these expressions are valid in the linear regime only. Theoretical models accounting for non-linearities have been developed and are used to extend the constraints to non-linear scales, see e.g.~\cite{1994MNRAS.267..927F}.}. These expressions are sufficiently accurate to place meaningful constraints on cosmological parameters with current data. They may however not be sufficient to analyse future surveys since they suffer from two important limitations: first they are based on the flat-sky (sometimes also called distant-observer) approximation. And second they take into account only density fluctuations and redshift-space distortions. They neglect lensing which is relevant especially when the redshifts $z_1$ and $z_2$ are significantly different. They also neglect all the relativistic projection effects which are relevant on large scales (close to horizon scale). These expressions are therefore only an approximate description of what we are observing, which is also reflected by the fact that they are gauge-dependent.

Due to these limitations, one would be tempted to use the angular power spectrum instead of Eqs.~\eqref{e:poldk} and~\eqref{e:corrflat} to analyse future redshift surveys. The gauge-invariant $C_\ell(z_1,z_2)$'s account indeed for all observable effects. They are directly observable and do not rely on the flat-sky approximation. And they can be determined numerically within a few seconds with sub-percent accuracy. Unfortunately they are not fully satisfactory for several reasons:

\begin{enumerate}[(1)]
\item If we want to profit optimally from {\it spectroscopic} redshift information from a survey like the one that will be generated by Euclid~\cite{Laureijs:2011gra}, DESI~\cite{Aghamousa:2016zmz} or the SKA~\cite{Carilli:2004nx},
we need several thousand redshift slices leading to several million $C_\ell(z,z')$ spectra. For an MCMC parameter estimation this is simply prohibitive. Even if one spectrum is calculated within a few seconds, calculating the millions of spectra $\sim 10^5$ times would take months even if highly parallelized.
\item In each spectroscopic redshift bin we then only have a few 1000 galaxies, less than one per square degree, and the observed spectra would have very large shot noise $\propto 1/N$, allowing only computation up to very low $\ell$.
\item One of the big advantages of  $\xi(r,\mu)$ and $P(k,\kcos)$ is that the growth rate $f(z)$ can be simply determined by isolating the monopole, quadrupole and hexadecapole components in an expansion of $P$ and $\xi$  in Legendre polynomials in  $\mu$ and $\kcos$ respectively. With the $C_\ell$'s on the other hand there is no simple way to isolate redshift-space distortions since each multipole $\ell$ is a non-trivial combination of density and velocity.
\end{enumerate}

Hence even though the $C_\ell$'s are very convenient theoretically, they are not fully satisfactory from an observational point of view. In this paper we therefore derive general expressions for the correlation function and the power spectrum, that can be used as theoretical models for future surveys. Our work builds on the result of several papers, which have studied the impact of some of the relativistic effects on the correlation function and on the power spectrum. In~\cite{Jeong:2011as, Yoo:2012se}, expressions for the flat-sky power spectrum including all non-integrated relativistic effects have been derived. In~\cite{Hui:2007cu, LoVerde:2007ke, Hui:2007tm}  the lensing contribution to the flat-sky power spectrum and the flat-sky correlation function has been studied in detail. Refs.~\cite{Szalay:1997cc,Szapudi:2004gh,Papai:2008bd} have derived full-sky expressions for density and redshift-space (RSD) contributions to the correlation function, which have then be further developed in~\cite{Raccanelli:2010hk, Samushia:2011cs,Bertacca:2012tp, Yoo:2013zga}. These expressions have been re-derived using an alternative method in~\cite{Campagne:2017wec}. Ref.~\cite{Reimberg:2015jma} has studied in detail the relation between the full-sky and flat-sky density and RSD for both the correlation function and the power spectrum. In~\cite{Bonvin:2013ogt} the full-sky calculation of~\cite{Szalay:1997cc,Szapudi:2004gh,Papai:2008bd} has been extended  to include gravitational redshift and Doppler terms, which are especially relevant in the case of multiple populations of galaxies. Ref.~\cite{Bertacca:2012tp} further expands the formalism  introduced in~\cite{Szalay:1997cc} by computing theoretical expressions for the wide-angle corrections including also the integrated terms and Ref.~\cite{Raccanelli:2013dza} numerically evaluates all the non-integrated relativistic terms in the full-sky. In~\cite{Raccanelli:2013gja} the integrated terms in the correlation function are plotted for the first time for two values of the angle $\theta$. The theoretical expressions in these works rely on an expansion of the correlation function in Tripolar Spherical Harmonics which on the one hand is a powerful tool to obtain simple expressions in the full-sky but on the other hand hides some properties of the correlation function enforced by isotropy.\footnote{Whether in flat-sky or full-sky the correlation function depends on three variables: two distances and one angle ($\xi(\chi_1,\chi_2, \theta)$ or $\xi(\bar \chi,r, \cos\alpha)$ in this work), one distance and two angles ($\xi(\theta,\gamma,r)$ in \cite{Szalay:1997cc}, $\xi(\chi_2, \theta,\phi)$ in \cite{Raccanelli:2013dza}) or three distances ($\xi(\chi_1,\chi_2,r)$). When $\xi$ is expanded in Tripolar Spherical Harmonics one obtains a function $\xi(\bx_1,\bx_2)$ and the three physical variables are in general not directly inferred.}

Here we generalise and complete these results. We first derive a full-sky expression for the correlation function including all local and integrated contributions, in which isotropy of the perturbations is explicit. In particular, we provide a detailed study of the gravitational lensing contribution to the correlation function which does not rely on the flat-sky or Limber approximation. We  discuss how these full-sky contributions modify the simple multipole expansion of Eq.~\eqref{e:corrflat}. This represents the first analysis of the full-sky lensing contributions to the multipoles of the correlation function, which is most relevant when extracting the growth factor. In this aspect as in several other ways, this analysis goes beyond the pioneering work of~\cite{Raccanelli:2013gja}.

In the second part of this work we use the correlation function to calculate the power spectrum, which we define as the Fourier transform of the full-sky correlation function. In this way the power spectrum does not rely explicitly on the flat-sky approximation. However, it has an unambiguous interpretation only in this limit. Comparing the full-sky and flat-sky derivations, we find that relativistic effects and wide-angle corrections~\footnote{Here we call wide-angle corrections the difference between the flat-sky and full-sky expressions.} are of the same order of magnitude and they have therefore to be treated in conjunction. This leads us to the conclusion that relativistic effects cannot be consistently studied in the flat-sky and that the correlation function is therefore more adapted than the power spectrum to investigate these effects.

This paper is the first part of this study where we present the theoretical derivation and some numerical results. An exhaustive numerical study, including also the effects of the new terms on cosmological parameter estimation, is deferred to a future publication~\cite{prep}. Of course, there are many studies estimating cosmological parameters using the $C_\ell(z_1,z_2)$, see for example~\cite{DiDio:2013sea,Raccanelli:2015vla,Montanari:2015rga,Cardona:2016qxn,DiDio:2016ykq}. However as argued above, these can mainly be used for large, photometric redshift bins while within such bins, in order to profit optically from spectroscopic redshift information, a correlation function or power spectrum analysis is required.

The remainder of the present work is structured as follows: in the next section we describe how we obtain the redshift-space correlation function from the angular correlation function. As already discussed above, the procedure of course depends on the cosmological model. We shall describe two possibilities: to go either over the  $C_\ell(z_1,z_2)$ spectra or to obtain 
$\xi(r,\mu,\bar z)$ directly from the density fluctuations, velocity fluctuations and the Bardeen potentials in Fourier space. In Section~\ref{s:ps} we study the power spectrum. In Section~\ref{s:dis} we discuss the implications of our findings for future surveys and  we conclude. Several technical derivations are relegated to 5 appendices.

\section{The correlation function}\label{s:Corr}

The galaxy number counts including relativistic corrections have been derived in~\cite{Bonvin:2011bg,Challinor:2011bk} with the following result
\bea
\De_g(\bn,z)=\Delta^{\rm den}+\Delta^{\rm rsd}+\Delta^{\rm len}+\Delta^{\rm d1}+\Delta^{\rm d2}+\Delta^{\rm g1}+\Delta^{\rm g2}
+\Delta^{\rm g3}+\Delta^{\rm g4}+\Delta^{\rm g5}\, ,
 \label{e:Degz}
\eea
where
\bea
\Delta^{\rm den}&=&b\de_c(\chi(z)\bn,z)\, ,\label{e:Ddens}\\
\Delta^{\rm rsd}&=&-\HH^{-1}\partial_rv_r\, ,\\
\Delta^{\rm len}&=&\frac{5s-2}{2\chi}\int_0^{\chi(z)}d\la\frac{\chi-\la}{\la}\De_\Om(\Phi+\Psi) ,\\
\Delta^{\rm d1}&=&-\left(\frac{\dot\HH}{\HH^2} +\frac{2-5s}{\HH\chi}+5s-f_{\rm evo}\right)v_r\, ,\label{e:Dd1}\\
\Delta^{\rm d2}&=& -(3-f_{\rm evo})\HH v\, ,\\
\Delta^{\rm g1}&=& \left(1+\frac{\dot\HH}{\HH^2} +\frac{2-5s}{\HH\chi}+5s-f_{\rm evo}\right)\Psi\, ,\\
\Delta^{\rm g2}&=&(5s-2)\Phi\, ,\\
\Delta^{\rm g3}&=&\HH^{-1}\dot\Phi\, \label{e:Dg3} \\
\Delta^{\rm g4}&=&\frac{2-5s}{\chi}\int_0^{\chi(z)}d\la(\Phi+\Psi)\, ,\\
\Delta^{\rm g5}&=&\left(\frac{\dot\HH}{\HH^2} +\frac{2-5s}{\HH\chi}+5s-f_{\rm evo}\right)\int_0^{\chi(z)}d\la(\dot\Phi +\dot\Psi)\, . \label{e:Dg5}
\eea
Here $\de_c$ is the matter density fluctuation in comoving gauge,  $v_r$ is the radial component of the velocity in longitudinal gauge, $v$ is the velocity potential such that $\bv=-\nabla v$,  $v_r=-\partial_rv$; hence $v$ has the dimension of a length (we later define $V$ via its Fourier transform, $\hat v=k^{-1}V(k)$, so that $V(\bx)$ is dimensionless). $\Phi$ and $\Psi$ are the Bardeen potentials and $\De_\Om$ denotes the Lapacian on the sphere of directions $\bn$. The galaxy bias is denoted by $b$, $s$ is the magnification bias and $f_{\rm evo}$ is the evolution bias. These biases generally depend on redshift. The magnification bias $s$ comes from the fact that in general we do not observe all galaxies but only those which are brighter than the flux limit of our instrument. Due to lensing and to some relativistic effects, some fainter galaxies may make it into our surveys. This is taken into account by $s$ which is proportional to the logarithmic derivative of the galaxy luminosity function at the flux limit of our survey, see~\cite{Challinor:2011bk,DiDio:2013bqa} for more details.

The terms $\Delta^{\rm den}$ and $\Delta^{\rm rsd}$ are the density and redshift-space distortion terms usually taken into account. In the following we call the sum of these two terms the 'standard terms'. $\Delta^{\rm len}$ represents the lensing term, also often called magnification. This term has already been measured with quasars at large redshift, see e.g.~\cite{Menard:2009yb}, but it is usually neglected in galaxy surveys, since it is subdominant at low redshift. $\Delta^{\rm d1}$ is the Doppler contribution. Note that here we have used Euler's equation to derive this term. In all generality this term contains a contribution from gravitational redshift, proportional to $\partial_r\Psi/\HH$, which can be rewritten in terms of the velocity $v_r$ using Euler equation, see e.g.~\cite{Bonvin:2013ogt}. $\Delta^{\rm d2}$ is a velocity term which comes from transforming the longitudinal gauge density into the comoving density. $\Delta^{\rm g1}, \Delta^{\rm g2}$ and $\Delta^{\rm g3}$ are relativistic effects, given by the gravitational potentials at the source. As such they are sometimes called 'Sachs-Wolfe' terms. $\Delta^{\rm g4}$ denotes the so-called Shapiro time-delay contribution and $\Delta^{\rm g5}$ is the integrated Sachs-Wolfe term. 

In the following we will sometimes group together the relativistic non-integrated terms (d1, d2, g1, g2, g3). The lensing term is treated separately since its calculation is different. The relativistic integrated terms (g4 and g5) are neglected in our numerical results since their contribution is largely subdominant with respect to the lensing term.

\subsection{Using $C_\ell$'s}
\label{s:Cls}
We start by deriving the correlation function of~\eqref{e:Degz}, using the angular power spectrum $C_\ell$. Using Eqs.~(\ref{e:xiCl}) and  (\ref{e:rofztheta}) we can write
\be\label{e:corrztheta}
\xi(r,\bar z, \theta)  =\frac{1}{4\pi}  \sum_\ell (2\ell+1)C_\ell(\bar z-\De z,\bar z+\De z)L_\ell(\cos\theta) \,,
\ee
where $\De z$ is given by ($\bar H=H(\bar z)$, $\bar\chi=\chi(\bar z)$)
\be\label{e:Dez}
\De z(r,\bar z, \theta)= \frac{\bar H\sqrt{r^2-2\bar\chi^2(1-\cos\theta)}}{\sqrt{2(1+\cos\theta)}} \in \left[0, r\bar H/2\right] \,.
\ee
This is a simple consequence of \eqref{e:rofztheta} setting $z_{1,2}=\bar z\pm\De z$ and approximating $\chi_{1,2}=\chi(\bar z\pm\De z)\simeq  \chi(\bar z)\pm\De z/H(\bar z)$. This function is the same full correlation function as the one given in Eq.~(\ref{e:xiCl}), but now expressed in terms of the variables $r,\bar z$ and $\theta$ instead of $z_1,z_2$ and $\theta$. We shall use the same symbol $\xi$ to denote it.

Usually, the correlation function is not considered as a function of $r,\bar z$ and the opening angle $\theta$ between the two directions which are correlated, but as a function of   $r,\bar z$ and the angle with a fictitious but fixed line-of-sight between the two directions of observation. If $\theta$ is small enough, redshift-space distortions
are proportional to the $\cos^2$ of the angle with this fictitious direction. To mimic this
situation we  introduce 
\bea
r_\parallel &=&\chi_2-\chi_1\; \simeq \; 2\De z/H(\bar z)\; \leq \; r \,, \label{e:rpa}\\
\mu &=& \frac{r_\parallel}{r}\, , \qquad -1\leq \mu\leq 1 \qquad  \mbox{and} \qquad
r_\perp \;=\;\sqrt{r^2-r_\parallel^2} \,. \label{e:mu}
\eea
Writing $\bar\chi=(\chi_1+\chi_2)/2$  and using Eq.~\eqref{e:rpa} we obtain
\be
\cos\theta  = \frac{2\bar\chi^2-r^2+\frac{1}{2}\mu^2r^2}{2\bar\chi^2-\frac{1}{2}\mu^2r^2}
= \frac{2\bar\chi^2-r_\perp^2-\frac{1}{2}r_\parallel ^2}{2\bar\chi^2-\frac{1}{2}r_\parallel ^2} \equiv c(\bar z, r,\mu) \,.
\label{e:ctheta}
\ee
Note that $\bar\chi$ and $\chi(\bar z)$ are not exactly the same but in what follows we neglect this difference which is of order $(\De z)^2/\HH(\bar z)$. 
With this, the correlation function, $\xi(r,\bar z, \theta)$ can be written as a function of $\bar z$, $r_\parallel$ and $r_\perp$ (or, equivalently, $\bar z, r$ and $\mu$)
\bea
\xi(r_\pa,r_\perp,\bar z) &=& \frac{1}{4\pi}\sum_\ell (2\ell+1)C_\ell\!\!\left(\bar z- \frac{r_\pa \bar H}{2},\bar z+ \frac{r_\pa \bar H}{2}\right)
L_\ell\left( c(\bar z,r,\mu)\right)  \label{xirrz}\\
&=&\langle\De(\bx_1,\bar z-\De z)\De(\bx_2,\bar z+\De z)\rangle\,. \label{e:xitot}
\eea
Note that, again, we have re-expressed $\xi$ in different variables.

\begin{figure}[th]
\centering
\includegraphics[width=13.5cm]{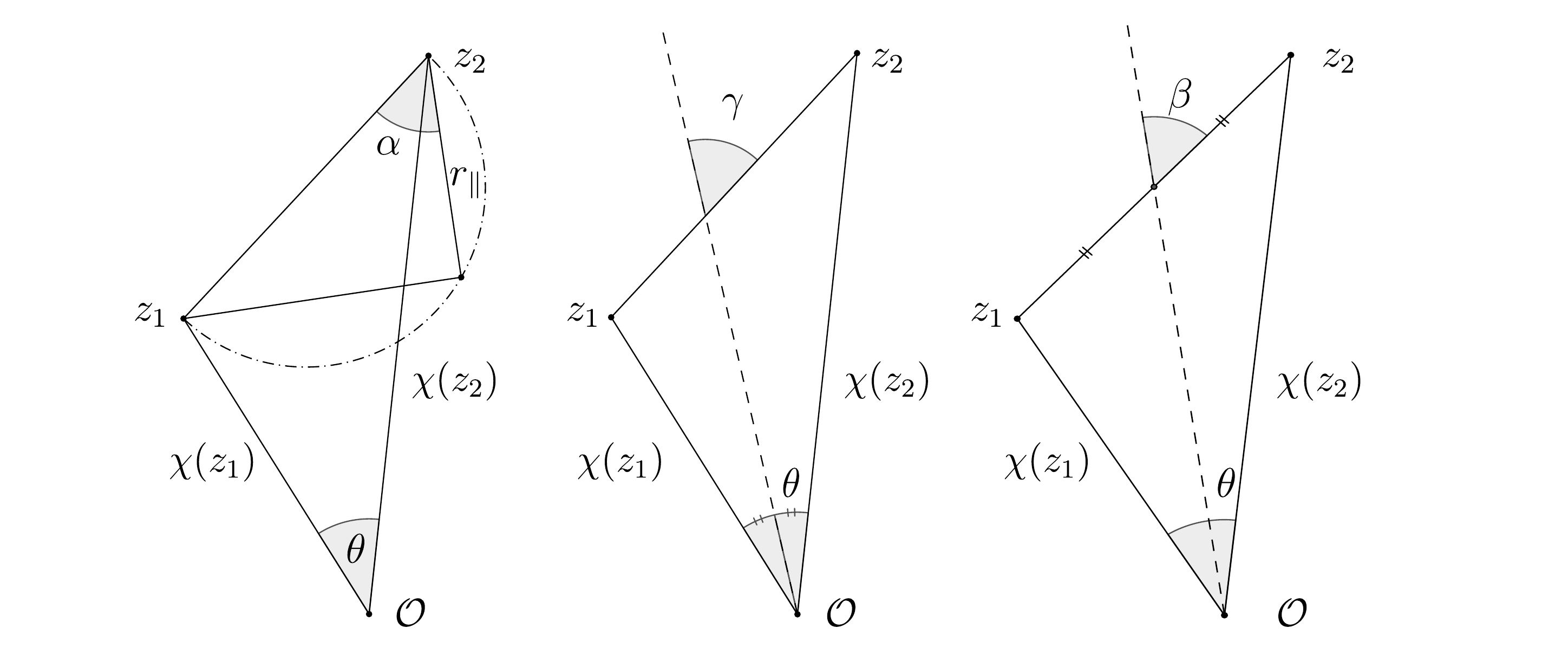}
\caption{\label{f:angles} The definitions of the angles $\al$ (left panel, $r_\pa=\chi_2-\chi_1$), $\ga$ (middle panel) and $\beta$ (right panel) as discussed in the text. }
\end{figure}

Expression \eqref{xirrz} is valid as long as $\De z$ is small so that $\De z\simeq  r_\pa H(\bar z)/2 = (\chi_2-\chi_1) H(\bar z)/2$ is a good approximation. Expression \eqref{e:xitot} however, is valid for all possible values of $r_\pa = \chi(\bar z+\De z)-\chi(\bar z-\De z)$ and $r=\sqrt{(\bx_1-\bx_2)^2}$, $r_\perp=\sqrt{r^2-r_\pa^2}$ where $\bx_1=\chi(\bar z-\De z)\bn_1$,  $\bx_2=\chi(\bar z+\De z)\bn_2$ such that $c(\bar z,r,\mu)=\bn_1\cdot\bn_2$. For a given cosmology, fixing $r_\pa$ and $\bar z$ is therefore equivalent to fixing $z_1$ and $z_2$ while $r_\perp$ then fixes $\cos\theta$.
Given a cosmological background model, there is a one-to-one correspondence between the model-independent angular correlation function \eqref{e:xiCl} and the model-dependent correlation function \eqref{e:xitot}.

The angle $\al$, given by $\mu=\cos\al$ defined by Eq.~\eqref{e:mu}, is the angle between the line $r$ connecting $\bx_1$ and $\bx_2$ and the line connecting the intersection of the circle around $\bx_2$ with radius $r_\pa=\mu r$ and the Thales circle over $r$ (see Fig.~\ref{f:angles}, left panel). This angle is not very intuitive and it is not what observers use. In practice the angles used are either $\beta$, the angle between $r$ and the line dividing $r$ into two equal halves (see Fig.~\ref{f:angles}, right panel)
or $\ga$, the angle between the line bisecting the angle $\theta$ and $r$ (see Fig.~\ref{f:angles}, middle panel). Using elementary geometry we can express the angles $\beta$ and $\ga$ in terms of $\theta$, $\chi_1$ and  $\chi_2$ (see Appendix~\ref{a:angles} for a derivation):
\begin{align}
&\cos\beta = \mu f_\beta(\theta,\chi_1,\chi_2)\,,  &  &\cos\gamma = \mu f_\ga(\theta,\chi_1,\chi_2)\,,\\
&f_\beta = \frac{\chi_1+\chi_2}{\sqrt{\chi_1^2+\chi_2^2+2\chi_1\chi_2\cos\theta}}\,, &  &f_\ga = \frac{\sqrt{1+\cos\theta}}{\sqrt{2}}\,.
   \label{e:theta-gamma}
\end{align}
In the small angle approximation, $\theta\ra 0$, both functions behave as
$$ f_{\beta,\gamma} = 1 + {\cal O}(\theta^2) \,.$$
 If $r_\pa\neq 0$, i.e. $\chi_1\neq \chi_2$, we can express $c(\bar z,r,\mu)$ in terms of $\bar z,r,\cos\beta$ as
 \be\label{e:costheta}
 c(\bar z,r,\cos\beta) = \frac{1}{2\chi_1\chi_2}\left[ \frac{(\chi_1^2-\chi_2^2)^2}{r^2\cos^2\beta} -\chi_1^2 -\chi_2^2
\right]\,. \ee
 Here $\chi_{1,2}$ are given in terms of $\bar\chi$ and $r$ by solving the equations 
 \bea
  \bar \chi &=&  (\chi_1+\chi_2)/2\qquad \mbox{and} \qquad
r^2\;=\; \chi_1^2+\chi_2^2-2\chi_1\chi_2\cos\theta\,. \label{e:rchi12} \label{e:barchi}
\eea
 If we want to express the correlation function in terms of $\bar z$, $r$ and $\cos\beta$, we have to solve the system  (\ref{e:costheta},\ref{e:barchi}). A short calculation gives
\bea
 \cos\theta &=&1-\frac{ 8r^2\bar\chi^2(1- \cos^2\beta)}{16  \bar\chi^4- r^2\cos^2\beta(8\bar\chi^2-r^2)}\,,\label{e:thetabeta} \qquad
 \chi_{1,2} \;=\; \bar\chi \pm \sqrt{\bar\chi^2 -\frac{4\bar\chi^2-r^2}{2(1+\cos\theta)}}\,,
\qquad  \label{e:chi12}\\
 r_\pa &=& \chi_2-\chi_1 = 2\sqrt{\bar\chi^2 -\frac{4\bar\chi^2-r^2}{2(1+\cos\theta)}}\, .
 \label{e:rpab}
 \eea
 Inserting $\cos\theta$ from \eqref{e:thetabeta} and $r_\pa$ from \eqref{e:rpab} in \eqref{xirrz}, we can express the correlation function as a function of $r,\bar z$ and $\cos\beta$. In terms of $\ga$ we find
\bea
\cos\theta &=& 1-\frac{r^2}{2\bar\chi^2}(1-\cos^2\ga) \,.
\eea
In the small angle limit, all three angles, $\al$, $\beta$ and $\ga$ coincide.
In Section~\ref{sec:corr_mult} we will see that the angle which gives the result closest to the flat-sky limit is the angle $\mu$. For this reason and
due to its simplicity in what follows  we express both, the correlation function and the power spectrum in terms of the projection along and transverse to the line-of-sight using the angle $\al$ with $\cos\al=\mu=(\chi_2-\chi_1)/r = r_\pa/r$. As explained above, for small angles this is equivalent to choosing $\beta$ or $\gamma$, but for large angles, the expressions in terms of $\mu$ are simpler.

\begin{figure}[t]
\centering
\setlength{\unitlength}{\textwidth} 
\vspace{-3.2cm}
\begin{picture}(1,0.5)       
\put(0,0){\includegraphics[width=1\unitlength]{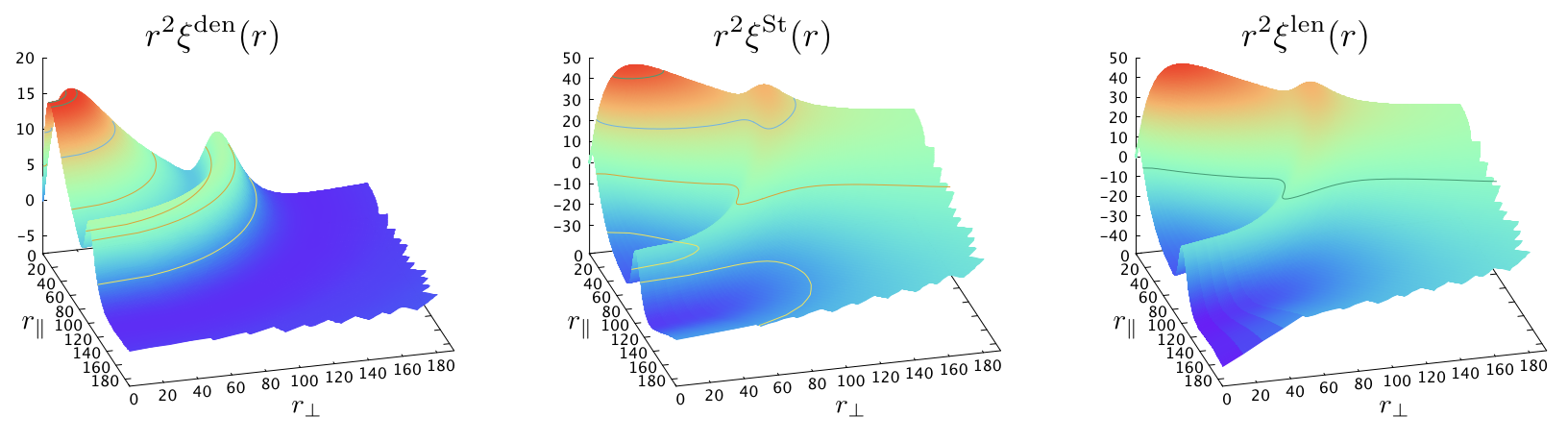}}
\end{picture}
\caption{\label{f:c1} The correlation function at redshift $\bar z=1$ as a function of $r_\parallel$ and $r_\perp$. The left panel contains only the density contribution, $\xi^{\text{den}}$, the middle panel contains also RSD, $\xi^{\text{st}}$, and the right panel contains also the lensing term, $\xi^{\text{st+len}}$.}
\end{figure}

\begin{figure}[t]
\centering
\includegraphics[scale=0.51]{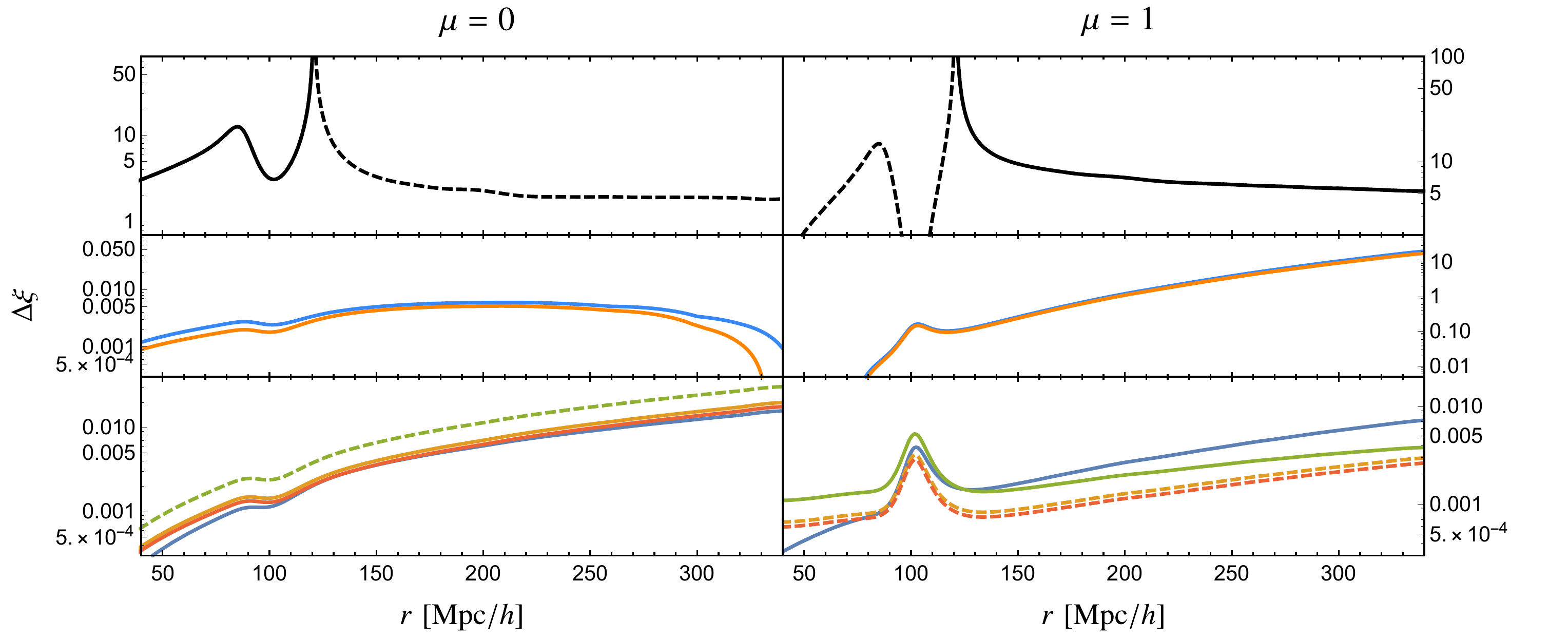}
\caption{\label{f:c2} The relative difference $\De \xi$ at redshift $\bar z=1$ for $\mu=0$ (left panels) and $\mu=1$ (right panels). \emph{Top panels:} $\De\xi^{\rm rsd}=(\xi^{\rm st}-\xi^{\rm den})/\xi^{\rm den}$. \emph{Middle panels:}  fractional difference induced by lensing  $\De \xi ^{\rm lensing}$ (full-sky in orange and flat-sky in blue). \emph{Bottom panels:} $\De \xi^A$ where $A=$: d1 (blue), d2 (orange), g1 (green) and g2 (red). The contribution g3 is subdominant (see Eqs.~\eqref{e:Degz} to \eqref{e:Dg5} for a definition of the various relativistic terms). Negative contributions are dashed.}
\end{figure}

In Fig.~\ref{f:c1} we show  the correlation function at $\bar z=1$ as a function of $r_\parallel$ and $r_\perp$. In all figures, we use the cosmological parameters: $h^2 \Omega_m= 0.14$, $h^2 \Omega_b= 0.022$, $h=0.676$, $A_s= 2.215 \times 10^{-9}$ at $k_*= 0.05 \,\, \text{Mpc}^{-1}$, $n_s=0.961$, $b(z)=1$, $f_{\text{evo}}=0$ and $s=0$ unless otherwise stated.
In the left panel of Fig.~\ref{f:c1} we include only the density, in the middle panel we also consider redshift-space distortions (RSD) and in the right panel we include also the lensing term.  While the pure density term is spherically symmetric with a well visible baryon acoustic oscillation (BAO) feature at $r\sim 100$Mpc, the RSD removes power for small $r_\perp$ and adds power at large $r_\perp$. Also the maximal amplitude has more than doubled due to RSD~\footnote{Note that we have chosen $b=1$. For larger values of $b$, the importance of redshift-space distortion with respect to the density contribution is reduced.}. Finally the lensing term adds a very significant amount of power for large $r_\pa$ and small $r_\perp$. This is the case when a foreground density fluctuations lenses a structure at higher redshift along its line of sight.  The additional relativistic contributions are very small and become visible only on very large scales, as we shall see in the rest of this paper and as has already been anticipated in several papers, e.g.~Refs.~\cite{Bonvin:2011bg,Challinor:2011bk}.

In Fig.~\ref{f:c2} we show  fractional differences for $\mu=0$ (left) and $\mu=1$ (right)
\be
\De \xi^A\equiv \frac{\xi^A-\xi^{\rm st}}{\xi^{\rm st}}\,,
\ee
where 
\be
\xi^A=\big\langle\big(\Delta^{\rm st}+\Delta^A\big)(\bn_1, z_1)\big(\Delta^{\rm st}+\Delta^A\big)(\bn_2, z_2) \big\rangle\, .
\ee
In this way we show separately the contribution of each correction $A$ with respect to the standard term, including its correlation with density and redshift-space distortion. The middle panel shows $\De \xi^A$ for $A={\rm lensing}$ and the lower panel for all the non-integrated relativistic effects, namely the terms d1, d2, g1, g2 and g3 (see Eqs.~\eqref{e:Degz} to \eqref{e:Dg5} for a definition of the various relativistic terms). Finally, as reference, we plot in the top panel the fractional difference due to redshift-space distortion, namely $\De\xi^{\rm rsd}=(\xi^{\rm st}-\xi^{\rm den})/\xi^{\rm den}$. 


Not surprisingly, for $\mu=0$ the lensing term is very small apart from a small effect on the acoustic peaks. For $\mu=1$ however, at large scales $r>150$\,Mpc, lensing becomes the dominant term. As also noted in~\cite{LoVerde:2007ke}, it increases linearly with distance. Comparing our full-sky calculation of the lensing (orange) with the flat-sky expression (blue) derived in~\cite{LoVerde:2007ke} and in Appendix~\ref{a:flat} (see Eq.~\eqref{e:xifinalflat}) we see that for $\mu=1$ the two expressions agree very well, which is not surprising because in this case $\bn_1=\bn_2$ and flat-sky is a good approximation. The only source of difference in this case comes from the fact that the flat-sky result uses Limber approximation whereas the full-sky result is exact. This difference is very small, showing that Limber approximation for $\mu=1$ is very good. For $\mu=0$ on the other hand we see a non-negligible difference between the flat-sky and full-sky result. We will discuss this in more detail in Section~\ref{s:mu-r}. 

From the bottom panel, we see that the non-integrated relativistic terms generate a correction of the order of the percent at large separation $r\sim 350\,$Mpc$/h$. Naively we would expect the Doppler term (d1: blue) to dominate over the other relativistic effects because it is proportional to the peculiar velocity and contains therefore one more factor $k/\HH$ than the terms proportional to the potentials (see e.g. Eqs.~\eqref{e:den} to \eqref{e:g5} below). However, as shown in~\cite{Bonvin:2013ogt} (see also Appendix~\ref{a:corf}), the correlation of this term with the standard term $\langle \Delta^{\rm d1}\Delta^{\rm st}\rangle$ exactly vanishes in the flat-sky because it is totally anti-symmetric. The contribution that we see in Fig.~\ref{f:c2} is therefore due to the correlation $\langle \Delta^{\rm d1}\Delta^{\rm d1}\rangle$, which is a factor $\HH/k$ smaller, hence $\sim \langle \Delta^{\rm st}\Psi\rangle$ and to the full-sky contributions to $\langle \Delta^{\rm d1}\Delta^{\rm st}\rangle$, which are of the order 
$r/\chi\langle \Delta^{\rm d1}\Delta^{\rm st}\rangle\sim \langle \Delta^{\rm d1}\Delta^{\rm d1}\rangle\sim\langle \Delta^{\rm st}\Psi\rangle$. Consequently, with one population of galaxies the Doppler contribution to the correlation function is of the same order of magnitude as the gravitational potential contributions (d2, g1 and g2). Only in the case where one cross-correlates two populations of galaxies, the Doppler contribution strongly dominates over the other relativistic contributions, because in this case $\langle \Delta^{\rm d1}\Delta^{\rm st}\rangle$ does not vanish in the flat-sky.

For $\mu=0$, the Sachs-Wolfe like term (g1) dominates over the other corrections at all scales. For $\mu=1$ this term still dominates at small separation, but at large separation the full-sky corrections to the Doppler term become important and dominates over g1.
Interestingly the second Sachs-Wolfe like term (g2) and the second Doppler term (d2) are nearly equal for both values of $\mu$. 
It is easy to derive from the continuity and the Poisson equations that in a matter dominated Universe $(\HH/k)V=-(2/3)\Phi$, hence $\De^\text{d2}_\ell=\De^\text{g2}_\ell$ if $s=0$, see Eqs.~\eqref{e:d2} and \eqref{e:g2}. At lower redshifts, when $\Lambda$-domination sets in, we expect this equality to be less precise. 
The relativistic terms not shown in Fig.~\ref{f:c2}  are the Shapiro time delay (g4) and the integrated Sachs-Wolfe term (g5). These integrated terms are always subdominant with respect to the lensing term.

Let us also note that the difference between the flat-sky standard term and the full-sky standard term is of the same order of magnitude as the relativistic terms depicted in the bottom panel of Fig.~\ref{f:c2}. It is therefore not consistent to use the flat-sky approximation for the standard terms when investigating relativistic effects.

Finally we should point out that in this work we present the \emph{theoretical} contributions of relativistic effects on the correlation function and the power spectrum (see Figs.~\ref{f:c2}, \ref{f:Delta_rel_mult}, \ref{f:Delta_lens_mult}, \ref{f:Pfracnonint} and \ref{f:Pfracint}). To estimate the \emph{observational} impact of these terms one should build a realistic estimator and  proceed with signal-to-noise analysis, forecasts and constraints for a specific survey. Such studies have been performed for the angular power spectrum $C_\ell$ in~\cite{DiDio:2013sea,Raccanelli:2015vla, Raccanelli:2016avd, Cardona:2016qxn, DiDio:2016ykq, Lorenz:2017iez} and for the antisymmetric part of the correlation function $\xi_g$ in~\cite{Bonvin:2015kuc}. In a future work~\cite{prep}, we will develop this for the multipoles of the correlation function and the power spectrum. This will allow us to compare the observational impact of the relativistic effects on the angular power spectrum with their impact on the multipoles of the correlation function and power spectrum, which are the standard observables currently used in large-scale structure surveys to measure the growth rate $f$.

\subsection{Direct determination of the correlation function}
\label{ss:direct}

In the calculation of the correlation function presented in the previous section, we still need all the $C_\ell(z_1,z_2)$ for an accurate calculation. Hence the reason (1) given in the introduction for the use of the correlation function and the power spectrum is not satisfied: the calculation is not simplified. To compute the correlation function for thousands of spectroscopic redshifts in an MCMC would still take months even if very highly parallelised. In this section we show how to improve this. The method explained in this section reduces the calculation of several thousand $C_\ell(z_1,z_2)$'s into 
just several terms. This results in a very significant speed up so that the computation becomes feasible.
 
We expand on a method introduced in~\cite{Campagne:2017wec} which avoids the computation of $C_\ell(z_1,z_2)$  but requires integrations in $k$-space and over the line-of-sight, as we shall see.  In this method, no flat-sky approximation is performed, and the correlation function is therefore exact, within linear perturbation theory. We start from expression \eqref{e:xiCl} for the correlation function and use that the $C_\ell(z_1,z_2)$ are of the form (see~\cite{DiDio:2013bqa}),
\bea
C_\ell(z_1,z_2) &=& \sum_{A,B}C_\ell^{AB} (z_1,z_2) \,, \qquad 
C_\ell^{AB} (z_1,z_2) \,=\, 4 \pi\hspace{-1.2mm} \int\! \frac{ dk }{k} \mathcal{P}_{\mathcal{R}}(k) \De_\ell^A(k,z_1)\De_\ell^B(k,z_2)\,. \qquad ~~
\eea
Here $ \mathcal{P}_{\mathcal{R}}$ denotes the primordial power spectrum, determined by the amplitude $A_s$ and the primordial spectral index $n_s$:
 $$ 
 \mathcal{P}_{\mathcal{R}}(k) = \frac{1}{2 \pi^2} A_s \left(\frac{k}{k_*}\right)^{n_s-1}\, ,
 $$ 
 and 
 $\De^A_\ell$, $\De^B_\ell$ are the Fourier-Bessel transforms of the terms defined in \eqref{e:Ddens} to  \eqref{e:Dg5}. More precisely
 \bea
\De_\ell^\text{den}&=&  b(z) S_D j_\ell( k \chi ) \,,\label{e:den}\\
\De_\ell^\text{rsd}&=&  \frac{k}{\HH} S_V j_\ell''(k \chi)\,, \label{e:rsd}\\
\De_\ell^\text{len}&=& \left(\frac{2 - 5 s}{2}\right)\frac{\ell(\ell+1)}{\chi} \int_0^\chi \dd \la  \frac{\chi-\la}{\la} (S_\phi+S_\psi) j_\ell( k \la) \,,\label{e:len}\\
\De_\ell^\text{d1}&=&  \left(\frac{\dot \HH}{\HH^2}+\frac{2-5s}{\chi \HH}+5 s - f_\text{evo} \right) S_V j_\ell'(k \chi) \,,\label{e:d1}\\
\De_\ell^\text{d2}&=& -(3-f_{\rm evo})\frac{\HH}{k} S_V j_\ell(k \chi) =\De^\text{d2}(z,k)j_\ell(k \chi)  \,,\label{e:d2}\\
\De_\ell^\text{g1}&=&  \left(\!\!1+\frac{\dot \HH}{\HH^2}+\frac{2-5s}{\chi \HH}+5 s - f_\text{evo} \!\right)\! S_\psi j_\ell(k \chi)  = \De^\text{g1}(z,k)j_\ell(k \chi) \,,\label{e:g1}\\
\De_\ell^\text{g2}&=& (-2+5s)S_\phi j_\ell(k \chi) =\De^\text{g2}(z,k)j_\ell(k \chi) \,,\label{e:g2} \\
\De_\ell^\text{g3}&=& \frac{1}{\HH} \dot S_\phi j_\ell(k \chi) =\De^\text{g3}(z,k)j_\ell(k \chi) \,,  \label{e:g3}\\
\De_\ell^\text{g4}&=&  \frac{ 2-5s}{\chi}  \int_0^\chi \dd \la(S_\phi +S_\psi) j_\ell(k \la)     \,, \label{e:g4}\\
\De_\ell^\text{g5}&=&  \left(\frac{\dot \HH}{\HH^2}+\frac{2-5s}{\chi \HH}+5 s - f_\text{evo} \right) \int_0^\chi \dd \la   (\dot S_\phi +\dot S_\psi) j_\ell(k \la) \,.\label{e:g5}
\eea
Here $j_\ell$ are the spherical Bessel functions and the functions $S_X(z,k)$ are the transfer functions for the variable $X$ which we specify in Appendix~\ref{a:corf}. Over-dots indicate derivatives with respect to conformal time. For the evolution bias $f_\text{evo}$, the magnification bias $s$ and the galaxy bias $b$ we follow the conventions of~\cite{DiDio:2013bqa}. From these expressions one also infers the scaling of the different terms with respect to the density term. On sub-Hubble scales, $k>\HH$, the scaling of these terms with powers of $\HH/k$ is a simple consequence of Newtonian physics. The continuity equation implies $S_V \sim (\HH/k)S_D$ and the Poisson equation yields $S_\phi \sim S_\psi \sim 
(\HH/k)^2S_D$, we see that the density, RSD and lensing terms dominate, while the Doppler term d1 is suppressed by one factor of $ (\HH/k)$, and all other terms are suppressed by $(\HH/k)^2$. For this reason all relativistic terms apart from lensing are strongly suppressed on sub-horizon scales and we call them 'large-scale contributions'. Most of them are relevant only on very large scales close to $\HH(z)^{-1}$. Exceptions to this rule are $\De_\ell^\text{d1}$ and $\De_\ell^\text{g1}$ which contain a pre-factor $1/(\chi\HH)$ which becomes large at very low redshift where $\chi$ is small.
On super horizon scales all the transfer functions $S_X$ are typically of the same order but they become gauge dependent.

Using these expressions, the correlation function $\xi$ can be written as
\be
\xi = \sum_{A,B}\xi^{AB} \qquad \mbox{with }\qquad \xi^{AB}(\theta,z_1,z_2) = \int \frac{dk}{k} \, \mathcal{P}_{\mathcal{R}} \,\,Q^{AB}_k (\theta,z_1,z_2) \label{corrQ} \,,
\ee
where we define 
\be
Q^{AB}_k (\theta,z_1,z_2) \equiv \sum\limits_\ell (2\ell+1) \De_\ell^A(k,z_1)\De_\ell^B(k,z_2) L_\ell(\cos \theta) \,.
\ee
In most of the terms $Q^{AB}_k$ we have a sum of the form
\bea
\sum_{\ell}(2\ell+1)L_\ell(\cos\theta)j_\ell(k\chi_1)j_\ell(k\chi_2) &=&j_0(kr)\, ,
\label{e:pljlj0}
\eea
where $r=\sqrt{\chi_1^2+\chi_2^2-2\chi_1\chi_2\cos\theta}$ (see e.g. \cite{Abram} (10.1.45)). Inserting~\eqref{e:pljlj0} into~\eqref{corrQ} we  can easily calculate the correlation function for these terms avoiding the numerically costly sum over the $C_\ell$'s. The redshift-space distortion and the Doppler term give rise to contributions that are slightly different because they contain first and second derivatives of the spherical Bessel functions with respect to $k\chi_1$ and $k\chi_2$. These terms can however be treated in a very similar way using recurrence relations for the spherical Bessel function. For this we define
\be 
\zeta^{ij} \equiv  \sum\limits_\ell (2\ell+1) j_\ell^{(i)}(k \chi_1) j_\ell^{(j)}(k \chi_2) L_\ell(\cos \theta) =  \sum\limits_\ell (2\ell+1) j_\ell^{(i)}(x_1) j_\ell^{(j)}(x_2) L_\ell(\cos \theta)\,,
\ee
where we have set $x_i = k \chi_i$ and $ j_\ell^{(i)}(x) = \frac{\pd^i}{\pd x^i} j_\ell(x)$. Using 
\be
\zeta^{ij}(x_1,x_2) = \zeta^{ji}(x_2,x_1) \quad\mbox{and}\quad \frac{\pd^{n+m}}{\pd x_1^n \pd x_2^m } \zeta ^{ij} = \zeta^{i+n,j+m} ,\nonumber
\ee
we can determine explicit expressions for the $\zeta^{ij}$ for $i,j\in \{0,1,2\}$. They are all given in Appendix~\ref{a:corf}.

The only coefficients that do not fall into this category are the ones in $\De_\ell^\text{len}$ which contain additional factors $\ell$ and $(\ell+1)$ (see Eq.~\eqref{e:len}). These terms can however be computed using the identity  
$$\bigtriangleup_\Omega L_\ell (\cos \theta)= -\ell(\ell+1) L_\ell (\cos \theta)\,.$$
They are given by
\be
\zeta^{\text{LL}} \equiv \sum\limits_\ell (2\ell+1) \ell^2 (\ell+1)^2 j_\ell (x_1) j_\ell(x_2) L_\ell(\cos \theta) = \bigtriangleup_\Omega^2 \,\zeta^{00} \,,
\ee
\be
\zeta^{i\text{L}} \equiv \sum\limits_\ell (2\ell+1)\ell(\ell+1) j_\ell^{(i)}(x_1) j_\ell (x_2) L_\ell(\cos \theta) = - \bigtriangleup_\Omega  \zeta^{i0} \,,
\ee
where LL denotes the correlation of lensing with itself and $i$L the cross-correlation of lensing with one of the other terms. With this we can build all the functions $Q^{AB}_k$ and hence, with Eq.~(\ref{corrQ}), the correlation function. The complete list of $Q^{AB}_k$ is given in Appendix~\ref{a:corf}. Here we just report the dominant contributions, i.e. the contributions which are not suppressed with additional powers of $\HH/k$ with respect to the density term:
\begin{flalign}
&Q^\text{den}(\theta,z_1,z_2) = b(z_1) b(z_2)  S_D(z_1) S_D (z_2) \,\zeta^{00}(k\chi_1,k\chi_2) \,,&& \nonumber\\
&Q^\text{rsd}(\theta,z_1,z_2) =  \frac{k^2}{\HH_1\HH_2} S_V(z_1)S_V(z_2)  \,\zeta^{22}(k\chi_1,k\chi_2)  \,,\nonumber\\
&Q^\text{len}(\theta,z_1,z_2) = \frac{\left(2 - 5 s\right)^2}{4 \chi_1\chi_2} \int_0^{\chi_1} \! \int_0^{\chi_2} \!\!\dd \la \dd \la'  \bigg[ \frac{(\chi_1-\la)(\chi_2-\la')}{\la\la'} S_{\phi+\psi}(\la) S_{\phi+\psi}(\la') \zeta^{LL}(k\la,k\la') \bigg] \,,  \nonumber\\
&Q^\text{den-rsd}(\theta,z_1,z_2) = \frac{k b(z_1)}{\HH_2}  S_D(z_1)S_V(z_2)  \,\zeta^{02}(k\chi_1,k\chi_2) \,, \nonumber\\
&Q^\text{den-len}(\theta,z_1,z_2) =  b(z_1) S_D(z_1) \left(\frac{2 - 5 s}{2\chi_2}\right) \int_0^{\chi_2} \dd \la \bigg[ \frac{\chi_2-\la}{\la}  \Bigl(S_\phi(\la)+S_\psi(\la)\Bigr)  \, \zeta^{0L}(k\chi_1,k\la) \bigg]\,,  \nonumber\\
&Q^\text{rsd-len}(\theta,z_1,z_2) =   \frac{k}{\HH_1}  S_V(z_1) \left(\frac{2 - 5 s}{2\chi_2}\right) \int_0^{\chi_2} \dd \la \bigg[  \frac{\chi_2-\la}{\la}  \Bigl(S_\phi(\la)+S_\psi(\la)\Bigr)  \, \zeta^{2L}(k\chi_1,k\la) \bigg]\,.  \nonumber
\end{flalign}
Note that here and in the following we suppress the argument $\theta$ in the functions $\zeta^{\rm AB}(k\chi_1, k\chi_2, \theta)$ for simplicity.
The correlation function is then given by Eq.~\eqref{corrQ}. For example, the correlation function including only the standard terms  is given by
\be\begin{split}
\xi^{\text{st}}=& \int \frac{dk}{k} \, \mathcal{P}_{\mathcal{R}} \bigg[Q^{\text{den}}(\theta,z_1,z_2) +Q^{\text{den-rsd}}(\theta,z_1,z_2) +Q^{\text{rsd-den}}(\theta,z_1,z_2) +Q^{\text{rsd}}(\theta,z_1,z_2)\bigg] \\
=&  \frac{2 A_s}{9 \pi^2 \Omega_m^2} D_1(z_1)D_1z_2)  \int \frac{dk}{k} \bigg[ b(z_1) b(z_2) \zeta^{00} (k\chi_1,k \chi_2) - b(z_1)f(z_2) \zeta^{02} (k\chi_1,k \chi_2) \\& - b(z_2)f(z_1) \zeta^{02} (k\chi_2,k \chi_1) + f(z_1)f(z_2) \zeta^{22} (k\chi_1,k \chi_2) \bigg] \left(\frac{k}{H_0} \right)^4 \left(\frac{k}{k_*}\right)^{n_s-1}T^2(k) \,. 
\end{split}\label{corrstme}\ee
For the second equal sign we made use of the transfer functions given in Appendix~\ref{a:corf}.  Eq.~\eqref{corrstme} is expressed in terms of the redshift $z_1$ and $z_2$ and the angle $\theta$. It can however easily be written in terms of a mean redshift $\bar z$, the separation of the galaxies $r$ and the orientation of the pair using Eqs.~(\ref{e:rpa}),(\ref{e:mu}),(\ref{e:ctheta}). 

The correlation function obtained in this way agrees with the full-sky expressions derived in~\cite{Szalay:1997cc,Szapudi:2004gh,Papai:2008bd} for the standard terms and in~\cite{Bonvin:2013ogt} for the Doppler term. This method has however the advantage that it can be used to calculate also expressions for the integrated terms valid in the full-sky. Since the lensing is the dominant correction, it is important to have an accurate expression for this term valid at all scales and not relying on the Limber approximation.

\subsubsection{$\mu$ and $r$ dependence of the correlation function}
\label{s:mu-r}

\begin{figure}[t]
\centering
\includegraphics[scale=0.49]{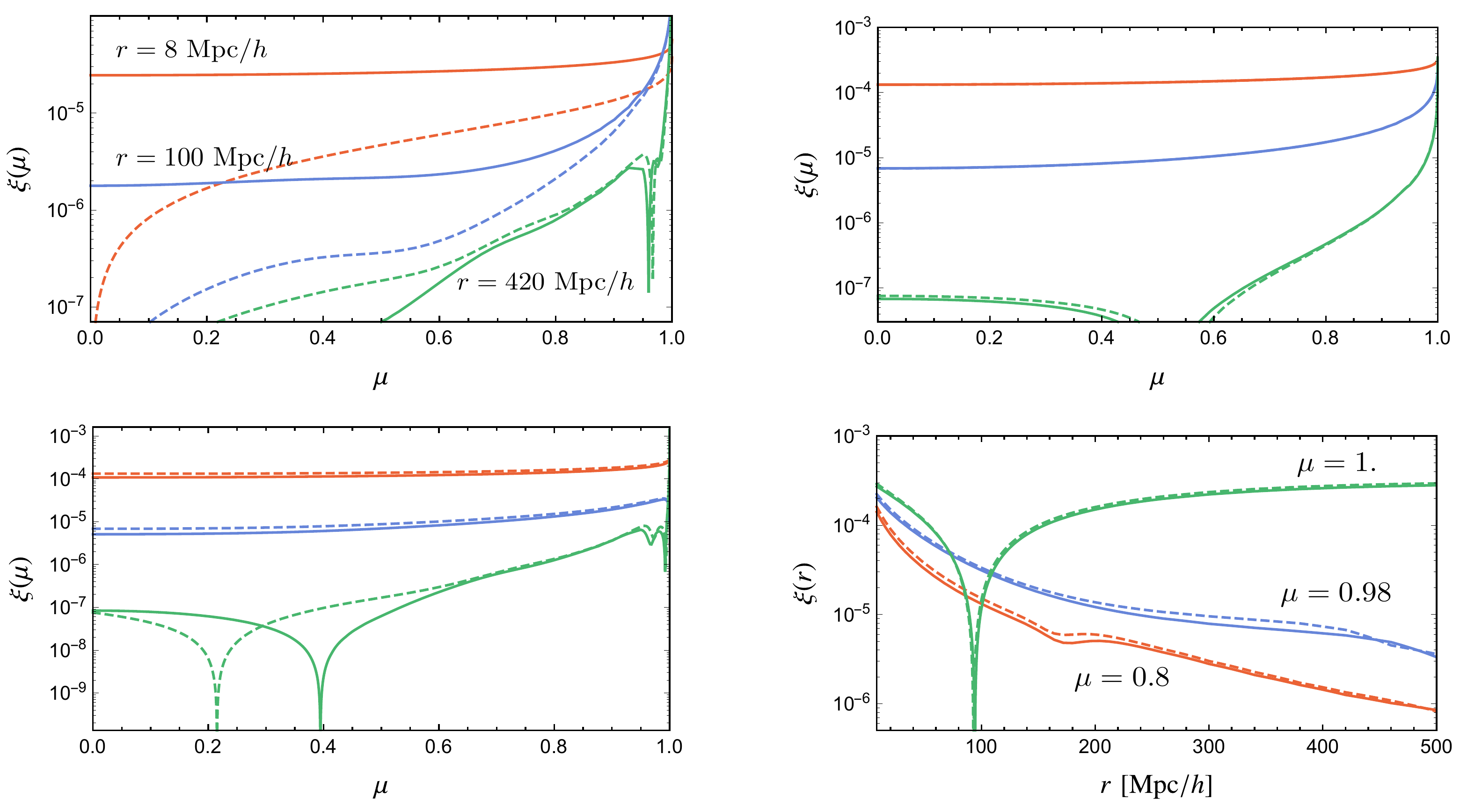}
\caption{\label{f:lensing_linear} Top panels: correlation between density and lensing (left) and lensing-lensing (right) at $\bar z=1$, as a function of $\mu$ and for fixed separation $r=8$\,Mpc/$h$ (orange), $r=100$\,Mpc/$h$ (blue) and $r=420$\,Mpc/$h$ (green). Solid lines show the full-sky result and dashed lines the flat-sky result using Limber approximation. Bottom panels: total lensing contribution as a function of $\mu$ (left) and $r$ (right) at $\bar z=1$.}
\end{figure}

Let us first discuss the full-sky correlation function as a function of $\mu$ and $r$. In Fig.~\ref{f:lensing_linear} we show the lensing contribution
\be
\xi^{\rm lensing}=\big\langle\big(\Delta^{\rm st}+\Delta^{\rm lensing}\big)(\bn_1, z_1)\big(\Delta^{\rm st}+\Delta^{\rm lensing}\big)(\bn_2, z_2) \big\rangle - \big\langle\Delta^{\rm st}(\bn_1, z_1)\Delta^{\rm st}(\bn_2, z_2) \big\rangle\, ,
\ee
as a function of $\mu$ and $r$. We compare the full-sky result (solid lines) with the flat-sky result (dashed lines) derived in~\cite{Hui:2007cu} and given in Eq.~\eqref{e:xifinalflat}. In the top left panel we show the cross-correlation between density and lensing, whereas in the top right panel we show the lensing-lensing correlation. We see that the flat-sky expression for the lensing-lensing agrees extremely well with the full-sky expression. The density-lensing cross-correlation is however significantly different in the flat-sky and full-sky, even at small separation. This can be understood in the following way. The flat-sky result assumes not only that $\bn_1=\bn_2$, but it also uses the Limber approximation, which implies that only correlations at the same redshift contribute to the correlation function. Hence instead of integrating the lensing along the line-of-sight as is done in the full-sky expression, the flat-sky expression correlates the density at position $z_2$ with the lensing from the same redshift. This can be seen by looking at Eq.~\eqref{e:xiif2}, where the integral along the line-of-sight has been replaced by the function $\delta(\chi_2-\lambda)$. This approximation is quite good for values of $\mu$ close to 1, i.e. when the galaxies are behind each other, but it is very bad when $\mu$ becomes small and for small separations $r$. In such cases, the density $\delta$ is correlated with the gravitational potentials generated by that same density $\Phi$ and $\Psi$ and therefore the correlation is non-negligible even when the two redshifts are not exactly the same. As a result the flat-sky expression, which ignores this direct correlation, strongly underestimates the density-lensing correlation. Since the density-lensing cross-correlation is negative whereas the lensing-lensing is positive, this means that the flat-sky result overestimates the total correlation function, as shown in the bottom left panel of Fig.~\ref{f:lensing_linear}. The bottom right panel shows the total lensing contribution as a function of separation for various values of $\mu$. In general we find that the relative difference between the flat-sky and full-sky result is of the order of 20 percents and it can become much larger in some configurations.

In all these plots we do not calculate the lensing contribution when $\mu$ is exactly equal to 1. This value is indeed not physical since it would correspond to a galaxy situated exactly behind the other, which we can of course not see. Numerically this value is also problematic because it requires the computation of the correlation function between points that are exactly at the same position. This correlation function diverges if one uses the linear power spectrum and it has to be regularised by non-linear effects which suppress the power spectrum on very small scales, where fluctuations are damped. The largest value that we take is therefore $\mu=0.9997895$. This value ensures us that the line-of-sight from the most distant galaxy passes sufficiently far away from the closest galaxy to avoid being absorbed by it. In the following when we discuss about the parallel correlation function or when we show plots for $\mu=1$, this has to be understood as $\mu=0.9997895$. Finally let us mention that we do not include the correlation between redshift-space distortion and lensing. This correlation is exactly zero in the flat-sky approximation and we do expect it to remain very small in the full-sky \footnote{We have checked numerically that at $\bar z\sim 1$ the RSD-lens contribution to the angular power spectrum is 3 to 4 orders of magnitude smaller than the $\de$-lens term.}.

So far we have calculated all the flat-sky and full-sky correlation functions using the linear power spectrum. Since we are mainly interested in correlations at large separations, this is a very well motivated approximation for all the non-integrated terms. We have indeed checked that all the large-scale relativistic contributions change by at most 2-3 percents at small scales if we use the halo-fit power spectrum instead of the linear one to calculate the correlation function. For the lensing contribution on the other hand, non-linearities are important even at large separation, as already pointed out in~\cite{Hui:2007cu, LoVerde:2007ke, Hui:2007tm}. This is due to the fact that lensing is sensitive not only to correlations between the two positions of the galaxy, but also to all correlations between the two lines-of-sight from these galaxies. When $\mu$ is large, these two lines-of-sight are close to each other at least in the vicinity of the observer, even when $r$ is large, and consequently non-linear effects are important. Lensing has the property to mix large and small separations and a full-sky non-linear treatment is therefore necessary. 

\begin{figure}[t]
\centering
\includegraphics[scale=0.5]{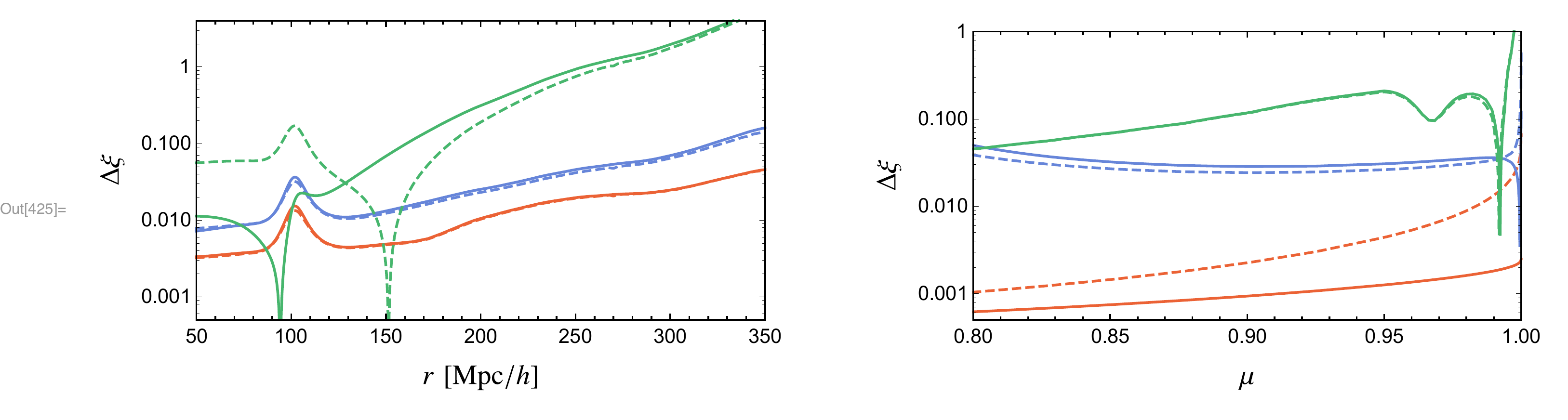}
\caption{\label{f:lensing_nonlinear} Fractional differences $\Delta\xi^{\rm lensing}$ at redshift $\bar z=1$ using the full-sky formalism. The solid lines show the fractional difference using  the linear transfer function and the dashed line is using halofit. In the \emph{left panel} we show $\Delta\xi^{\rm lensing}$ as a function of separation $r$ for fixed values of $\mu$: $\mu=1$ (green), $\mu=0.98$ (blue) and $\mu=0.8$ (orange), and in the \emph{right panel} we show it as a function of $\mu$ for fixed separation: $r=8$Mpc$/h$ (orange), $r=100$Mpc$/h$ (blue) and $r=420$Mpc$/h$ (green).}
\end{figure}

The simplest way to calculate the full-sky lensing non-linearly is to use the Poisson equation to relate the gravitational potentials along the line-of-sight to the density (this equation is indeed valid also in the non-linear regime) and to use halo-fit to calculate the non-linear density power spectrum. This procedure can however not be implemented exactly because the full-sky lensing requires the density power spectrum at different redshifts along the two lines-of-sight $P_m(k, z, z')$ where $z$ and $z'$ can take any values between $0$ and $z_1$ and $z_2$. Halo-fit gives an expression for the power spectrum only when $z=z'$. Note that this problem does not arise in the calculation of the flat-sky expression which uses Limber approximation and therefore neglects correlations coming from $z\neq z'$. In order to overcome this problem we use the following approximate procedure: we calculate the non-linear power spectrum at a middle redshift along the line-of-sight $z_{*}$ and then evolve it using the linear growth rate $D_1(z)$ along the photon trajectory. This is of course not completely correct because in the non-linear regime density does not evolve with the linear growth rate, but it gives us a good approximation of the true non-linear lensing contribution. To determine which $z_*$ is the most appropriate, we use the flat-sky approximation~\footnote{More precisely we do the following: we calculate the flat-sky contribution using the correct non-linear power spectrum integrated along the line-of-sight (remember that in the flat-sky we can do that since we have only one line-of-sight). Then we use the same approximation as in the full-sky to calculate also the flat-sky and we compare the correct flat-sky result with the approximate flat-sky result for various values of $z_*$. This allows us to find the best $z_*$. For $z=1$ we find $z_*=0.42$ and for $z=2$, $z_*=0.73$.}. We checked that our result behaves in a consistent way when we vary $z_*$, which gives us confidence in this approximation (see Fig.~\ref{f:checkzNL} in Appendix~\ref{app:lensNL} for more detail).

In Fig.~\ref{f:lensing_nonlinear} we show the fractional difference with respect to the standard term due to the full-sky lensing in the linear and non-linear regime $\Delta \xi^{\rm lensing}$.
Contrary to Fig.~\ref{f:c2} where the fractional difference of all the terms was calculated with respect to the full-sky standard term, here we show the fractional difference with respect to the flat-sky standard term given in Eq.~\eqref{e:corrflat}. In this way Fig.~\ref{f:lensing_nonlinear} can be directly interpreted as the fractional error that one makes when using the standard flat-sky correlation function instead of the full-sky observable correlation function containing lensing~\footnote{Note that to calculate the flat-sky  standard  expression in the non-linear regime we use the linear continuity equation to relate the velocity to the density and then we use halo-fit for the density power spectrum. This procedure is not completely correct as the continuity equation is also modified in the non-linear regime. Current data analyses use a more sophisticated procedure to calculate the non-linear redshift-space distortions, based on~\cite{1994MNRAS.267..927F}. Our procedure is however conservative since it tends to overestimate the impact of non-linearities on redshift-space distortions and therefore to underestimate the relative importance of lensing.}.
Clearly, lensing becomes very important at large separation and large $\mu$. Neglecting it in this regime can therefore impact the determination of cosmological parameters in a significant way. We defer a detailed study of this impact to a future work~\cite{prep}. Comparing  linear and non-linear results, we find that for $\mu=1$, the non-linear result is very different from the linear one at all separations up to $250\,{\rm Mpc}/h$. For $r\leq 150$\,Mpc/$h$, the non-linear lensing is significantly enhanced with respect to the linear regime. At larger separation however, the tendency is reversed. This reflects the fact that non-linearities move power from small to large $k$. On the right panel we see that at small separation, $r=8$\,Mpc/$h$, the non-linear lensing is significantly larger than the linear one for all $\mu$. In summary, Fig.~\ref{f:lensing_nonlinear} shows that lensing cannot be neglected at redshift 1 and that it has to be calculated in the full-sky non-linear regime, because it mixes small scales (where non-linearities are important) and large scales (where full-sky effects are important).

\subsubsection{Multipole expansion of the correlation function}

\label{sec:corr_mult}

The correlation function is in general a function of separation $r$ and orientation $\mu$. However, the dependence in $\mu$ of the standard flat-sky expression~\eqref{e:corrflat} is very simple, since it is given by $L_2(\mu)$ and $L_4(\mu)$ only. This simple dependence has been exploited to measure directly the growth rate $f$. In practice this means that each pair of galaxies is weighted either by $L_0(\mu)=1$, $L_2(\mu)$ or $L_4(\mu)$. The average over all orientations is then performed, allowing one to measure the coefficient in front of each of the $L_\ell$, i.e. the monopole, quadrupole and hexadecapole.

In the full-sky regime the dependence of redshift-space distortions on $\mu$ becomes more complicated, first due to the fact that $\bn_1$ and $\bn_2$ are not parallel (wide-angle effects) and second because the growth rate and bias are evolving with time $f(z_1)\neq f(z_2)$. In addition, the large-scale relativistic effects and the integrated effects have their own $\mu$-dependence, which cannot be simply expressed in terms of $L_2(\mu)$ and $L_4(\mu)$ as we saw in Fig.~\ref{f:lensing_linear}. As a consequence the multipole expansion of the full-sky observable correlation function differs from the flat-sky standard expansion. Firstly the monopole, quadrupole and hexadecapole of the full-sky standard term differ from the flat-sky ones. Secondly, these multipoles get corrections from the relativistic and lensing contributions. And finally, due to wide-angle effects and lensing, the multipoles beyond $\ell=4$ no longer vanish. 

\begin{figure}[t]
\centering
\includegraphics[scale=0.6]{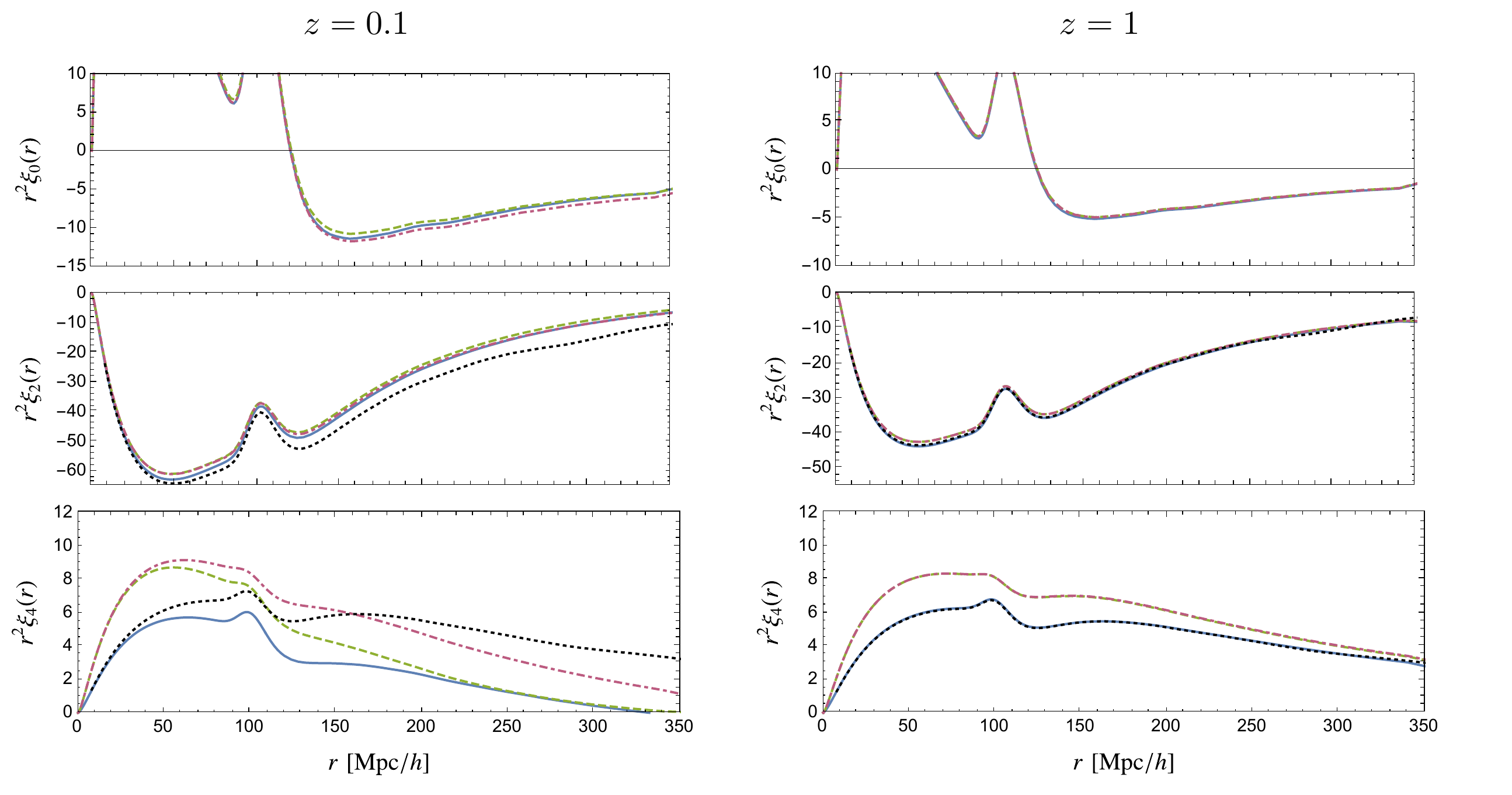}
\caption{\label{f:c3} The multipoles from density and redshift-space distortions, $\xi^{\text{st}}$, at redshift $\bar z=0.1$ (left) and $\bar z=1$ (right).  We show the monopole (top) quadrupole (middle) and hexadecapole (bottom) for different definitions of the angle in the full-sky: $\mu$ (blue, solid), $\cos\ga$ (purple, dash-dotted) and $\cos\beta$ (green, dashed) and we compare this with the flat-sky multipoles obtained from~\eqref{e:corrflat} (black, dotted).}
\end{figure}

In Fig.~\ref{f:c3} we show the impact of wide-angle effects on the monopole, quadrupole and hexadecapole. Since the standard terms are almost not affected by non-linearities above 20\,Mpc/$h$, we calculate these multipoles using the linear power spectrum. In black we show the flat-sky multipoles from density and redshift-space distortions, that are simply given by the coefficients in front of $L_\ell(\mu)$ in Eq.~\eqref{e:corrflat}. In blue, purple and green we show the full-sky multipoles from density and redshift-space distortions obtained from expression~\eqref{corrstme}, which we multiply by the appropriate Legendre polynomial and numerically integrate over directions~\footnote{Note that the multipoles defined in Eq.~\eqref{e:defmult} completely differ from the multipoles defined in~\cite{Raccanelli:2013gja}  (see their Eq.~(17)). The multipoles in~\eqref{e:defmult} are defined at fixed galaxy separation $r$ and they correspond to what observers are measuring in redshift surveys. The multipoles in~\cite{Raccanelli:2013gja} are on the contrary defined at fixed angular separation $\theta$ (see their Fig. 1). As a consequence they mix different separations $r$ and have completely different properties.}
\be 
\xi_\ell(r,\bar z) = \frac{1}{2\ell+1}\int_{-1}^1\xi(r,\bar z,\si)L_\ell(\si)d\si\, . \label{e:defmult}
\ee
As discussed in Section~\ref{s:Cls}, in the full-sky there is no unique way to define the orientation of the pairs of galaxies. We therefore calculate the multipoles for different choices: 
$\si=\cos\beta$, $\si=\cos\ga$ and $\si=\mu$.
The amplitude of the multipoles depends on this choice, as can be seen from the different colours in Fig.~\ref{f:c3}. At redshift $\bar z=1$ (right), we find that the monopole differs only at very large scales by a few percent, while the quadrupole also differs at intermediate scales by a few percent. The hexadecapole is significantly different at most scales. At redshift $\bar z=0.1$ (left) the difference is much more important, up to 10\% on the quadrupole at intermediate scales already. And the hexadecapole is very different at most scales. As already pointed out in~\cite{Szalay:1997cc,Szapudi:2004gh,Papai:2008bd,Raccanelli:2010hk, Samushia:2011cs,Bertacca:2012tp, Yoo:2013zga} it is therefore important to account for wide-angle effects when interpreting the multipoles. We also see in Fig~\ref{f:c3} that the angle 
which is closest to the flat-sky result is nearly always $\mu$ and especially it is always $\mu$ for $\bar z=1$. Note that in~\cite{Reimberg:2015jma}, expressions for the dominant wide-angle corrections to the monopole, quadrupole and hexadecapole have been derived for various choices of angles.

\begin{figure}[t]
\centering
\includegraphics[scale=0.45]{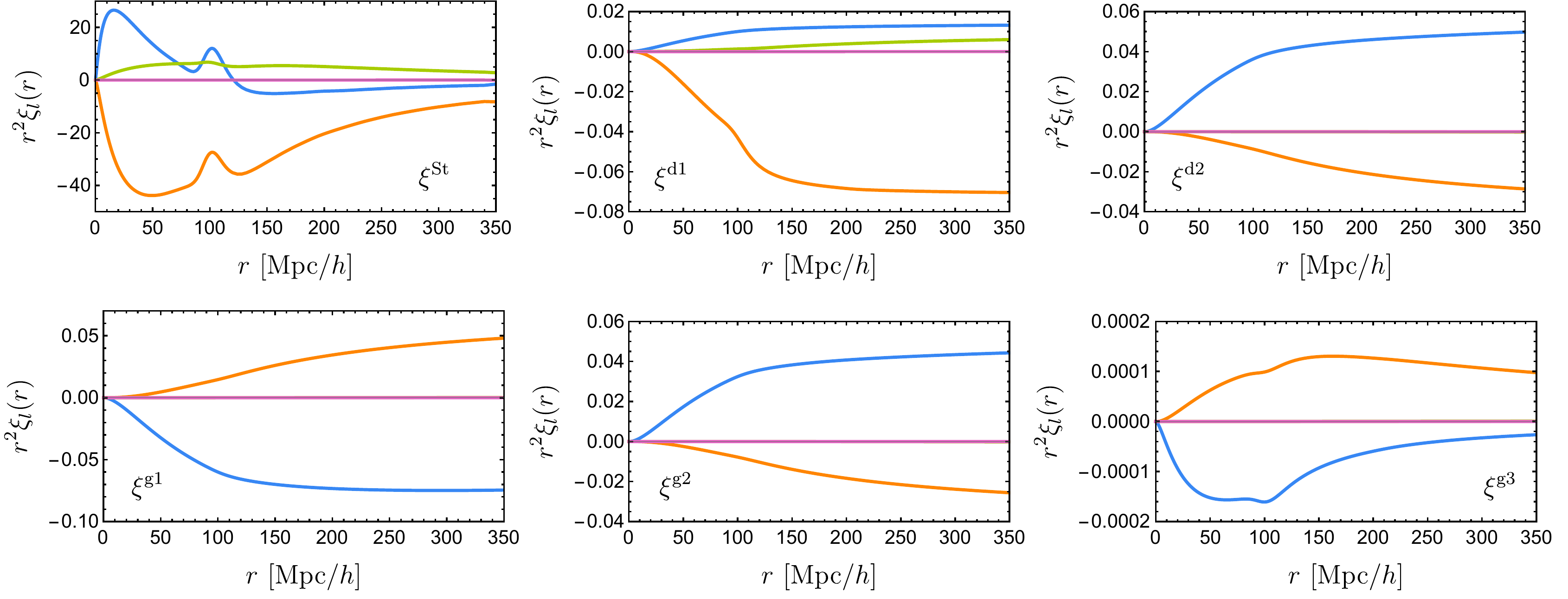}
\caption{\label{f:cmult} We show the multipoles of different contributions to the full-sky correlations function at $\bar z=1$. The monopole (blue), quadrupole (orange), hexadecopole (green) and  $\ell=6$ purple.}
\end{figure}

In Fig.~\ref{f:cmult} we show the multipoles from all the non-integrated contributions in the full-sky linear regime. We use  the angle $\mu$  for this figure. Each plot represents a different relativistic contribution (see Eqs.~\eqref{e:Ddens} to \eqref{e:Dg5} for a definition of the terms).  As in Figs.~\ref{f:c2} and \ref{f:lensing_nonlinear}, this encompasses the correlation of the term with itself as well as its cross-correlation with the standard term (density and redshift-space distortion). One would naively expect that the dominant contribution would come from the Doppler term d1 correlated with the standard term. However, as discussed in~Section~\ref{s:Cls}, this contribution exactly vanishes in the flat-sky approximation. It would contribute only to a dipole, which cannot be seen with one population of galaxies, due to its anti-symmetry (indeed only even multipoles exist in this case). As a consequence to measure the dominant dipole one needs to cross-correlate two populations of galaxies, as discussed in~\cite{Bonvin:2013ogt, Bonvin:2015kuc,Gaztanaga:2015jrs,Hall:2016bmm}.

However, as discussed in Section~\ref{s:Cls}, in the full-sky the Doppler-standard correlation does not exactly vanish and it contributes to the even multipoles. The amplitude of this term is then of the same order of magnitude as the d1-d1 correlation and as the other relativistic terms (for example g1 correlated with density). This is evident from the various panels in Fig.~\ref{f:cmult}, where we see that all the non-integrated relativistic terms generate multipoles of the same order of magnitude. The only exception is g3 which is much smaller. This is not surprising since at $z=1$ the universe is still matter dominated and the gravitational potential is nearly constant. For the same reason also d2 and g2 are very similar.

The Doppler contribution is the only one which generates a non-negligible hexadecapole. This comes from the correlation of d1 with redshift-space distortions which contains 3 gradient of the potential. In the flat-sky this gives rise to a $\mu^3$-dependence, which again vanishes for symmetry reason, but in the full-sky one obtains an additional factor $\mu\cdot r/\chi$ which leads to an hexadecapole~\footnote{This can been seen for example by expanding $\alpha_1-\alpha_2$ in powers of $r/\chi$ in the expression $\zeta^{12}$ in Appendix~\ref{a:corf}.}. 
In the flat-sky the other relativistic terms (d2, g1, g2 and g3) generate only a monopole and quadrupole, due to their correlation with redshift-space distortion. In the full-sky they do generate higher multipoles, but again those are suppressed by powers of $r/\chi$ and are consequently negligible. 

\begin{figure}[t]
\centering
\includegraphics[scale=0.51]{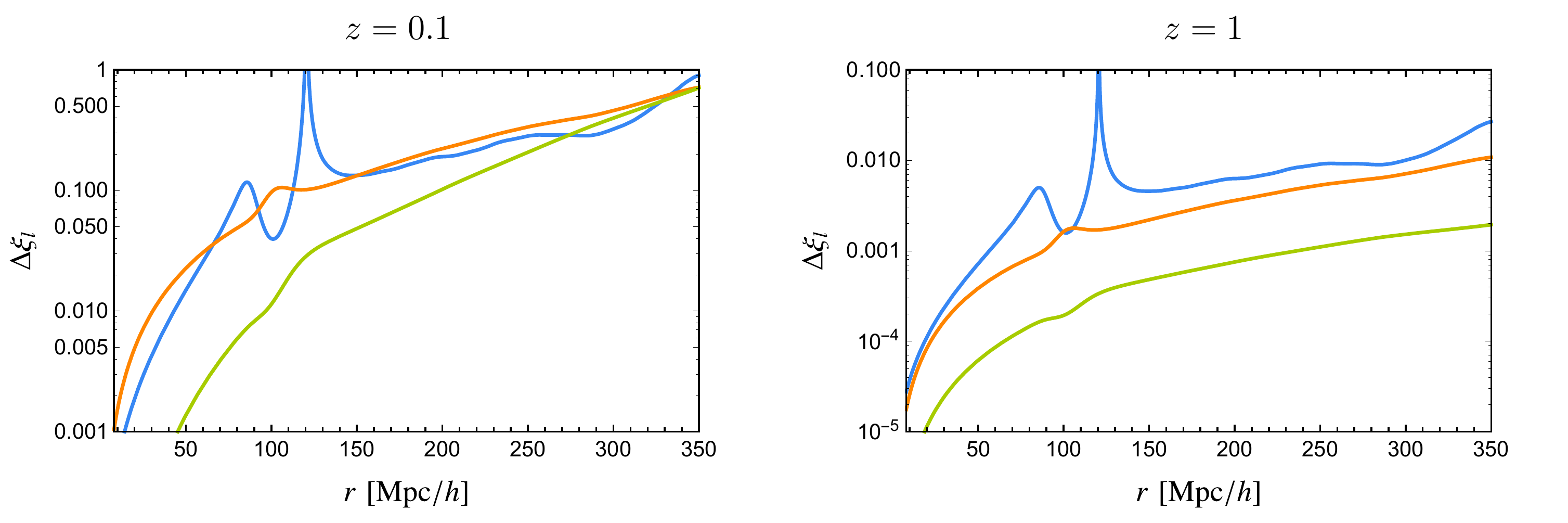}
\caption{\label{f:Delta_rel_mult} Fractional difference generated by the sum of the non-integrated relativistic effects on the monopole (blue), quadrupole (orange) and hexadecapole (green). The relativistic multipoles are calculated in the full-sky linear regime, whereas the standard multipoles are calculated in the flat-sky linear regime, to reproduce the theoretical prediction currently used.}
\end{figure}

In Fig.~\ref{f:Delta_rel_mult} we plot the fractional difference due to all non-integrated effects with respect to the standard flat-sky multipoles
\be
\Delta \xi_\ell^{\rm rel}=\frac{\xi_\ell^{\rm rel}}{\xi_\ell^{\rm st, flat-sky}}\, ,
\ee
where $\xi_\ell^{\rm rel}$ contains the correlation of all the non-integrated relativistic terms with themselves as well as their correlation with the standard term, i.e. they come from
\be
\langle \De^{\rm st}\De^{\rm rel}\rangle  +\langle \De^{\rm rel}\De^{\rm st}\rangle  + \langle \De^{\rm rel}\De^{\rm rel}\rangle  = \langle \De^{\rm st+rel}\De^{\rm st+rel}\rangle-\langle \De^{\rm st}\De^{\rm st}\rangle\,. \label{defrel}
\ee

At $\bar z=1$ (right panel), the relativistic terms modify the monopole by a few percent at separations $\geq 300$\,Mpc/$h$. The impact of these terms on parameter estimation is therefore probably negligible at high redshift. At $\bar z=0.1$ however (left panel) the relativistic contribution to the multipoles is non-negligible at most scales. The contributions to the monopole and quadrupole are already of a few percent at $50$\,Mpc/$h$. At $100$\,Mpc/$h$ these contributions reach 10\% and they quickly increase with separation. 

The large amplitude of the relativistic terms at small redshift is due to one specific term in the Doppler contribution, namely the one proportional to $1/(\HH\chi)$ (see Eq.~\eqref{e:Dd1}). The correlation of the Doppler term with itself has roughly the following amplitude: $$1/(\HH\chi)^2(\HH/k)^2\langle \Delta^{\rm den} \Delta^{\rm den}\rangle\sim (r/\chi)^2\langle \Delta^{\rm den} \Delta^{\rm den}\rangle\,,$$where we have used that $k$ corresponds to $1/r$. At small redshift and large separation, this suppression is not very strong. For example at $\bar z=0.1$, $\chi=433$\,Mpc/$h$ and therefore the amplitude of the Doppler term at $r=200$\,Mpc/$h$ is roughly $(r/\chi)^2\langle \Delta^{\rm den} \Delta^{\rm den}\rangle\sim 0.2\langle \Delta^{\rm den} \Delta^{\rm den}\rangle$, i.e. 20\% of the standard term. The same argument applies to the full-sky Doppler-standard correlation which contributes at the same level. The other relativistic terms on the other hand are more strongly suppressed. For example, the correlation g1-standard has the following amplitude: $1/(\HH\chi)(\HH/k)^2\langle \Delta^{\rm den} \Delta^{\rm den}\rangle\sim (r/\chi) r\HH\langle \Delta^{\rm den} \Delta^{\rm den}\rangle$. At $\bar z=1$, $\HH\sim 1/\chi$ and the Doppler contribution is similar to the g1 contribution, as already discussed. At $\bar z=0.1$ however, $\HH$ is significantly smaller than $1/\chi$ and therefore the Doppler contribution is enhanced with respect to the g1 contribution. Note that the importance of this Doppler effect on the correlation function has already been studied in detail in~\cite{Papai:2008bd} and further discussed in~\cite{Raccanelli:2010hk, Samushia:2011cs}. These references, however, do not include the other Doppler terms or lensing.

This result is especially relevant for a survey like the SKA that will cover wide parts of the sky from $z=0$ to 2 and will therefore be strongly affected by the Doppler term at low redshift. In a forthcoming publication we will study the impact of this effect on the measurement of cosmological parameters, in particular on the measurement of the growth rate $f$ from the monopole and quadrupole. Note that, as discussed above, such a study has to be performed using the full-sky formalism, since full-sky effects (from the Doppler-density correlation) contribute at the same level.

\begin{figure}[ht]
\includegraphics[scale=0.43]{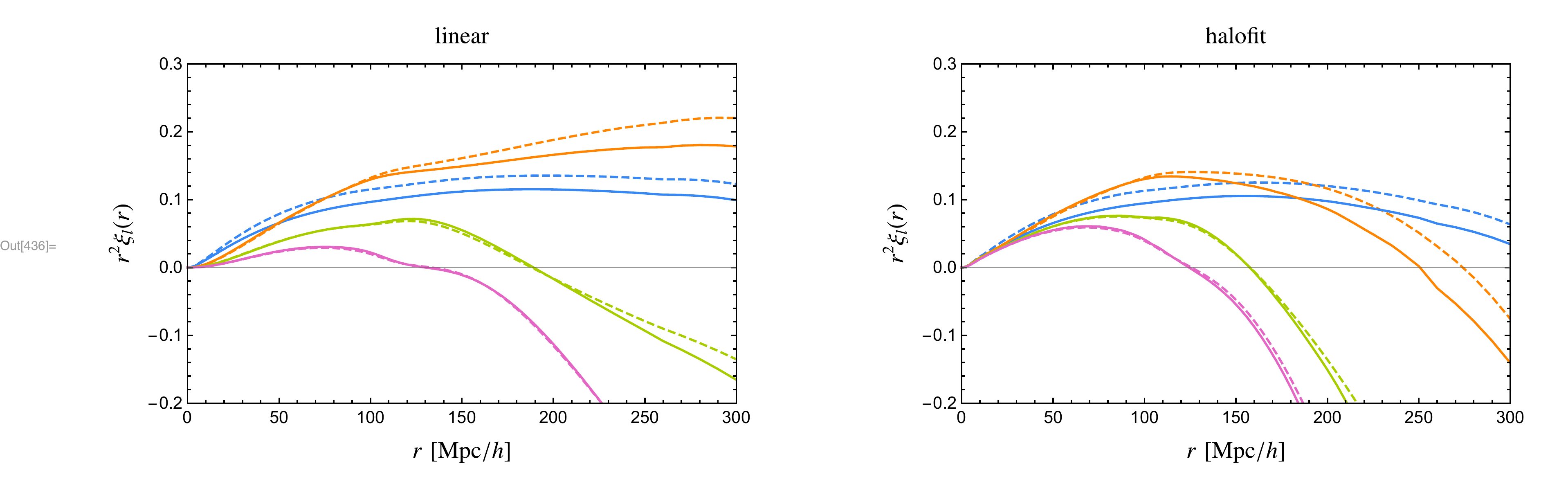}
\caption{\label{f:L_mult} Multipoles of the lensing contribution (including its correlation with the standard term) at $\bar z=1$. In the left panel we show the linear full-sky (solid) and linear flat-sky (dashed) result and in the right panel the non-linear full-sky (solid) and flat-sky (dashed) result. The monopole is shown in blue, the quadrupole in orange, the hexadecapole in green and the $\ell=6$ in purple.}
\end{figure}

\begin{figure}[ht]
\centering
\includegraphics[scale=0.36]{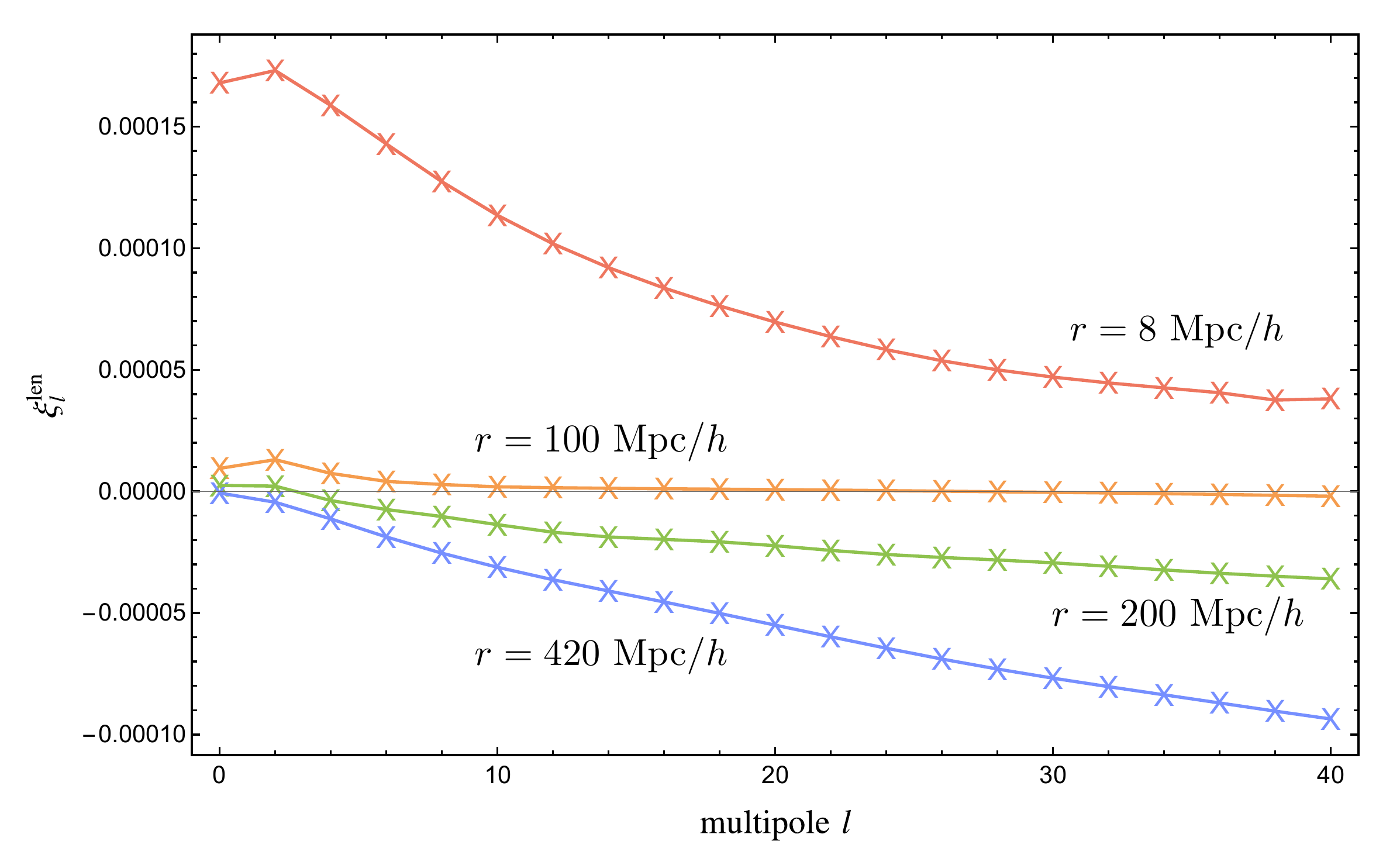}
\caption{\label{f:len_multipoles} The full-sky non-linear lensing multipoles as a function of $\ell$ for different separations at $\bar z=1$.}
\end{figure}

In Fig.~\ref{f:L_mult} we show the lensing contribution to the multipoles at $\bar z=1$. In the left panel we show the linear result, using the flat-sky and Limber approximation (dashed) and the full-sky calculation (solid); and in the right panel we show the non-linear result. The flat-sky systematically overestimates the lensing contribution. As explained in Section~\ref{s:mu-r} this is due to the fact that the Limber approximation underestimates the correlation between density and lensing, which is negative, and consequently it overestimates the total in most configurations. Above $r\sim 50h^{-1}$Mpc the lensing contribution is 10\% and more. Hence it has to be included for an accurate estimation of the growth rate $f$. Contrary to the non-integrated relativistic effects, lensing generates non-negligible $\ell=4$ and $\ell=6$. Actually, as is shown in Fig.~\ref{f:len_multipoles} the amplitude of the multipoles remains large for large values of $\ell$. Measuring $\ell>4$ will therefore provide a way of isolating the lensing contribution from the standard terms.

\begin{figure}[t]
\centering
\includegraphics[scale=0.44]{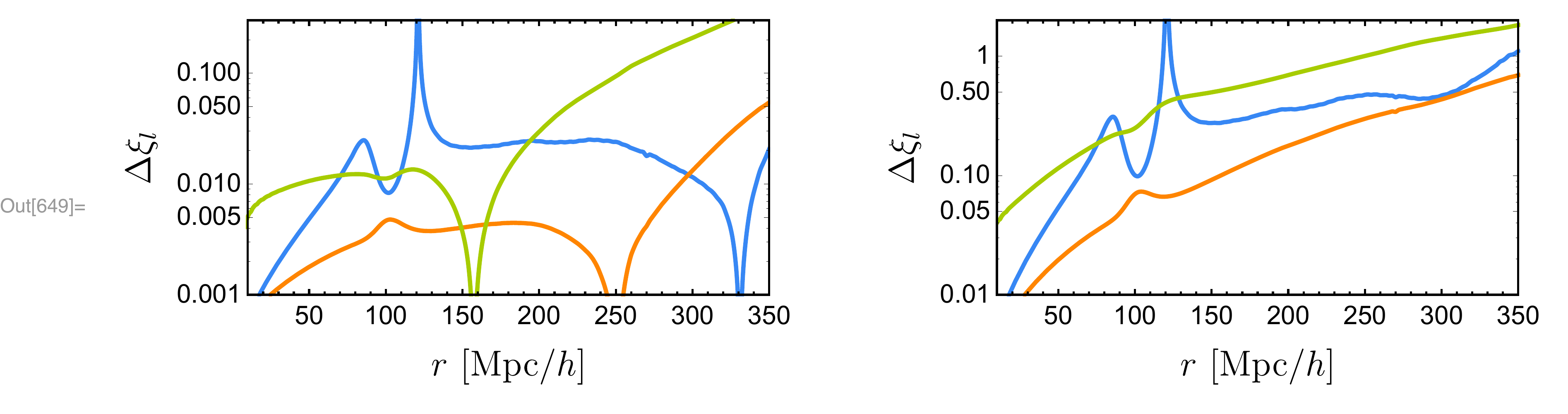}
\caption{\label{f:Delta_lens_mult} Fractional difference generated by lensing on the monopole (blue), quadrupole (orange) and hexadecapole (green). The lensing multipoles are calculated in the full-sky non-linear regime, whereas the standard multipoles are calculated in the flat-sky non-linear regime, to reproduce the theoretical prediction currently used. The left panel is for $\bar z=1$ and the right panel for $\bar z=2$.}
\end{figure}

In Fig.~\ref{f:Delta_lens_mult} we show the fractional difference of the monopole, quadrupole and hexadecapole generated by lensing at $\bar z=1$ and $\bar z=2$. At $\bar z=1$ we see that lensing modifies the monopole by a few percent at intermediate scales. The quadrupole is less affected, apart from at very large scales $r\sim 350$\,Mpc/$h$ where lensing contributes by 5\%. The hexadecapole is the one that is the most affected by lensing, up to 10-20\% above 250\,Mpc/$h$. At $\bar z=2$ the lensing contribution is significant for all multipoles. The monopole is modified by 30\% already at a 150\,Mpc/$h$ and this increases to 50\% at 300\,Mpc/$h$. The contribution to the quadrupole is slightly smaller, but it still reaches 10\% at 150\,Mpc/$h$ and 40\% at 300\,Mpc/$h$. And the hexadecapole is strongly affected at all scales. Surveys like Euclid and the SKA, that will observe up to high redshift should therefore include lensing in their modelling of the multipoles of the correlation function.

In this Section we have only discussed the contribution from even multipoles to the correlation function. As stated before, in the flat-sky approximation only even multipoles exist, even in the presence of relativistic effects and lensing~\footnote{Note that this is not the case with the alternative definition of multipoles used in~\cite{Raccanelli:2013dza} which mixes different scales.}. This follows directly from the fact that the correlation function is symmetric $\xi(\br)=\xi(-\br)$ and that the flat-sky angle goes from $\mu$ to $-\mu$ when $\br$ goes to $-\br$. In the full-sky, the existence of odd multipoles depend on the choice of angle used to measure them. If the cosine of the angle simply changes sign when $\br$ goes to $-\br$, then odd multipoles exactly vanish also in the full-sky. This is the case for the angles $\beta$, $\gamma$ and $\alpha$ defined in Fig.~\ref{f:angles}. However if one uses instead the angle $\alpha_1$ (see Fig.~\ref{f:a-gammabeta}) to measure the multipoles, then the correlation function contains odd multipoles in the full-sky because $\alpha_1$ goes to $\pi+\alpha_1-\theta$ when $\br$ goes to $-\br$. Hence even if the correlation function is symmetric, its expansion in terms of $\alpha_1$ contains odd multipoles due to the fact that the angle itself breaks the symmetry of the configuration~\cite{Reimberg:2015jma}. Note that the dipole of the correlation function using the angle $\alpha_1$ has been measured in~\cite{Gaztanaga:2015jrs}. Finally let us stress that if we cross-correlate different populations of galaxies, then the correlation function is not symmetric anymore $\xi_{AB}(\br)\neq\xi_{BA}(-\br)$ (where $A$ and $B$ denote the two populations under considerations) and it contains therefore odd multipoles already in the flat-sky approximation, as demonstrated in~\cite{Bonvin:2013ogt}.

\section{From the correlation function to the power spectrum}\label{s:ps}

As discussed in the introduction, an alternative observable which is routinely used to analyse redshift surveys is the power spectrum. Here we discuss the impact of the large-scale relativistic effects and of the lensing on this observable. 

Of course, since galaxies are seen on our background light-cone and not in 3D physical space, a galaxy position is fixed by a redshift $z$ and a direction $\bn$. But we can split the distance vector between two galaxies,  $\br$ (which is the argument of the galaxy correlation function $\xi(\br,\bar z)$) in a sufficiently small redshift bin into a radial, $r_\pa$ and a transverse, $r_\perp$ component and express $\xi$ in the variables $\xi(r_\pa,r_\perp,\bar z)$. We can then define the power spectrum simply as the Fourier transform of the correlation function,
\bea
P(k_\pa,k_\perp,\bar z) &=& \int d^3r \xi(r_\pa,r_\perp,\bar z)e^{i(r_\pa k_\pa + r_\perp k_\perp\cos\phi)} \\
&=& 2\pi\int_{-\infty}^\infty dr_\pa \int_0^\infty dr_\perp  \xi(r_\pa,r_\perp,\bar z)e^{i(r_\pa k_\pa)}J_0(k_\perp r_\perp)\,. \label{e:pofk}
\eea
In this expression $r_\pa = r\sigma$ and $r_\perp = r\sqrt{1-\sigma^2}$ where
\be
[-1,1]~\ni~\sigma = \left\{\begin{array}{c}
\mu=\cos\alpha  \\ \cos\beta \\ \cos\ga \\ \cos\al_2 \end{array}\right.
\ee
depending on the angle used to split the survey into a radial and a transversal component. Note that $\br_\perp = r_\perp(\cos\phi,\sin\phi)$ is a 2D vector in the plane normal to the parallel direction and we have performed the $\phi$ integration choosing the $x$-axis in the $\br_\perp$ plane parallel to $\bk_\perp$. 
For the case $\sigma=\mu$, $r_\pa=\chi_2-\chi_1$ the expression for the correlation function is given in Appendix~\ref{a:corf} and Section~\ref{s:Corr}, \eqref{corrQ}. For the other angles, one has to use the relations given in Appendix~\ref{a:angles}.

 However,  we must  consider that while the correlation function as given e.g. in Eq.~\eqref{xirrz} can be defined for all values $r_\pa\in [0,\chi(\infty)] \simeq [0,14h^{-1}$Gpc$]$ and $r_\perp\in [0,2\chi(\infty)]$, and is correct for $|r_\pa H(\bar z)| \ll 1$, this is no longer so for its Fourier transform\footnote{Here $\chi(\infty)\simeq14h^{-1}$Gpc represents the comoving size of our horizon today.}. To compute it we have to integrate the correlation function over all space, but as we just said, we cannot observe the correlation function outside of our horizon and the result is not reliable if $|r_\pa H(\bar z)| \gtrsim 1$. It is well defined only for a range of $(r_\pa,r_\perp)$. This situation is further complicated by the fact that this range depends on redshift. Therefore, the simple Fourier transform given above gives a physically sensible result only for
$$k_\pa\gg \frac{1}{\chi(\bar z+\De z)-\chi(\bar z-\De z)} \sim 2\De z H(\bar z) =\frac{1}{r_{\pa\max}(\bar z,\De z)}\,, \qquad \De z\ll 1.$$
For these values of $k_\pa$, contributions from radial distances such that the two galaxies are not in a thin shell around $\bar\chi= \chi(\bar z)$ are cancelled by the rapid oscillations of the exponential in the Fourier transform. 

With this word of caution we now simply Fourier transform the correlation function to obtain the power spectrum.  We can either use the correlation function obtained via the $C_\ell(z_1,z_2)$'s or the one from the direct computation.
Here we present the details for the latter.

As stated above, for the 'true' power spectrum, the integral over $r_\pa$ should extend from $-\infty$ to $+\infty$ and  the integral over $r_\perp$ should extend from $0$ to $+\infty$. The correlation function is however not observable outside the horizon and the integral must therefore be truncated by a window function which removes these scales. In practice galaxy surveys do not observe the whole horizon but only part of it and therefore the range of integration is even more reduced. The true window function of the observation patch leads to a convolution in the correlation function and therefore to a multiplication of the Fourier transform of the window in the power spectrum

From Eq.~\eqref{e:pofk} we see that there is another reason to truncate the integral. The arguments  $k_\pa$ and $k_\perp$ (or equivalently $k$ and $\kcos=\hat\bk\cdot\hat\bn$) of the power spectrum are parallel to $r_\pa$ and $r_\perp$ respectively. Now the direction of $r_\pa$, for example, depends on the direction of the pair of galaxies we consider. If the domain of integration in~\eqref{e:pofk} is sufficiently small, then a mean direction $\bn$ can be introduced and this splitting is well defined: one can identify one line-of-sight for the whole patch of sky we are observing and split parallel and transverse directions with respect to this line-of-sight. If the patch is too large however, this procedure is no longer valid~\footnote{Note however the work of~\cite{Yamamoto:2005dz} which proposes methods to account for different lines-of-sight in the measurement of the power spectrum.}. The integral~\eqref{e:pofk} can still be done mathematically, but its physical interpretation becomes unclear. This illustrates the fact that the power spectrum is truly well defined only in the flat-sky. In practice this means that we can consider the Fourier transform of the correlation function in a sphere of radius $\De z/\HH(\bar z)$ for values $k\gg \HH(\bar z)/\De z$. 

Similar to what is done for the correlation function, in the standard analysis, the $\nuP$ dependence of $P(k,\nuP,\bar z)$ is used to extract the growth rate $f(\bar z)$. Indeed as seen in Eq.~\eqref{e:poldk}, the standard power spectrum takes the simple form
\be
P(k,\nuP,\bar z) = p_0(k,\bar z) +  p_2(k,\bar z)L_2(\nuP) +  p_4(k,\bar z)L_4(\nuP)\, ,
\ee
where the coefficients $p_n$ are given by:
\bea
 p_0(k,\bar z) &=& D_1^2(\bar z)P_m(k)\left[b^2+\frac{2bf}{3} +\frac{f^2}{5}\right]\, ,
 \label{e:P0}\\
 p_2(k,\bar z) &=& D_1^2(\bar z)P_m(k)\left[\frac{4bf}{3}+\frac{4f^2}{7} \right]\, , \label{e:P2}\\
 p_4(k,\bar z) &=& D_1^2(\bar z)P_m(k)\frac{8f^2}{35} \,.  \label{e:P4}
\eea
The multipoles $p_0$ and $p_2$ contain different combinations of the bias and of the growth rate $f(\bar z)$ and can be used to measure these two quantities. If $p_4$ can be measured as well it can be used as an additional consistency check. Furthermore, this quantity is independent of galaxy bias which renders it especially valuable.

The large-scale relativistic effects and the gravitational lensing are however expected to modify this simple multipole expansion. In principle to calculate the contribution of these effects to the multipoles, one would need to calculate Eq.~\eqref{e:pofk} for all values of $k_\pa$ and $k_\perp$ and then integrate over all directions, weighting by the appropriate Legendre polynomial
\bea
p_\ell(k,\bar z) \;=\;\frac{1}{2\ell+1}\int_{-1}^1\!\!d\nuP P(k,\nuP,\bar z) L_\ell(\nuP) \,.
\eea
As the correlation function is a symmetric function of $\mu$, $\xi(\bx_1,\bx_2)=\xi(\bx_2,\bx_1)$, the power spectrum will be symmetric in $\nuP$ so that only even $\ell$'s are non-zero. This is no longer the case when one correlates different tracers, e.g. bright and faint galaxies~\cite{McDonald:2009ud,Yoo:2012se}.

The procedure to obtain the multipoles of the power spectrum can however be simplified by using directly the multipoles of the correlation function $\xi_\ell(r)$ (see Appendix \ref{app:theorem} for a proof of this relation)
\be
\label{e:pn}
p_\ell(k) =4\pi i^\ell\int_0^\infty dr r^2j_\ell(kr)\xi_n(r) \,.\\
\ee
As discussed before, the integral over $r$ cannot run until infinity because the correlation function (and consequently its multipoles) is not observable over the whole space. For simplicity we assume that we observe galaxies within a sphere of radius $r_{\rm max}$, centred at redshift $\bar z$. This corresponds to introducing a window function in Eq.~\eqref{e:pn} which removes scales larger than $r_{\rm max}$. For the standard terms, the multipoles $p_\ell(k)$ 
are relatively insensitive to the choice of $r_{\rm max}$ since $r^2 \xi^{\text{st}} \rightarrow 0$ as $r\rightarrow \infty$. The large-scale relativistic effects scale however as $r^2 \xi^{\text{rel}} \rightarrow \text{constant}$ as $r\rightarrow \infty$ and  consequently their multipoles depend on the choice of $r_{\rm max}$. This reflects the fact that these terms diverge when $k\rightarrow 0$ as we will see in section~\ref{s:flatskyP}. The situation for the lensing term is even worse: the correlation function scales as $r^2 \xi^{\text{len}} \rightarrow \infty$ and the dependence in $r_{\rm max}$ is even stronger. The lensing power spectrum is therefore strongly dependent on the geometry of the survey, as already noticed in~\cite{LoVerde:2007ke}. 

\subsection{The flat-sky approximation}
\label{s:flatskyP}

In the previous section we obtained the power spectrum by integrating over the full-sky correlation function, weighted by a window function to restrict the range of integration to the observed patch of the sky. Here we would like to compare this procedure with a flat-sky direct calculation of the power spectrum~\footnote{Note that the relation between the flat-sky and full-sky power spectrum of density and RSD has been studied in detail in~\cite{Reimberg:2015jma}.}. The power spectrum for the non-integrated terms has been derived previously in~\cite{Jeong:2011as, Yoo:2012se}. 
It can be easily obtained by Fourier transforming the non-integrated relativistic contributions to the number counts, namely $\Delta^{\rm d1}, \Delta^{\rm d2}, \Delta^{\rm g1}, \Delta^{\rm g2}$ and $\Delta^{\rm g3}$ (see Eqs.~\eqref{e:Dd1} to~\eqref{e:Dg3}). Note that in principle this procedure does not generate an observable, because the Fourier transform of a function $f(\bk, \eta)$ at a given conformal time $\eta$ requires the knowledge of the function over the whole hypersurface of constant $\eta$~\footnote{In principle we do not observe at constant conformal time $\eta$ but rather at constant redshift $z$. However the difference between $\eta$ and $z$ has been consistently included in the derivation of $\Delta$ so that a constant $z$ can now be seen as a constant $\eta$.}. An observer cannot observe this hypersurface, but only its intersection with her past light-cone. However, due to the statistical homogeneity and isotropy of our Universe, the properties of the function are the same everywhere, and the Fourier transform can be performed. We obtain (in agreement with~\cite{Jeong:2011as} where only the non-integrated terms are considered) 
\be
\label{e:Pflatnonint}
P_\De^{\rm flat,\,non-int}(k,\nuP,z)=\left|A + B\,\frac{\HH}{k} + C\left(\frac{\HH}{k}\right)^2\right|^2D^2_1(z)P_m(k)\, ,
\ee
where
\bea  \label{e:AB}
A(\nuP,z) &=& (b-\nuP^2f)\,,\\
B(\nuP,z)  &=& -i\nuP\left(\!\frac{\dot\HH}{\HH^2} +\frac{2-5s}{\HH\chi}+5s-f_{\rm evo}\!\right)\,,\\
C(z)  &=& \left[3f +\frac{3}{2}\Om_m(1+z)\frac{H_0^2}{\HH^2}\left(1-5s -\frac{\dot\HH}{\HH^2} -\frac{2-5s}{\HH\chi}-5s+f_{\rm evo}\right)\right]  \,.  \label{e:C}
\eea
$A$ represents the standard terms, density and redshift space distortions. $B$ is the Doppler term which is suppressed by a factor $\HH/k$ and $C$ represents the additional relativistic contributions which are suppressed by $(\HH/k)^2$.
To arrive at this result we have set $\Psi=\Phi$ and we have neglected the term containing the time derivative of the potential, since it is relevant only at late time and at very large angular scales where the flat sky approximation is not valid.

\begin{figure}[t]
\centering
\includegraphics[scale=0.46]{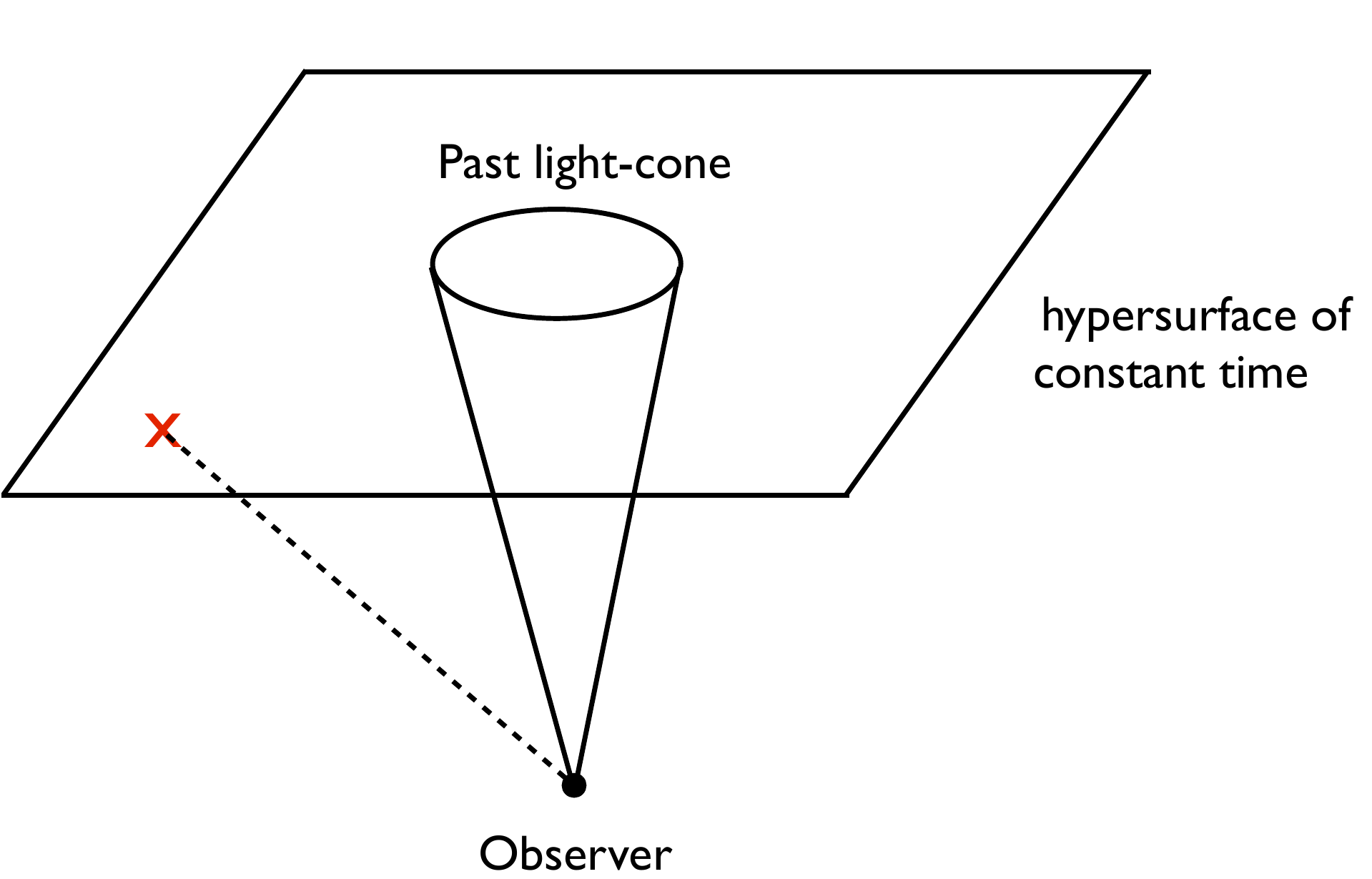}
\caption{\label{f:pastlightcone} To calculate the Fourier transform of the lensing term $\Delta^{\rm lens}(\bk, \eta)$, one needs to know the value of $\Delta^{\rm lens}(\bx, \eta)$ for all $\bx$ on the hypersurface of constant time $\eta$. However for a given observer, $\Delta^{\rm lens}(\bx, \eta)$ is well defined only on her past light-cone. Calculating $\Delta^{\rm lens}(\bx, \eta)$ outside of the past light-cone, like for example at the position of the cross would require to integrate the gravitational potential along the dashed trajectory, which is not physical, and would lead to wrong results.}
\end{figure}

The contribution of the integrated terms to the flat-sky power spectrum are more complicated to calculate and have been neglected in~\cite{Jeong:2011as, Yoo:2012se}. The reason is that integrated terms, like for example the lensing $\Delta^{\rm lens}(\bn, \eta)$, depend on the value of the gravitational potential along the photon trajectory in direction $\bn$. As a consequence $\Delta^{\rm lens}(\bn, \eta)$ is well defined only on the past light-cone of the observer and not on the whole hypersurface of constant conformal time $\eta$. Calculating $\Delta^{\rm lens}(\bn, \eta)$ for a point which is not on the past light-cone of the observer would require to calculate the lensing signal along arbitrary trajectories that have nothing to do with the trajectories followed by photons, as depicted in Fig.~\ref{f:pastlightcone}. 

To calculate the power spectrum of the integrated terms, we need therefore to go through the correlation function.

In Appendix~\ref{a:flat} we show how this can be done in the flat-sky approximation. 
To calculate the integrated terms in the flat sky approximation, we define a sky direction $\bn_*$ and split the observation directions as $\bn_1=\bn_*+\De\bn/2$, $\bn_2=\bn_*-\De\bn/2$. We also split $\br=\br_\perp +\bn_*r_\pa$ with $ \br_\perp=\chi(z)\De\bn$. Representing the correlation function as the Fourier transform of the power spectrum, we can then perform the integral over $k_\pa$ by neglecting the slow dependence of the power spectrum and taking into account only the fast oscillations of the exponential. This leads to the $\de(k_\pa)$ and $\de^P(k_\pa)$ defined below. All details are given in Appendix~\ref{a:flat}.
We obtain  
\begin{align}
&P_\De^{\rm flat, int}(k,\nuP,z) =-3\pi\frac{\Om_mH_0^2(1+z)D_1(z)(2-5s(z))}{\chi} 
P_m(k_\perp)\al(k_\perp,0,z)\left[\chi \de^P(k_\pa)+\frac{2}{k_\perp^{2}}\de(k_\pa)\right]\nonumber\\
&   +\frac{\pi}{2} \left(\frac{3\Om_mH_0^2(2-5s(z))}{\chi}\right)^2\de(k_\pa) 
\int_0^\chi d\la P_m(k\chi/\la)\left[\frac{(\chi-\la)\chi^2}{\la}+\frac{2}{k^2}\right]^2\hspace{-1.5mm}D_1^2(z(\la))(1\!+\!z(\la))^2 \,.
\label{e:Pflatint}
\end{align}
The first line comes from the correlation of the integrated terms with density and the second line is the correlation of the integrated terms with themselves. The distribution $\de^P$ is defined by (see Appendix~\ref{a:flat} for more detail)
\be
\de^P(k)= \frac{1}{2\pi}\int_{-\infty}^{\infty}dx |x|e^{ikx}\,.
\ee
The lensing terms are proportional to the distributions $\de(k_\pa)$ and $\de^P(k_\pa)$. They have to be understood as formal expressions. 
Physical power spectra are obtained by smoothing the signal with a longitudinal window function.
Let us briefly explain this: we assume that our galaxies are all inside a radial window, $W(r_\pa)$, with which the correlation function has to be convolved. Its Fourier transform, the power spectrum is then multiplied by the Fourier transform of the window, $\widehat W(k_\pa)$. As an example, for the cross term involving $\de^P(k_\pa)$, denoting the pre-factor of $\de^P(k_\pa)$ by $P_\times$ and the result by  $P_{\times\, {\rm obs}}$,
we obtain an integral of the form 
\be
P_{\times\, {\rm obs}}(\bk,z)=  P_\times(k_\perp,z) \frac{1}{2\pi} \int dr_\pa dk_\pa|r_\pa|e^{ik_\pa r_\pa}|\widehat W(k_\pa)|^2\,.
\ee
More details with examples of Gaussian and top hat windows can be found in~\cite{Hui:2007tm}.

\subsection{Numerical results: comparison of the flat-sky and full-sky expressions}

\begin{figure}[ht]
\includegraphics[scale=0.46]{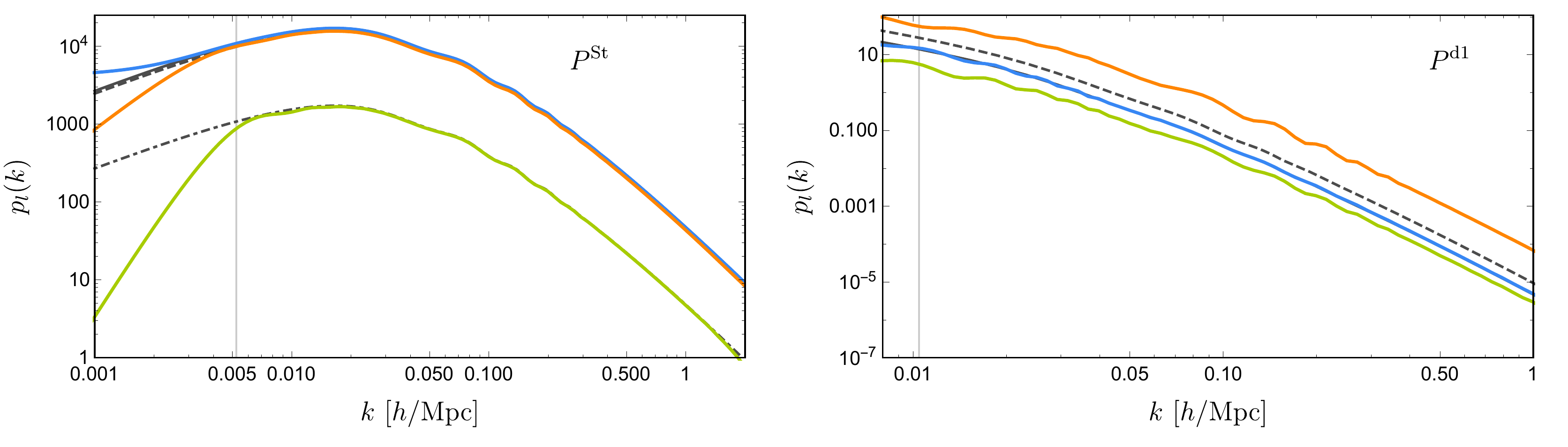}
\caption{\label{f:p1} The multipoles of the power spectrum $p_\ell(k) $ at redshift $\bar z=1$. The coloured lines show the multipoles obtained from Eq.~\eqref{e:pn}: blue for the monopole  ($\ell=0$), orange for the quadrupole  ($\ell=2$) and green for the hexadecapole ($\ell=4$). The black lines show the flat-sky result from Eqs.~\eqref{e:P0}-\eqref{e:P4} and~\eqref{e:Pflatnonint}: solid for the monopole, dashed for the quadrupole and dot-dashed for the hexadecapole. The grey vertical line shows the smoothing scale of the window function. In the left panel we plot the well known density and redshift-space distortions, and in the right panel we plot the Doppler contribution d1.}

\end{figure}
\begin{figure}[ht]
\includegraphics[scale=0.46]{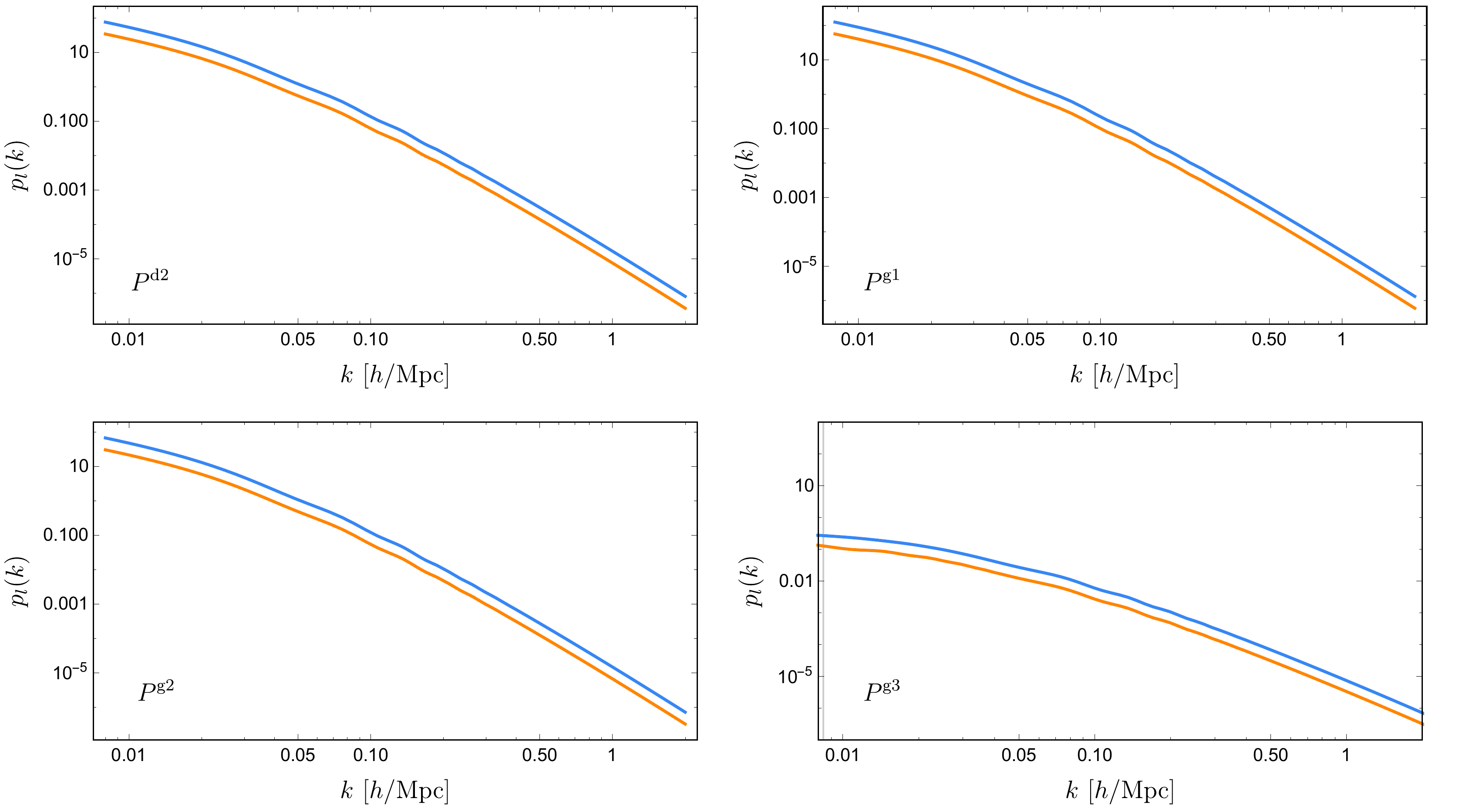}
\caption{\label{f:p2} The monopole (blue) and quadrupole (orange) for the other non-integrated relativistic terms, d2 to g3 at $\bar z=1$. The flat-sky results are indistinguishable from the full-sky ones and are therefore not indicated. }
\end{figure}

In Figs.~\ref{f:p1} and~\ref{f:p2} we show the multipoles of the power spectrum at $\bar z=1$ as a function of $k$ for all the non-integrated terms. We compare the results obtained from the full-sky correlation~\eqref{e:pn} (coloured lines) with the flat-sky results given in~\eqref{e:Pflatnonint} (black lines). In principle, one could use a sharp cut-off in~\eqref{e:pn} to reflect the fact that the correlation function outside of the observed patch of the sky is zero. However, it is well-know that such a cut-off introduces spurious oscillations to the power spectrum. We therefore use the following window function to smoothly remove scales outside of the observed patch of the sky 
\be 
\label{e:windowni}
W(r)=\frac{1}{2} \left(1-\text{tanh}\left[ \frac{r-\la_1+3\la_2}{\la_2}\right] \right)
\ee 
with $\la_1= 1000 \text{Mpc}/h$ and $\la_2= 50 \text{Mpc}/h$ which gives $\la_\text{smooth} \simeq 700 \text{Mpc}/h$ or $k_\text{smooth} \simeq 0.005 h/\text{Mpc}$.

The multipoles of the standard terms are shown in the left panel of Fig.~\ref{f:p1}. We see that for $k$ larger than the smoothing scale (depicted by the grey vertical line), the full-sky multipoles agree extremely well with the flat-sky expression. For $k$ smaller than the smoothing scale, the full-sky multipoles differ from the flat-sky ones, due to the presence of the window function which removes large scales.  

The right panel of Fig.~\ref{f:p1} shows the multipoles of the Doppler term d1. The full-sky quadrupole (orange) is significantly larger than the flat-sky quadrupole (black dashed). This is due to the fact that in the flat-sky, the contribution coming from the correlation of the Doppler term with the standard terms exactly vanishes, as it gives rise only to odd multipoles, which are exactly zero if one has only one population of galaxies. As a consequence the only contribution to the quadrupole comes from the correlation of the Doppler term with itself. In the full-sky, this is no longer the case. A quadrupole is induced  from the correlation of the Doppler term with the density. This contribution is suppressed by a power $r/\chi\sim \HH/k$ and becomes therefore of the same order of magnitude as the Doppler-Doppler correlation function. This situation again reflects the fact that to properly evaluate the impact of relativistic effects it is not consistent to use the flat-sky approximation, because full-sky corrections generate effects that are of the same order of magnitude as the relativistic terms.

\begin{figure}[t]
\includegraphics[scale=0.53]{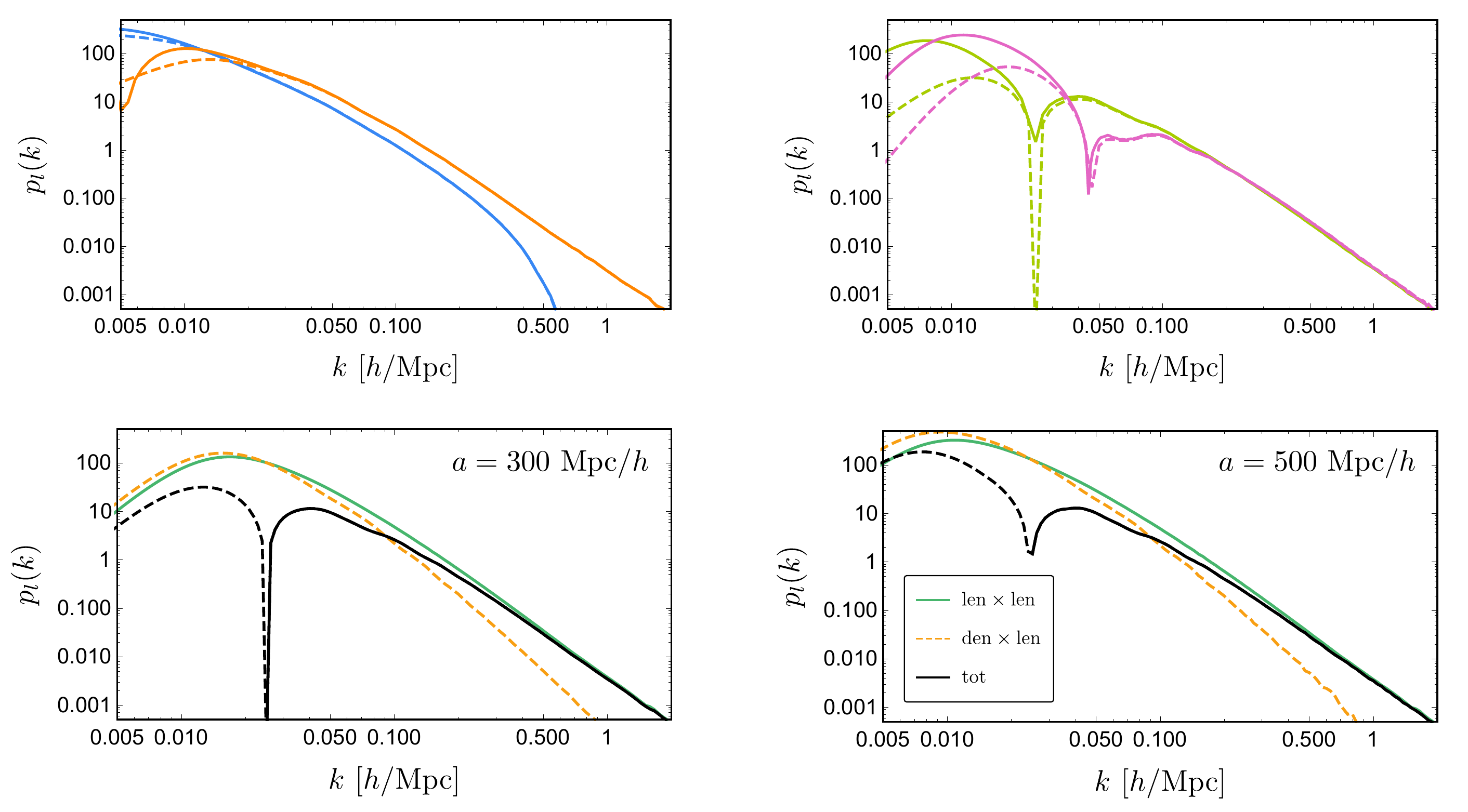}
\caption{\label{f:pk_multipoles_len} The lensing multipoles in the linear regime at $\bar z=1$, calculated from Eq.~\eqref{e:pn} with the window function~\eqref{e:Wgauss}, with $a=300$\,Mpc/$h$ (dashed lines) and $a=500$\,Mpc/$h$ (solid lines). The left panel shows the monopole (blue) and quadrupole (orange). The right panel shows the hexadecapole (green) and the $\ell=6$ multipole (purple).
In the lower panels the hexadecapole contributions from density$\times$lensing and lensing$\times$ lensing are shown separately for clarity.}
\end{figure}

In Fig.~\ref{f:p2} we show the other non-integrated relativistic effects. In this case the full-sky and flat-sky multipoles agree very well. This is due to the fact in this case the difference between the flat-sky and full-sky result is of the order of $(r/\chi)^2$ and not $r/\chi$ and is therefore not visible at $\bar z=1$~\footnote{This can be understood by noting that full-sky corrections to the correlation function bring terms of the form $r/\chi\cdot \mu$. Since the cross-correlation between the standard terms and the Doppler term d1 contains a contribution proportional to $\mu$ in the flat-sky, the first non-zero even multipole in the full-sky will be given by $\mu\times r/\chi\cdot \mu$. On the other hand the flat-sky expression for the other relativistic effects contains even powers of $\mu$ and their full-sky correction must therefore contain at least two  powers of $\mu$, i.e. two powers of $r/\chi$.}.

In Fig.~\ref{f:pk_multipoles_len} we show the multipoles of the lensing contribution.  Here we only calculate the full-sky multipoles given by Eq.~\eqref{e:pn} since the flat-sky expression~\eqref{e:Pflatint} is not well defined for $k_\pa\neq 0$. As discussed before, the lensing power spectrum is extremely sensitive to the cut-off because the correlation function increases with $r$. As a consequence the window function defined in~\eqref{e:windowni} and used for the non-integrated terms is too sharp and not well adapted for the lensing term. It gives rise to large unphysical oscillations in the power spectrum. We therefore use instead a Gaussian window function which is smoother
\be
\label{e:Wgauss}
W(r)=\exp^{-r^2/a^2}\, ,
\ee 
where we consider two different values for $a$: $a=300$\,Mpc/$h$ (dashed lines) and $a=500$\,Mpc/$h$ (solid lines). All multipoles from the lensing term, monopole and quadrupole (left panel) as well as the hexadecapole and $\ell=6$ multipole (right panel) are of the same order of magnitude. This is very different from the standard expression which is dominated by the monopole and quadrupole. The hexadecapole and the $\ell=6$ multipole depend more strongly on the value of $a$ than the monopole and quadrupole, which differ only for $k\leq 0.03 h$/Mpc.  Nevertheless, the passage through zero is independent of the window size. This zero of the hexadecapole and of the $\ell=6$ multipole is due to the competition between the positive lensing-lensing correlation which dominates at large $k$ and the negative density-lensing correlation which dominates at small $k$, as can be seen from the bottom panels of Fig.~\ref{f:pk_multipoles_len}. 

\begin{figure}[t]
\centering
\includegraphics[scale=0.6]{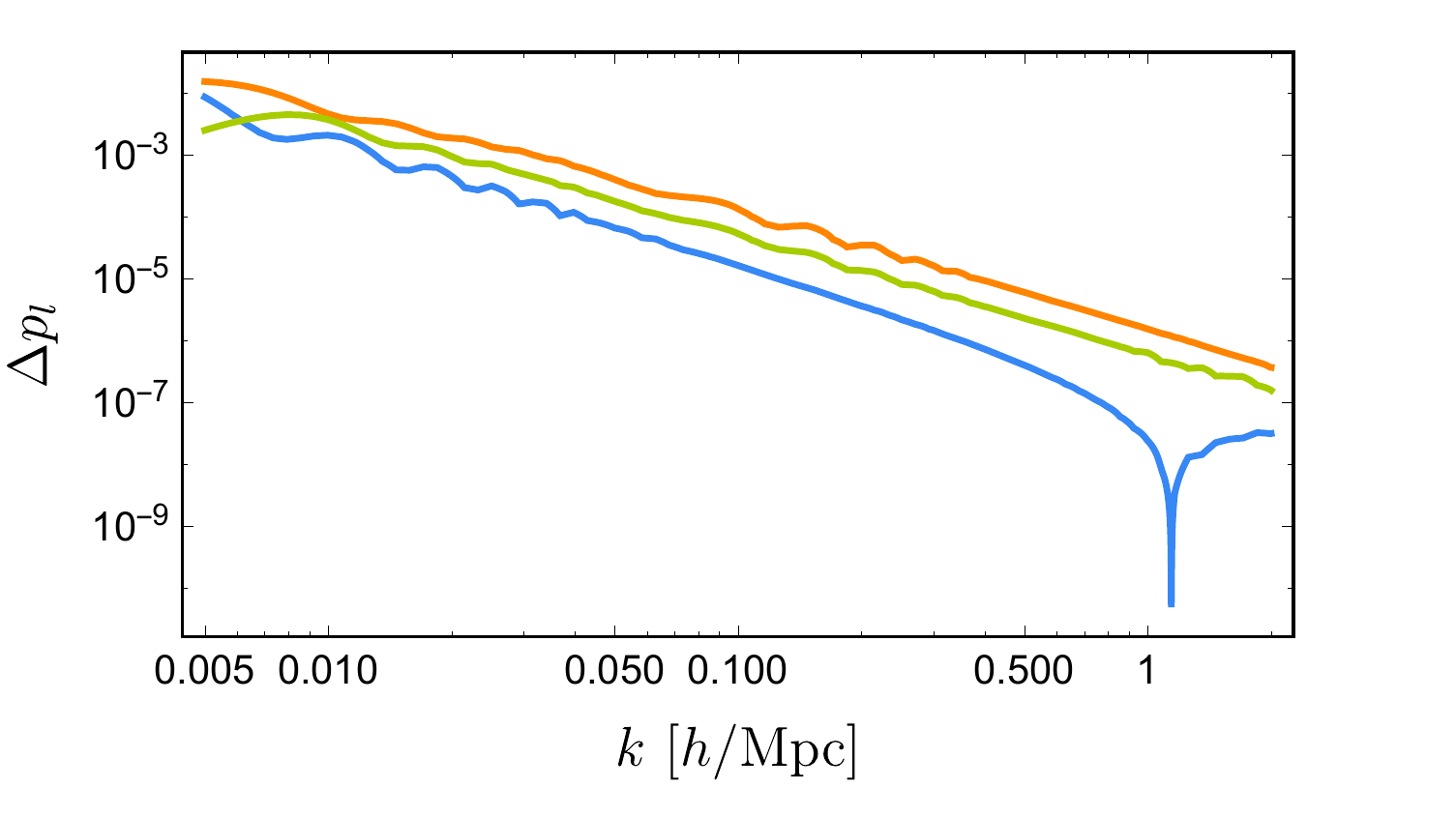}
\caption{\label{f:Pfracnonint} Fractional difference at $\bar z=1$ generated by the large-scale relativistic effects on the monopole (blue), quadrupole (orange) and hexadecapole (green) of the power spectrum.}
\end{figure}

The standard monopole, quadrupole and hexadecapole of the power spectrum are used to measure the growth rate and constrain cosmological parameters. Since large-scale relativistic effects and gravitational lensing contribute to these multipoles, they can in principle contaminate this estimation. In Fig.~\ref{f:Pfracnonint} we show the fractional difference between the full-sky non-integrated relativistic  multipoles and the flat-sky standard multipoles at $\bar z=1$
\be
\Delta p_\ell^{\rm rel}=\frac{p_\ell^{\rm rel}}{p^{\rm st}_\ell}\, ,
\ee
where $p_\ell^{\rm rel}$ denotes the multipoles from all the non-integrated relativistic effects and their correlation with the standard terms, similarly to~\eqref{defrel}. We see that the correction generated by the relativistic effects is less than a percent at all scales and can therefore be neglected. At small redshift $\bar z=0.1$ we expect a larger contribution, similar to the one that affects the multipoles of the correlation function, see Fig.~\ref{f:Delta_rel_mult}. However we found that this contribution strongly depends on the window function and we defer therefore a careful study of this effect to a future publication~\cite{prep}.

\begin{figure}[t]
\centering
\includegraphics[scale=0.44]{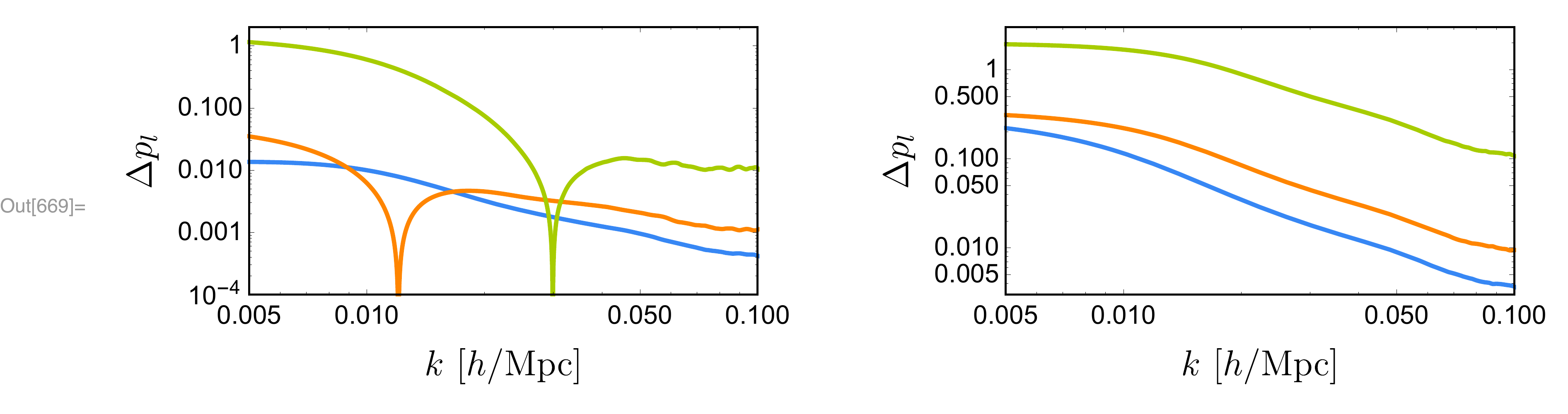}
\caption{\label{f:Pfracint} Fractional difference generated by the non-linear full-sky lensing on the monopole (blue), quadrupole (orange) and hexadecapole (green) of the power spectrum. Here a Gaussian window with $a=300$\,Mpc/$h$ has been used. The left panel is for $\bar z=1$ and the right panel for $\bar z=2$.}
\end{figure}

In Fig.~\ref{f:Pfracint} we show the fractional difference between the lensing multipoles and the flat-sky standard multipoles at $\bar z=1$ and $\bar z=2$
\be
\Delta p_\ell^{\rm lens}=\frac{p_\ell^{\rm lens}}{p^{\rm st}_\ell}\, ,
\ee
where $p_\ell^{\rm lens}$ denotes the multipoles from the lensing and its correlation with the standard terms. We see that above 0.01\,$h/$Mpc the lensing contribution to the monopole and quadrupole is less than a percent. Only on very small $k$ does it reach a few percents. The hexadecapole is more strongly affected at all scales. At $\bar z=2$, the monopole and quadrupole get corrections of 10-20\% at small $k$ and the corrections remain above 1\% at all scales. These numbers seem to be in broad agreement with the flat-sky results of~\cite{Hui:2007tm}~\footnote{Note that the fractional differences in~\cite{Hui:2007tm} are with respect to the BBKS power spectrum which contains no baryons and no redshift-space distortions and is linear, whereas our result is with respect to the non-linear standard power spectrum which contains density and redshift-space distortions. It is therefore expected that our fractional difference be smaller.}. Again the hexadecapole is strongly affected by lensing at all scales.

Note that as mentioned previously, the multipoles of the power spectrum strongly depend on the window function chosen to integrate~\eqref{e:pn}, especially for the lensing contribution which grows with separation. In addition the multipoles of the power spectrum depend on the minimal separation we use in the integral~\eqref{e:pn}, which in practice is given by the size of the pixels in which we measure the number counts. In particular, we have found that a window function which is too sharp leads to strong oscillations in the power spectrum. Similarly, the lower cutoff leads to oscillations for $k>\pi/r_{\rm min}$. In our case we choose $r_{\rm min}=8$\,Mpc/$h$ leading to oscillations around $k\sim 0.2\,h$/Mpc. The results shown in Figs.~\ref{f:pk_multipoles_len},~\ref{f:Pfracnonint} and~\ref{f:Pfracint} should therefore be taken with some caution as they will depend on the form of the window, the smoothing scale and the minimum separation used in the Fourier transform. We defer a more detailed study of these parameters to a future publication, where we will also analyse the impact of the large-scale relativistic effects and of the lensing on the determination of cosmological parameters~\cite{prep}.

\section{Discussion and Conclusions}\label{s:dis}

In this paper we have studied the redshift-space correlation function and the power spectrum of galaxy number counts. Even though these functions depend on the cosmological model used to convert angles and redshifts into distances~\footnote{Note that deviations from the fiducial model can be accounted for in a consistent way by introducing correction parameters that rescale the correlation function, see e.g.~\cite{Xu:2012fw}.}, they are useful for several reasons. First they are well adapted to describe the 3-dimensional information present in large-scale structure. This is not the case for the observable $C_\ell(z_1,z_2)$ angular-redshift power spectrum for which we cannot employ very fine redshift binning due to under-sampling. Second, the multipoles of the correlation function and of the power spectrum contain important information about the growth of perturbations which is difficult to isolate in the angular-redshift power spectrum. We therefore propose to use the redshift-space correlation function to analyse thin shells in redshift space, $\De z\sim 0.2$ and the power spectrum to analyse small (a few 100 Mpc)  patches of sky.

Computing these quantities within linear perturbation theory and with the halofit approximation, we have shown how they are affected by large-scale relativistic effects and by lensing. The large-scale relativistic effects are important mainly at small redshifts. At $z=0.1$ they introduce corrections to the monopole and quadrupole of the correlation function of the order of 10\% at a separation of 100\,Mpc/$h$ and they quickly increase with separation. The hexadecapole is less affected at intermediate scales, but at large scales the correction becomes similar to the other multipoles. We have seen that this large correction is due to the Doppler effect, which contains a term proportional to $1/(\HH\chi)$ which is enhanced at small redshift. This term has previously been identified in~\cite{Papai:2008bd,Raccanelli:2010hk,Samushia:2011cs}. At large redshift however,  this Doppler term contributes to the multipoles at the same level as the other relativistic effects and generates corrections that are never larger than about 1\%. We have also seen that full-sky corrections to the correlation function are of the same order as relativistic corrections. It is hence inconsistent to take onto account only one or the other. They have to be discussed together as we do it in this work.

At large redshift the lensing term becomes much more relevant than the large-scale relativistic contributions. Furthermore, the importance of  lensing strongly depends on the orientation of the pair of galaxies. In particular it is most important along the line-of-sight, when $\mu\sim 1$. In this case on large scales, $r>200$\,Mpc/$h$, the lensing term even dominates over the standard terms (see Fig.~\ref{f:c2}). We have also studied the contribution of lensing to the multipoles of the correlation function and of the power spectrum and we have seen that at $z=1$ lensing modifies the monopole and quadrupole of the correlation function and of the power spectrum by a few percents. At larger redshift $z=2$ these corrections amount to 10-30\% at intermediate scales and quickly increase with separation. This clearly shows that lensing cannot be neglected in the analysis of future galaxy surveys at high redshift. Moreover we have seen that the hexadecapole of the correlation function and of the power spectrum are strongly affected by lensing at $z=1$ and $z=2$. This comes from the fact that the hexadecapole from the standard terms is significantly smaller than the monopole and quadrupole, whereas the hexadecapole of lensing is of the same order as the monopole and quadrupole (as can be seen from Fig.~\ref{f:len_multipoles}). Measuring the hexadecapole is expected to provide a clean way of measuring the growth rate $f$ since it is independent of bias. Here we see however that such a measurement would require a careful modelling of the lensing contribution. Furthermore, we have found that lensing generates significant higher multipoles $\ell>4$ in the correlation function and in the power spectrum, see Figs.~\ref{f:L_mult}, \ref{f:len_multipoles}, \ref{f:Delta_lens_mult} and \ref{f:pk_multipoles_len}. 

In our work, contrary to previous studies on the subject, we have derived an expression for the lensing correlation function which is exact, i.e. which does not  rely on the flat-sky and Limber approximation. By comparing our result with the flat-sky result, we have found that the flat-sky approximation is only good in forward direction, $\mu=1$, see Fig.~\ref{f:lensing_linear}. The full-sky lensing multipoles differ from the flat-sky one by 20-40\%, see Fig.~\ref{f:L_mult}. 
Finally, we have seen that due to the mixing of scales, non-linearities in the matter power spectrum are relevant for lensing even for large separations out to $r>200$\,Mpc/$h$ for $\mu\sim 1$ where lensing is most relevant, see Fig.~\ref{f:lensing_nonlinear}. A correct treatment of lensing requires therefore the use of the full-sky non-linear expressions.

The presence of higher multipoles in both, the correlation function and the power spectrum, might represent  an ideal observational target to identify the lensing term. As it has been discussed previously~\cite{Montanari:2015rga}, measuring the convergence $\ka$ via the lensing of number counts is a promising alternative to shear measurements. On the other hand, it has been shown that neglecting lensing in the analysis of future surveys, at least for photometric surveys induces significant errors in parameter estimation~\cite{Cardona:2016qxn}. It will be important to investigate whether this is also the case when precise spectroscopic redshifts are available. We shall study this in a forthcoming paper~\cite{prep} using the methods outlined in this work.

\acknowledgments{We thank Chris Clarkson, Roy Maartens and Mariele Motta for interesting discussions and comments. This work is supported by the Swiss National Science Foundation.}

\appendix

\section{Relations between the angles \label{a:angles}}
\begin{figure}[ht]
\centering
\includegraphics[scale=0.5]{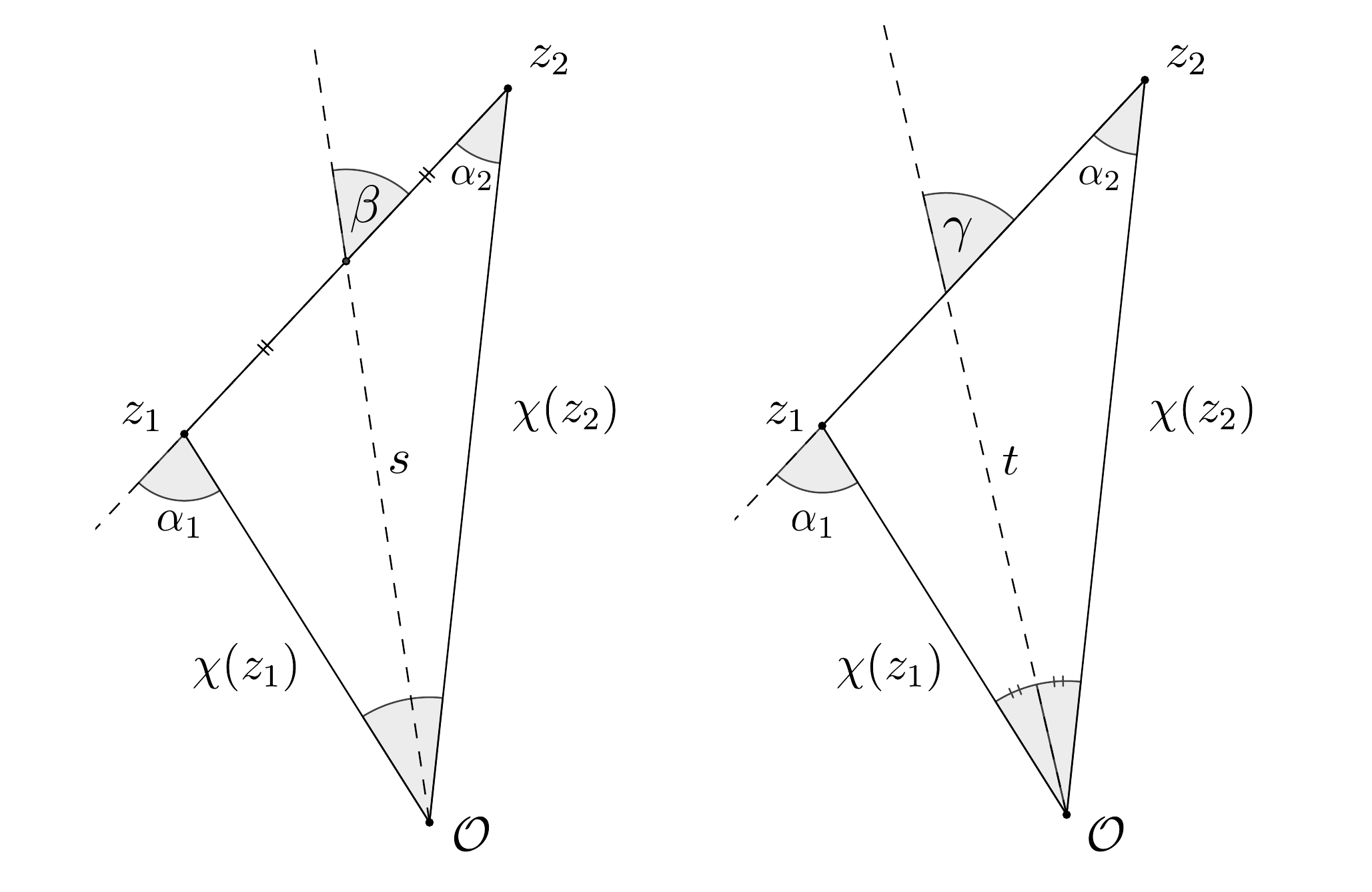}
\caption{\label{f:a-gammabeta} The angles $\al_1$, $\al_2$, $\beta$, $\ga$ and the lengths $s$ and $t$ used to determine respectively $\beta$ and $\gamma$ are indicated.}
\end{figure}

In this appendix we derive in detail the relation between the angles $\theta$, $\al$, $\beta$ and $\gamma$, see 
Fig.~\ref{f:a-gammabeta}. More precisely, we give expressions for $\cos\al$, $\cos\beta$ and $\cos\ga$ in terms of $r$, $\cos\theta$ and $\bar z= (z_1+z_2)/2$ or rather $\bar\chi =\chi(\bar z)$. Note that $(\chi_1+\chi_2)/2$ and $\chi(\bar z)$ differ by a term of order $(\De z)^2/H(\bar(z)$ which we neglect.

As defined in the main text, $\al$ is the angle between the line of length $r$ connecting the two positions at redshifts $z_1$ and $z_2$ which span an angle $\theta$ at the observer and the line connecting $z_2$ and the intersection or the circle or radius  $r_\pa$ around $z_2$ with the Thales circle over $\br$ (see Fig.~\ref{f:angles}, left panel). Evidently $\al$ is given by 
\bea
\cos\al &=& r_\pa/r =  \frac{2}{r}\sqrt{\bar\chi^2 -\frac{4\bar\chi^2-r^2}{2(1+\cos\theta)}} \,.
\eea
Here we have used eq.~\eqref{e:rpab} to express $r_\pa$ in terms of $(r,\bar\chi,\cos\theta)$.

The angle $\beta$ is obtained as follows: We denote by $s$ the length of the line from the observer $O$ to the middle of $r$ and by $\al_2$ the angle of the triangle $(O,z_2,z_1)$ at $z_2$, see Fig.~\ref{f:a-gammabeta}. 
The cosine law gives the following relations
\bea
\chi_2^2 &=&s^2 +(r/2)^2 +rs\cos\beta\,, \qquad
s^2 \;=\; (r/2)^2 +\chi_2^2-r\chi_2\cos\al_2
\eea
Eliminating $s$ and solving for $\cos\beta$ we find
\be
\hspace*{-1cm}\cos\beta = \frac{-r^2/2+r\chi_2\cos\al_2}{r\sqrt{(r/2)^2+\chi_2^2-r\chi_2\cos\al_2}}
\ee
Using furthermore
$$ \cos\al_2=\frac{\chi_2-\chi_1\cos\theta}{r}$$
we obtain after some simplifications
\be
\cos\beta = \frac{\chi_2^2-\chi_1^2}{r\sqrt{\chi_1^2+\chi_2^2+2\chi_1\chi_2\cos\theta}}= \frac{2\bar\chi}{r}\sqrt{\frac{2r^2-4(1-\cos\theta)\bar\chi^2}{8\bar\chi^2\cos\theta^2+(1\cos\theta)r^2}} \,.
\ee
For the second line we used expressions \eqref{e:chi12} for $\chi_{1,2}$.

Considering the angle $\ga$ and using $t$ as indicated in Fig.~\ref{f:a-gammabeta} and $\al_2$ as before we see that $\ga=\theta/2+\al_2$ hence
$$\cos\ga=\cos(\theta/2)\cos\al_2-(1-\cos^2\theta/2)^{1/2}(1-\cos^2\al_2)^{1/2}$$

Inserting 
$$\cos\theta/2=\left(\frac{1+\cos\theta}{2}\right)^{1/2}$$
and the expressions for $\cos\al_2$ we obtain
\bea
\cos\ga &=&\frac{(1+\cos\theta)^{1/2}(\chi_2-\chi_1)}{\sqrt{2}r} 
 \;=\; \frac{\sqrt{r^2-2(1-\cos\theta)\bar\chi^2}}{r} \,.
\eea

Again we have inserted the expressions \eqref{e:chi12} for $\chi_{1,2}$ in the last equal sign.

We shall also use the expressions for $\cos\al_i$ which are easily derived from the cosine theorem:
\bea
\cos\al_2=&\hat\br\cd\bn_2 &=~ \frac{\chi_2-\chi_1\cos\theta}{r}\,,\qquad
\cos\al_1=\; \hat\br\cd\bn_1 \;=~ -\frac{\chi_1-\chi_2\cos\theta}{r} \,.
\eea

\section{The full angular--redshift correlation function\label{a:corf}}
The 'full angular redshift correlation function' is $\xi(\theta,z_1,z_2)$ when we include all the relativistic terms. It can be computed as follows.\\
We first write down derivatives of Eq.~\eqref{e:pljlj0} wrt $\chi_1$ and $\chi_2$ which are encoded in the functions $\zeta^{ij}(k \chi_1,k\chi_2)$. Using $r=\sqrt{\chi_1^2+\chi_2^2-2\chi_1\chi_2\cos\theta}$ and the recurrence relations for derivatives of spherical Bessel functions
$$j_\ell' = \frac{1}{2\ell+1}\left(\ell j_{\ell-1} -(\ell+1) j_\ell\right)\quad 
\mbox{ and }\quad
\frac{j_\ell(x)}{x} = \frac{1}{2\ell+1}\left( j_{\ell-1} + j_\ell\right)(x)$$ 
we find

\bea
\zeta^{00} &=& j_0 ( k r)\\ 
\zeta^{01} &=& \frac{\chi_1 \cos \theta -\chi_2}{r} j_1(k r)=- j_1(k r) \cos\al_2 \\
\zeta^{11} &= &\left(\frac{2 }{k r } j_1(k r) - j_0 (k r) \right) \left(\frac{\chi_1-\chi_2 \cos \theta }{r}\right)\left(\frac{\chi_2 -\chi_1 \cos \theta}{r} \right)+j_1( k r) \frac{\chi_1 \chi_2 \sin^2 \theta}{k r^3} \nonumber\\
&=& -\frac{(\chi_2-\chi_1\cos\theta)(\chi_1-\chi_2\cos\theta)}{r^2}j_2(kr)+\frac{\cos\theta}{3}\left(j_0(kr) -j_2(kr)\right) \nonumber\\
 &=& j_2(k r) \cos\al_2\cos\al_1 +\frac{\cos(\al_2-\al_1)}{3}\left[j_0(kr) -j_2(kr)\right] 
 \\
\zeta^{02}&=& \left(\frac{2 }{k r } j_1(k r) - j_0 (k r) \right)\left(\frac{\chi_2 -\chi_1 \cos \theta}{r} \right)^2 -j_1( k r) \frac{\chi_1^2 \sin^2 \theta}{k r^3} \nonumber \\
&=&  \left(\frac{2}{3} -(1-\cos^2\theta)\frac{\chi_1^2}{r^2}\right)j_2(kr) -\frac{1}{3}j_0(kr) \nonumber\\
&=&  \left(\frac{2}{3} -\sin^2\al_2\right)j_2(kr) -\frac{1}{3}j_0(kr)  \\
\zeta^{12} &=& \frac{(1+2\cos^2\theta)\chi_1-3 \chi_2\cos\theta}{5r}j_1(kr)  +\nonumber\\
&&\frac{(1-3\, \cos^2\theta)\chi_1^3 +\cos\theta(5+\cos^2\theta)\chi_1^2\chi_2-2(2+\cos\theta^2)\chi_1\chi_2^2 +2\chi_2^3\cos\theta}{5r^3}j_3(kr)\nonumber\\
&=& -\frac{\left[2\cos(\al_2-\al_1)\cos\al_2+\cos\al_1\right]}{5}j_1(kr) +
\Big[\cos\al_1 \sin ^2\al_2-\frac{2}{5} \cos\al_2 \cos (\al_1-\al_2)\Big]j_3(kr)
 \nonumber\\  &&
\\
\zeta^{22} &=& \frac{1+2\cos^2\theta}{15 } j_0(kr)-\frac{1}{21}\left[1+11\cos^2\theta +\frac{18\cos\theta(\cos^2\theta-1)\chi_1\chi_2}{r^2}\right]\ j_2(kr)  + \nonumber \\ &&
 \left[\frac{4(3\cos^2\theta-1)(\chi_1^4+\chi_2^4)}{35r^4} +  \chi_1\chi_2 (3+\cos^2\theta)\frac{3 (3+\cos^2\theta)\chi_1\chi_2-8(\chi_1^2+\chi_2^2)\cos\theta }{35r^4} \right]j_4(kr ) \nonumber \\ 
&=&  \frac{1+2\cos^2(\al_1-\al_2)}{15 } j_0(kr)-\frac{1}{42}\left[4+9\cos(2\al_1)+9\cos(2\al_2)+2\cos(2(\al_1-\al_2))\right] j_2(kr)+ \nonumber\\
&&\frac{\left[3 \cos (2 (\al_1-\al_2))+35 \cos (2 (\al_1+\al_2))+10 \cos (2 \al_1)+10 \cos (2 \al_2)+6\right]}{280}  j_4(kr)  \,.  
\eea
The coefficients $\zeta^{21}(x_1,x_2)$ etc. are obtained from $\zeta^{12}$  etc. via the symmetry relation
$$ \zeta^{ij}(x,y) =\zeta^{ji}(y,x) \,.$$
The flat sky limit of the above function is obtained by setting $\al_1=\al_2\equiv \al$. In this case all the terms in front of a $j_\ell$ are a multiple of the  Legendre polynomial $L_\ell(\cos\al)$. More precisely, denoting the flat sky limit of $ \zeta^{ij}$ by $ \bar\zeta^{ij}$ we obtain

\bea
\bar\zeta^{00} &=&j_0 ( k r)\,, \hspace{4.7cm}\\\
\bar\zeta^{01} &=& -L_1(\cos\al) j_1(k r)\,, \qquad \qquad
\bar\zeta^{11} \;=\; \frac{2}{3}L_2(\cos\al) j_2(k r)+\frac{1}{3} j_0 ( k r)\,, \\ 
\bar\zeta^{02} &=&  \frac{2}{3}L_2(\cos\al) j_2(k r)-\frac{1}{3} j_0 ( k r)\,, \\
\bar\zeta^{12} &=& -3L_1(\cos\al) j_1(k r)-\frac{2}{5}L_3(\cos\al)\,,\\
\bar\zeta^{22} &=&  \frac{8}{35}L_4(\cos\al) j_4(k r)+\frac{4}{7}L_2(\cos\al) j_2(k r)+\frac{1}{5} j_0 ( k r)\,.
\eea
The terms $\bar\zeta^{00}$,  $\bar\zeta^{02}$ and $\bar\zeta^{22}$ give rise to the standard flat sky result~\eqref{e:P0} to \eqref{e:P4}.  The flat sky results $\bar\zeta^{01} $ and $\bar\zeta^{12}$ are more subtle. Since we always have to add 
$\bar\zeta^{ij}+ \bar\zeta^{ji}$ and $ \bar\zeta^{ij}(\cos\al)= \bar\zeta^{ji}(\cos(\pi-\al))=\ \bar\zeta^{ji}(-\cos\al)$ these odd terms actually cancel and do not contribute
in the case of a single population of galaxies. They do contribute to a multi tracer signal, see~\cite{Bonvin:2013ogt}.

The only coefficients that do not fall into this category, as explained in the main text, are the lensing terms which are computed using the identity  $$-\ell(\ell+1) L_\ell (\cos \theta)=\bigtriangleup_\Omega L_\ell (\cos \theta)= \frac{1}{\sin\theta}\partial_\theta\left(\sin\theta\partial_\theta L_\ell(\cos\theta)\right) \,.$$
They are given explicitly  by

\bea
\zeta^{0\text{L}} &=& 2\frac{k \chi_1\chi_2 \cos\theta}{r} j_1(kr) -\left(k^2 \frac{\chi_1^2\chi_2^2\sin^2\theta}{r^2}\right) j_2(kr) \nonumber\\
&= &k^2 \left[\frac{2}{3} \chi_1 \chi_2 \cos\theta j_0(kr) +\frac{\chi_1 \chi_2}{3}\left(2 \cos\theta -3 \chi_1 \chi_2\frac{\sin^2\theta}{r^2} \right) j_2(kr) \right]\\
&=& \frac{(kr)^2}{3}\Bigg[2\frac{\sin\al_1\sin\al_2\cos(\al_1-\al_2)}{\sin^2(\al_1-\al_2)}j_0(kr)
+\frac{ \sin\al_1\sin\al_2}{\sin^2(\al_1-\al_2)}\times
 \nonumber \\ && 
 \left[\cos (\al_1-\al_2)+\cos\al_1\cos\al_2)\right]  j_2(kr)\Bigg]  \\
\zeta^{1\text{L}}& =&k^2 \left[\frac{2}{3} \chi_2 r\cos\theta j_{-1}\left(kr\right)+
\frac{2\chi_2(\chi_1\cos\theta-\chi_2)(\chi_1-2\chi_2\cos\theta)}{5r}  j_1(kr)-
\right.   \nonumber\\   & & 
\frac{1}{15 r^3}\Bigl(\chi_2(4\chi_1^4\cos\theta-(9+\cos^2\theta) \chi_1^3 \chi_2+\cos\theta \left(\cos^2\theta+5\right) \chi_1^2 \chi_2^2+ \nonumber\\&& 2(3-2\cos^2\theta)\chi_1 \chi^3_2 -2 \chi_2^4\cos\theta) \Bigr)j_3\left(k r\right)\biggr] \,, \\ 
&=& (kr)^2 \left[\frac{2}{3}\frac{\cos(\al_1-\al_2)\sin\al_1}{\sin(\al_1-\al_2)}j_{-1}\left(kr\right)-\frac{2}{5} \frac{(2\sin \al_1- \sin \al_2)(\sin\al_1-\cos(\al_1\!-\!\al_2)\sin\al_2)}{ \sin^3 (\al_1-\al_2)}\times
\right.   \nonumber\\   & & 
\left. j_1(kr) -\frac{1}{120}\frac{\sin\al_1\left[6\sin(2\al_1)+\sin(2(\al_1-\al_2))-15\sin(2(\al_1+\al_2))\right]}{\sin^2(\al_1-\al_2)}j_3(kr)\right] 
\eea    \bea
\zeta^{2\text{L}}& =& -k^2\bigg\{\frac{2}{15} \chi_2 \left(3 \chi_1\cos\theta+(1-3\cos^2\theta)\chi_2\right) j_0\left(kr\right) + \nonumber\\   &&   \hspace{-1cm}
\bigg[ \frac{6 \chi_1^3\chi_2 \cos\theta -\left(9 \cos^2\theta+11\right) \chi_1^2 \chi^2_2}{21r^2} 
+\frac{2 \cos\theta \left(3 \cos^2\theta+8\right) \chi_1 \chi_2^3+4 \left(1-3 \cos^2\theta\right) \chi_2^4}{21r^2}\bigg] j_2\left(kr\right)
  \nonumber\\   &&  \left.  
+\left[\frac{\chi_2 \left(2 \left(1-3 \cos^2\theta\right) \chi_2^5   +6 \cos\theta (3-\cos^2\theta) \chi_1 \chi_2^4  +(\cos^4\theta+12 \cos^2\theta-21) \chi_1^2 \chi_2^3\right)}{35r^4}
\right. \right.   \nonumber\\   &&  \left.  \left.
- \;\frac{\chi_2 \left(2 \cos\theta \left(\cos^2\theta+3\right) \chi_1^3 \chi_2^2-12 \chi_1^4 \chi_2+4\chi_1^5 \cos\theta \right)}{35r^4}\right] j_4(kr)
\right\} 
\\
&=& -(kr)^2\Bigg\{\frac{1}{15} \frac{\sin\al_1 (2 \sin\al_1-3 \sin (2 (\al_1-\al_2))\cos\al_1 )}{ \sin ^2(\al_1-\al_2)}j_0(kr)  + \frac{\sin (\al_1)}{84\sin ^4(\al_1-\al_2) }\times
\nonumber \\
&&
\left[3 \sin (3 \al_1) (\cos (2 \al_2)+3)-12 \cos ^3\al_1 \sin (2 \al_2)-\sin \al_1 (3 \cos (2 \al_2)+1)\right] j_2(kr) \nonumber \\
&&
+\frac{\sin\al_1}{560 \sin^2(\al_1-\al_2)} 
  \bigg[ 5 \sin (\al_1\!+\!2 \al_2)-35 \sin (3 \al_1\!+\!2 \al_2)  \nonumber\\   && +\sin (\al_1\!-\!2 \al_2)+\sin (3 \al_1\!-\!2 \al_2)+2 \sin\al_1+10 \sin (3 \al_1)\bigg] j_4(kr) \Bigg\}  \\
\zeta^{\text{LL}} &=& -\sin^2\theta(k^2\chi_1\chi_2)^2 \left[\!\left(\!\frac{6(r^2+5\chi_1\chi_2\cos\theta)}{35r^2} -\frac{\chi_1^2\chi_2^2\sin^2\!\theta}{r^4}\!\right) j_4\left(kr\right)
\right.   \nonumber\\
&&\left.
+\frac{2  \left(2r^2+3\chi_1\chi_2\cos\theta\right)}{7r^2}j_2(k r) \ +\frac{2}{5}j_0\left(kr\right)\right]  \nonumber\\
&&+ 4 \cos\theta k^3\chi_1\chi_2 \bigg[\left(\frac{r^2+6\chi_1\chi_2\cos\theta}{15r}-\frac{\chi_1^2 \chi_2^2\sin^2\theta}{ 2r^3}\right) j_3\left(k r\right)  \nonumber \\&&
+\frac{2 \left(r^2+ \chi_1 \chi_2\cos\theta\right)}{5 r} j_1\left(kr\right)+\frac{r}{3} j_{-1}\left(kr\right)\bigg]
\\
&=& (kr)^3\Bigg\{\;\frac{4}{3}\frac{ \sin\al_1\sin \al_2 \cos ( \al_1- \al_2)}{ \sin^2(\al_1-\al_2)}j_{-1}(kr) \nonumber\\
&&  -\frac{2}{5}\frac{ \sin\al_1\sin\al_2\cot (\al_1-\al_2) [\cos (2 (\al_1-\al_2))+\cos (2 \al_1)+\cos (2 \al_2)-3]  }{\sin ^3(\al_1-\al_2)}j_{1}(kr) +\nonumber\\
&& \hspace{-0.93cm}\frac{\sin\al_1 \sin\al_2 \cos (\al_1\!-\!\al_2)}{60\sin^4(\al_1\!-\!\al_2)}  \left[2+6\cos(2\al_1)+\cos(2(\al_1\!-\!\al_2))+6\cos(2\al_2) \right.
 \nonumber\\   && \left.
 -15\cos(2(\al_1\!+\!\al_2))\right]j_3(kr)
 \Bigg\}
+(kr)^4\Bigg\{-\frac{2}{5}\frac{\sin^2\al_1\sin^2\al_2}{\sin^2(\al_1-\al_2)}j_0(kr) -
 \frac{2\sin ^2\al_1 \sin ^2\al_2}{7 \sin ^4(\al_1-\al_2)}\times \nonumber\\ && \left[2\sin^2  (\al_1-\al_2)+3 \cos ( \al_1-\al_2)\sin\al_1\sin\al_2\right]j_2(kr) +\frac{ \sin ^2\al_1 \sin ^2\al_2}{280\sin^4(\al_1-\al_2)} \times \nonumber\\
 &&\hspace{-0.3cm} \bigg[35 \cos (2(\al_1+\al_2))-10 \cos (2\al_2)- \cos (2 (\al_1-\al_2))-10\cos(2\al_1)-14
 \bigg]j_4(kr)
  \Bigg\}
\,.
\eea

For the lensing terms the flat sky limit cannot be obtained by setting $\al_1=\al_2$ since the terms $\xi^{iL}$ diverge in this limit. We discuss the flat sky approximation of lensing in Appendix~\ref{a:flat}.

We now give explicit expressions for the $Q^{AB}_k$ in terms of the $\zeta^{ij}$, to be inserted in eq.~(\ref{corrQ}) to build the correlation function:

\begin{flalign*}
\label{e:Qfirst}
&Q^\text{den}(\theta,z_1,z_2) = b(z_1) b(z_2)  S_D(z_1) S_D (z_2) \,\zeta^{00}(k\chi_1,k\chi_2,\theta)  \,,\\
&Q^\text{rsd}(\theta,z_1,z_2) =  \frac{k^2}{\HH_1\HH_2} S_V(z_1)S_V(z_2)  \,\zeta^{22}(k\chi_1,k\chi_2,\theta)  \,, \\
&Q^\text{len}(\theta,z_1,z_2) =\!  \frac{\left(2 - 5 s\right)^2}{4 \chi_1\chi_2} \!  \int_0^{\chi_1} \!\! \!\!\int_0^{\chi_2} \!\!\!\dd \la \dd \la'  \frac{(\chi_1\!-\!\la)(\chi_2\!-\!\la')}{\la\la'}S_{\phi+\psi}(\la)S_{\phi+\psi}(\la') \zeta^{LL}(k\la,k\la',\theta)   \,,\\
&Q^\text{den-rsd}(\theta,z_1,z_2) = \frac{k b(z_1)}{\HH_2}  S_D(z_1)S_V(z_2)  \,\zeta^{02}(k\chi_1,k\chi_2,\theta)  \,,\\
&Q^\text{den-len}(\theta,z_1,z_2) = b(z_1) S_D(z_1) \left(\frac{2 - 5 s}{2\chi_2}\right) \int_0^{\chi_2} \dd \la  \frac{\chi_2-\la}{\la} S_{\phi+\psi}(\la)  \, \zeta^{0L}(k\chi_1,k\la,\theta)   \,,\\
&Q^\text{rsd-len}(\theta,z_1,z_2) =  \frac{k}{\HH_1}S_V(z_1) \left(\frac{2 - 5 s}{2\chi_2}\right) \int_0^{\chi_2} \dd \la  \frac{\chi_2-\la}{\la}S_{\phi+\psi}(\la) \, \zeta^{2L}(k\chi_1,k\la,\theta)  \,, \\
&Q^\text{d1}(\theta,z_1,z_2) = \left[\left(\frac{\dot \HH}{\HH^2}+\frac{2-5s}{\chi \HH}+5 s - f_\text{evo} \right) S_V\right]\!\!(z_1)   \\
& \hspace{2.22 cm}\times \left[\left(\frac{\dot \HH}{\HH^2}+\frac{2-5s}{\chi \HH}+5 s - f_\text{evo} \right) S_V\right]\!\!(z_2)\, \zeta^{11}(k\chi_1,k\chi_2,\theta)  \,,  \\   
&Q^\text{X}(\theta,z_1,z_2) =\De^\text{X}(z_1,k)\De^\text{X}(z_2,k)\zeta^{00}(k\chi_1,k\chi_2,\theta)    \hspace{2cm}   X \in   \{\text{d2, g1, g2, g3}\}  \,,  \\  
&Q^\text{g4}(\theta,z_1,z_2) = \frac{( 2-5s)^2}{\chi_1\chi_2}  \int_0^{\chi_1} \dd \la  \int_0^{\chi_2} \dd \la'S_{\phi+\psi}(\la,k) S_{\phi+\psi}(\la',k) \zeta^{00}(k\la,k\la',\theta)   \,, \\
&Q^\text{g5}(\theta,z_1,z_2) =\left(\frac{\dot \HH}{\HH^2}+\frac{2-5s}{\chi \HH}+5 s - f_\text{evo} \right)\!(z_1)\left(\frac{\dot \HH}{\HH^2}+\frac{2-5s}{\chi \HH}+5 s - f_\text{evo}\right)\!(z_2) \\
& \hspace{2.22 cm}\times  \int_0^{\chi_1} \dd \la  \int_0^{\chi_2} \dd \la'\,\dot S_{\phi+\psi}(\la,k) \dot S_{\phi+\psi}(\la',k) \zeta^{00}(k\la,k\la',\theta)  \,, \\
&Q^\text{den-d1}(\theta,z_1,z_2) = b(z_1) S_D(z_1)\left[\left(\frac{\dot \HH}{\HH^2}+\frac{2-5s}{\chi \HH}+5 s - f_\text{evo} \right) S_V\right]\!(z_2) \zeta^{01}(k\chi_1,k\chi_2,\theta)  \,, \\
&Q^\text{den-X}(\theta,z_1,z_2) = b(z_1) S_D(z_1)\De^\text{X}(z_2,k)\zeta^{00}(k\chi_1,k\chi_2,\theta)  \,,\\
&Q^\text{den-g4}(\theta,z_1,z_2) = b(z_1) S_D(z_1) \frac{ 2-5s}{\chi_2}  \int_0^{\chi_2} \dd \la S_{\phi +\psi}(\la,k) \zeta^{00}(k\chi_1,k\la,\theta)   \,, \\
&Q^\text{den-g5}(\theta,z_1,z_2) = b(z_1) S_D(z_1)\left(\frac{\dot \HH}{\HH^2}+\frac{2-5s}{\chi \HH}+5 s - f_\text{evo} \right)\!(z_2)   \\
& \hspace{2.22 cm}\times \int_0^{\chi_2} \dd \la  \dot S_{\phi +\psi}(\la,k) \zeta^{00}(k\chi_1,k\la,\theta)  \,, \\
&Q^\text{rsd-d1}(\theta,z_1,z_2) =  \frac{k}{\HH_1}S_V(z_1)\left[\left(\frac{\dot \HH}{\HH^2}+\frac{2-5s}{\chi \HH}+5 s - f_\text{evo} \right) S_V\right]\!(z_2)\zeta^{21}(k\chi_1,k\chi_2,\theta)  \,, \\
&Q^\text{rsd-X}(\theta,z_1,z_2) =  \frac{k}{\HH_1}S_V(z_1)\De^\text{X}(z_2,k)\zeta^{20}(k\chi_1,k\chi_2,\theta)  \,, \\
&Q^\text{rsd-g4}(\theta,z_1,z_2) =  \frac{k}{\HH_1}S_V(z_1) \frac{ 2-5s}{\chi_2}  \int_0^{\chi_2} \dd \la S_{\phi +\psi}(\la,k) \zeta^{20}(k\chi_1,k\la,\theta)  \,,\\
&Q^\text{rsd-g5}(\theta,z_1,z_2) =  \frac{k}{\HH_1}S_V(z_1)\left(\frac{\dot \HH}{\HH^2}+\frac{2-5s}{\chi \HH}+5 s - f_\text{evo} \right)(z_2)    \\
&  \hspace{2.22 cm}\times  \int_0^{\chi_2} \dd \la  \dot S_{\phi +\psi}(\la,k) \zeta^{20}(k\chi_1,k\la,\theta) \,,  \\
&Q^\text{len-d1}(\theta,z_1,z_2) =\!  \left[\!\left(\!\frac{\dot \HH}{\HH^2}+\frac{2-5s}{\chi \HH}+5 s - f_\text{evo} \!\right) \!S_V\!\right](z_2)   \\
& \hspace{2.22 cm}\times  \frac{2 - 5 s}{2 \chi_1} \!  \int_0^{\chi_1} \hspace{-0.2cm}\dd \la \frac{\chi_1\!-\!\la}{\la}S_{\phi+\psi}(\la)\zeta^{L1}(k\la,k\chi_2,\theta)  \,, \\
&Q^\text{len-X}(\theta,z_1,z_2) =\!  \De^\text{X}(z_2,k)\frac{2 - 5 s}{2 \chi_1} \!  \int_0^{\chi_1} \!\!\dd \la \frac{\chi_1\!-\!\la}{\la}S_{\phi+\psi}(\la)\zeta^{L0}(k\la,k\chi_2,\theta)  \,,  \\
&Q^\text{len-g4}(\theta,z_1,z_2) =\!  \frac{(2 - 5 s)^2}{2 \chi_1\chi_2} \!  \int_0^{\chi_1} \hspace{-0.2cm}\dd \la \frac{\chi_1\!-\!\la}{\la}\!\int_0^{\chi_2}\hspace{-0.2cm} \dd \la'S_{\phi +\psi}(\la,k)  S_{\phi +\psi}(\la',k)\zeta^{L0}(k\la,k\la',\theta)   \,,\\
&Q^\text{len-g5}(\theta,z_1,z_2) =\! \left(\frac{\dot \HH}{\HH^2}+\frac{2-5s}{\chi \HH}+5 s - f_\text{evo} \right)\!(z_2) \frac{2 - 5 s}{2 \chi_1}  \\
& \hspace{2.22 cm}\times  \int_0^{\chi_1} \!\!\dd \la   \int_0^{\chi_2} \!\!\dd \la' \frac{\chi_1\!-\!\la}{\la}S_{\phi+\psi}(\la)\dot S_{\phi +\psi}(\la',k)\zeta^{L0}(k\la,k\la',\theta)   \,,\\
&Q^\text{d1-X}(\theta,z_1,z_2) = \left[\!\left(\!\frac{\dot \HH}{\HH^2}+\frac{2-5s}{\chi \HH}+5 s - f_\text{evo} \right)\! S_V\right]\!(z_1) \De^\text{X}(z_2,k)\zeta^{10}(k\chi_1,k\chi_2,\theta)  \,,\\
&Q^\text{d1-g4}(\theta,z_1,z_2) =  \left[\left(\frac{\dot \HH}{\HH^2}+\frac{2-5s}{\chi \HH}+5 s - f_\text{evo} \right) S_V\right](z_1)   \\
& \hspace{2.22 cm}\times \frac{ 2-5s}{\chi_2}  \int_0^{\chi_2} \dd \la S_{\phi +\psi}(\la,k) \zeta^{10}(k\chi_1,k\la,\theta)  \,,\\
&Q^\text{d1-g5}(\theta,z_1,z_2) =  \left[\!\left(\!\frac{\dot \HH}{\HH^2}+\frac{2-5s}{\chi \HH}+5 s - f_\text{evo}\! \right)\! S_V\right]\!(z_1)\left(\!\frac{\dot \HH}{\HH^2}+\frac{2-5s}{\chi \HH}+5 s - f_\text{evo}\right)\!(z_2)  \\
&  \hspace{2.22 cm}\times  \int_0^{\chi_2} \dd \la \dot S_{\phi +\psi}(\la,k) \zeta^{20}(k\chi_1,k\la,\theta)  \,,  \\
&Q^\text{X-Y}(\theta,z_1,z_2) = \De^\text{X}(z_1,k)\De^\text{Y}(z_2,k)\zeta^{00}(k\chi_1,k\chi_2,\theta)    \hspace{2cm} X\,,\, Y \in   \{\text{d2, g1, g2, g3}\}  \,, \\
&Q^\text{X-g4}(\theta,z_1,z_2) = \De^\text{X}(z_1,k)\frac{ 2-5s}{\chi_2}  \int_0^{\chi_2} \dd \la S_{\phi +\psi}(\la,k) \zeta^{00}(k\chi_1,k\la,\theta)  \,,    \\
&Q^\text{X-g5}(\theta,z_1,z_2) = \De^\text{X}(z_1,k)\left(\frac{\dot \HH}{\HH^2}+\frac{2-5s}{\chi \HH}+5 s - f_\text{evo} \right)\!(z_2)  \\
&\hspace{2.22 cm}\times  \int_0^{\chi_2} \dd \la  \dot S_{\phi +\psi}(\la,k) \zeta^{10}(k\chi_1,k\la,\theta) \,,\\
&Q^\text{g4-g5}(\theta,z_1,z_2) =\left(\frac{\dot \HH}{\HH^2}+\frac{2-5s}{\chi \HH}+5 s - f_\text{evo} \right)\!(z_2) \frac{ 2-5s}{\chi_1}  \\
& \hspace{2.22 cm}\times  \int_0^{\chi_1} \dd \la\int_0^{\chi_2} \dd \la'  S_{\phi +\psi}(\la,k) \dot S_{\phi +\psi}(\la,k) \zeta^{00}(k\la,k\la',\theta)  \;. 
\end{flalign*}

The correlators $Q^{BA}(z_1,z_2)$ are obtained from $Q^{AB}(z_1,z_2)$  using the identity $Q^{BA}(z_1,z_2)= Q^{AB}(z_2,z_1)$. 
The functions $S_X$ and $\De^X$ are given in terms of the transfer function $T(k)$ and the density growth function $D_1(a)$ as
\bea
S_D &=&-\frac{3}{5}\frac{k^2}{\Om_m\HH_0^2}\frac{D_1(a)}{a}T(k) \,,\\
S_V &=& \frac{3}{5}\frac{k\HH}{\Om_m\HH_0^2}\frac{d D_1(a)}{da}T(k) =-f\frac{\HH}{k}S_D\,,\\
S_\phi &=&\frac{9}{10}\frac{D_1(a)}{a}T(k)\,,  \qquad
S_{\phi +\psi} = 2S_\phi \,, \qquad \eea  \bea
\De^\text{d2} &=&-\frac{9}{5}\frac{\HH^2}{\Om_m\HH_0^2}\frac{d D_1(a)}{da}T(k)\,, \\
\De^\text{g1} &=&\left(\frac{\dot \HH}{\HH^2}+\frac{2-5s}{\chi \HH}+5 s - f_\text{evo} \right)S_\phi \,,\\
\De^\text{g2} &=&-(2-5s)S_\phi\,, \qquad 
\De^\text{g3} =\HH^{-1}\dot S_\phi  \, .
\eea
Here we have set $\Phi=\Psi$ and the transfer function $T(k)$ as well as the growth function $D_1(a)$ have to be determined either with a Boltzmann solver like {\sc class} or using an analytic approximation like  the one derived in Ref.~\cite{Eisenstein:1997ik}. We have normalized the growth function as well as the scale factor to unity today, $D_1(1)=1$. For the numerical results shown in our figures we used the Boltzmann solver {\sc class}.
We have checked analytically and numerically that our correlation functions for the standard and (d1)-terms agrees with the full sky results of~\cite{Bonvin:2013ogt}. 

\section{Approximation for the non-linear full-sky lensing}\label{app:lensNL}

\begin{figure}[ht]
\centering
\includegraphics[scale=0.4]{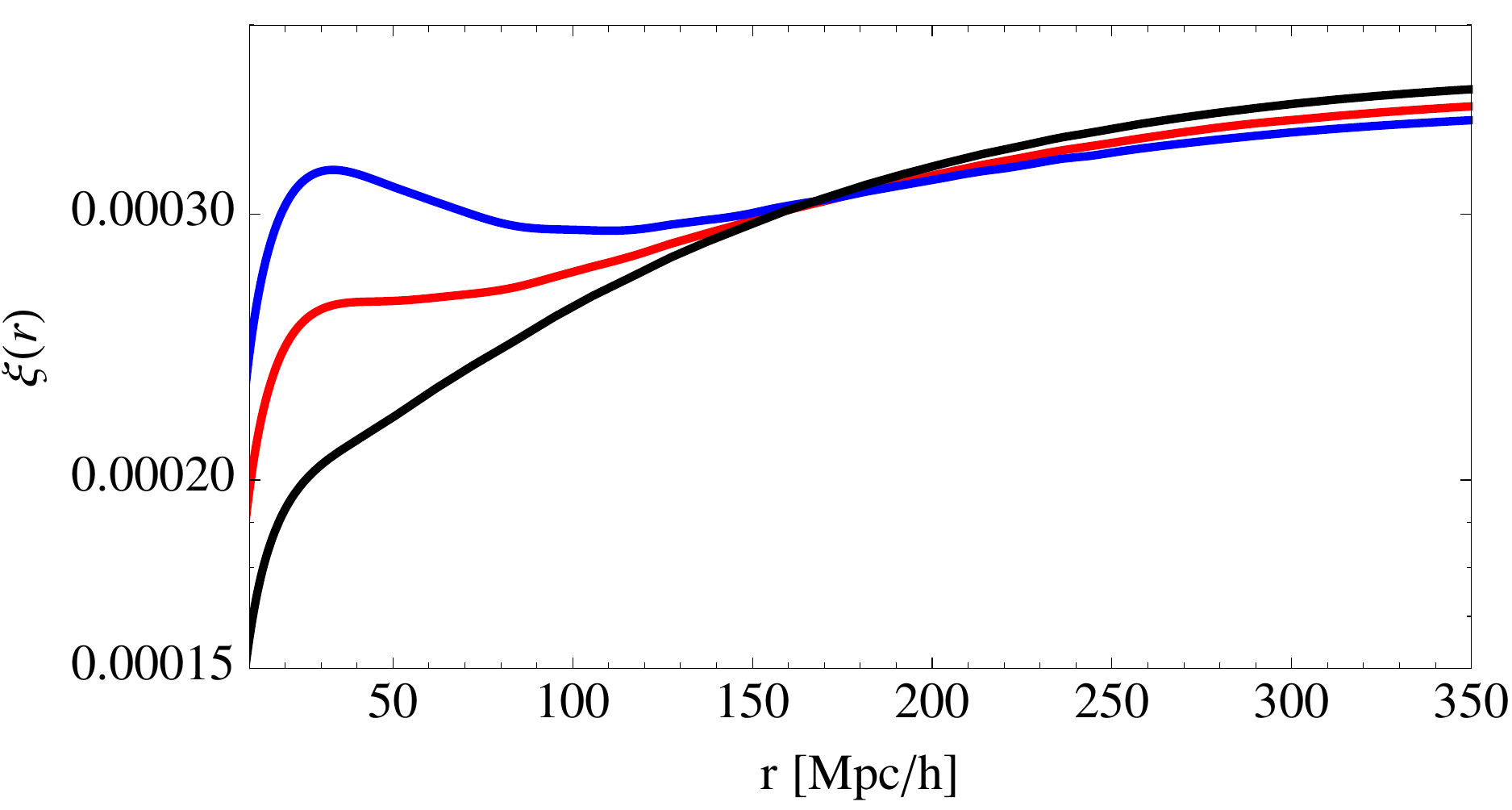}\includegraphics[scale=0.4]{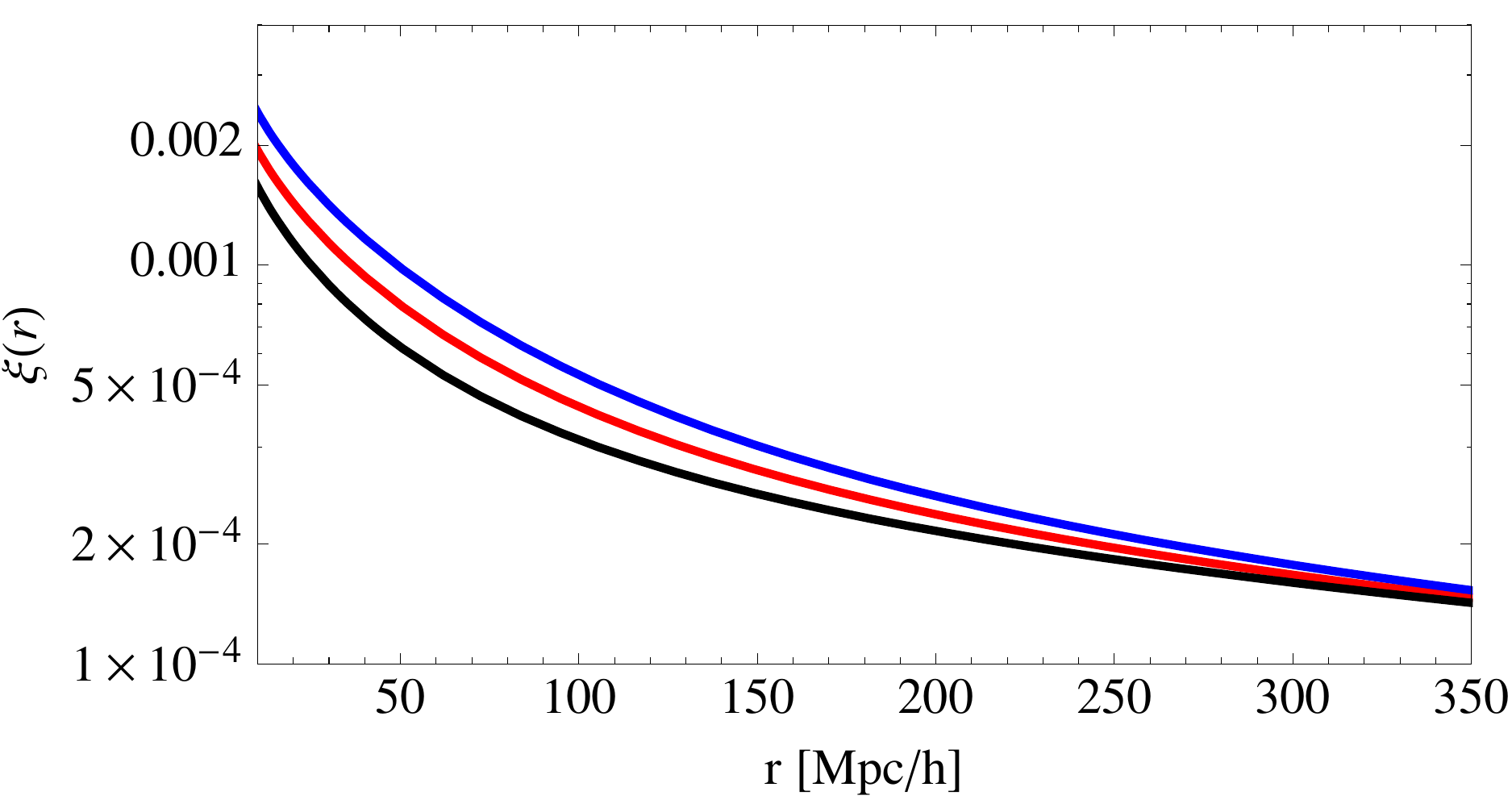}
\caption{\label{f:checkzNL} We show the full-sky  non-linear density-lensing correlation function (left) and lensing-lensing correlation function (right) at $\bar z=1$ as a function of separation, for $\mu=1$. The black solid line shows the calculation with $z_*=1$, the blue line with $z_*=0$ and the red line with $z_*=0.42$.}
\end{figure}

As discussed in Section~\ref{s:mu-r}, to calculate the non-linear full-sky lensing we calculate the halo-fit power spectrum at a fixed redshift $z_*$ and then evolve it along the line-of-sight using the linear growth rate. To choose $z_*$ we use the flat-sky non-linear result, that we calculate first without approximation and second with the same approximation as in the full-sky. We find that when $z_*=0.42$ the approximate solution is in extremely good agreement with the correct solution. We use therefore the same $z_*$ to calculate the full-sky result, for which it is not possible to do an exact integration (see discussion in Section~\ref{s:mu-r}).

In Fig.~\ref{f:checkzNL} we compare the non-linear full-sky lensing calculated with different values for $z_*$. In red we show the result for $z_*=0.42$ (best fit from the flat-sky), and in black and blue we show the two extreme cases: $z_*=1$ (black) and $z_*=0$ (blue). We see that the lensing terms behave as expected: a smaller $z_*$ gives rise to a larger result, since in this case we overestimate the power spectrum along the line-of-sight. The curve $z_*=0.42$ is well situated between the two extreme cases, as was the case in the flat-sky. This gives us confidence that the approximation works well also for the full-sky lensing.

\section{Direction dependent power spectra}\label{app:theorem}
In this appendix we prove a simple property of direction dependent power spectra which is often used. This result is of course not new but it is usually used without derivation and mainly in special cases. Here we prove it in full generality.\vspace{0.2cm}

\begin{theorem-non}
$\xi(\br)$ is a correlation function which depends on the orientation of $\br$ only via its scalar product with one fixed given direction $\bn$ (e.g. the line of sight). Denoting the corresponding direction cosine by $\mu$ and expanding $\xi$ in Legendre polynomials,
we have
\be\label{e:xinr}
\xi(\br) =\sum_n \xi_n(r)L_n(\mu) \,, \qquad \mu=\hat\br\cd\bn \,.
\ee
In this situation the Fourier transform of $\xi$, the power spectrum, is of the form
\bea
P(\bk) &=& \sum_n p_n(k)L_n(\kcos) \,, \quad \kcos=\hat\bk\cd\bn \quad\mbox{where} \quad \\
\label{e:pnk}
p_n(k) &=&4\pi i^n\int_0^\infty dr r^2j_n(kr)\xi_n(r) \,, \quad\mbox{and } \\
\label{e:xin2}
\xi_n(r) &=& \frac{(-i)^n}{2\pi^2}\int_0^\infty dk k^2j_n(kr)p_n(k)  \,.
\eea
\end{theorem-non}

\begin{proof}
The Fourier transform of $\xi$ is defined as
\be\label{e:pdef} 
P(\bk) = \int d^3r e^{i\br\cd\bk}\xi(\br) \,.
\ee
We use that 
$$e^{i\br\cd\bk}=\sum_\ell i^\ell(2\ell+1)j_\ell(kr) L_\ell(
\hat\bk\cd\hat\br)$$ and 
\bean 
L_\ell(\hat\bk\cd\hat\br) &=& \frac{4\pi}{2\ell+1}\sum_{m=-\ell}^\ell Y_{\ell m}(\hat\bk)Y^*_{\ell m}(\hat\br) 
\;=\; \frac{4\pi}{2\ell+1}\sum_{m=-\ell}^\ell Y_{\ell m}(\hat\br)Y^*_{\ell m}(\hat\bk)\,.
\eean 
Here $Y_{\ell m}$ are the spherical harmonics as given e.g. in~\cite{mybook}. Inserting
these identities in \eqref{e:pdef} using the ansatz~\eqref{e:xinr} for the correlation function, we obtain
\bea
 P(\bk) &=&\sum_{\ell m}\sum_{n m'}\frac{(4\pi)^2 i^\ell}{2\ell+1} \int d^3r \xi_n(r)j_\ell(kr)Y_{\ell m}(\hat\bk)
  Y^*_{\ell m}(\hat\br)Y_{n m'}(\hat\br)Y^*_{n m'}(\bn)\,.
\eea
Using the orthogonality relation of spherical harmonics, the integration over directions gives
\bea
 P(\bk) =4\pi\sum_{n} i^n\int_0^\infty dr r^2\xi_n(r)j_n(kr)L_n(\kcos) \,.
 \eea
 Identification of the expansion coefficients yields~\eqref{e:pnk}.
 Eq.~\eqref{e:xin2} is obtained in  the same way using the inverse Fourier transform,
$$ \hspace{2cm}\xi(\br)=\frac{1}{(2\pi)^3}\int d^3 ke^{-i\bk\cd\br}P(\bk) \,. \hspace{7cm} \qed $$ 
\end{proof} 

Clearly, if $\xi(\br)=\langle\De(\bx)\De(\bx+\br)\rangle$ is independent of $\bx$ ($\De$ is statistically homogeneous), $\xi$ does not depend on the sign of $\br$ and in the sum above only $\xi_n$ with even $n$'s can contribute so that $P(\bk)$ is real.

Inserting the expressions for the $Q^\text{AB}$ in \eqref{corrQ} to obtain the correlation function, we realize that in the flat sky limit ($\bn_1\ra\bn_2$), all our terms $\xi^{AB}$ where the corresponding $Q^{AB}$ do not contain integrated terms, are actually of this form. This also shows that in this limit $\zeta^{01}+\zeta^{10}$
and $\zeta^{12}+\zeta^{21}$ must vanish since they contain $j_1(kr)$ and $j_3(kr)$ and would yield imaginary contributions to the power spectrum.

For wide angles $\bn_1\neq \bn_2$ the correlation function depends on two directions. Furthermore, for large $\br$ it is not translation invariant as it depends on the redshift on our background light-cone at which $\br$ is placed. In this case, the Fourier transform of the correlation function is no longer simply given by the power spectrum of the fluctuations.

The theorem proven above has a simple but useful corollary which is sometimes called the \emph{closure relation} of spherical Bessel functions~\cite{ArfkenWebver}. Inserting the expression \eqref{e:pnk} into \eqref{e:xin2} and using that it holds for arbitrary functions $p_n(k)$,
we find
\be\label{e:sbes-de}
\frac{2}{\pi}\int_0^\infty j_n(rk)j_n(rk')r^2dr =\de(k-k')k^{-2} \, ,
\ee
for positive $k$ and $k'$. Using $$j_n(x)=\sqrt{\frac{\pi}{2x}}J_{n+1/2}(x)$$ we can convert \eqref{e:sbes-de} into  an equation for ordinary Bessel functions $J_m$:
\be
\int_0^\infty J_{n+1/2}(rk)J_{n+1/2}(rk')rdr =k^{-1}\de(k-k') \, ,
\ee
This identity  also holds for $J_m$ with integer $m$, see~\cite{GradRyz}, No 6.512-8.

\section{The flat sky approximation}\label{a:flat}
To derive expression \eqref{e:Pflatint} we consider the observed galaxy density fluctuation in real space given in Eq.~\eqref{e:Degz}.
We neglect the integrated Sachs Wolfe term and the $\dot\Phi$ term in the first line; they are very small and relevant mainly on very large angular scales where
the flat sky approximation breaks down. The remaining integrated term is then only the lensing term and the subdominant Shapiro time delay. Furthermore, we set $\Psi=\Phi$ which is a very good approximation in $\La$CDM at late times.
Denoting the power spectrum of the comoving density contrast $\de_c$ at redshift $z=0$ by $P_\de$
and using the perturbed Einstein and continuity equations we find
\bea
\Phi &=&\Psi ~=~ -\frac{3}{2}\frac{\Om_mH_0^2(1+z)D_1(z)}{k^2}\de_c \\
V &=& -\frac{\HH}{k}f(z)D_1(z)\de_c \,,
\eea
where $f(z)$ is the growth rate as given in \eqref{e:growth}, $D_1(z)$ is the growth function such that $\de_c(k,z)= D_1(z)\de_c(k)\equiv D_1(z)\de_c(k,0)$ 
and $\Om_m$ is the matter density parameter today.

Neglecting first the integrated terms we can simply Fourier transform this expression
from $\chi(z)\bn \equiv \bx$ to $\bk$ and use that the power spectrum is the square of the Fourier transform amplitude. This yields
\be\label{e:Pni}
P_{n.i}= \left|A +B/(k\HH) + C/(k\HH)^2\right|^2P_\de(k) \,,
\ee
where $A$, $B$ and $C$ are given in \eqref{e:AB} and \eqref{e:C}.

To derive the cross term of the non-integrated with the  integrated terms, it is more useful to start with the correlation function. Let us denote $A +B/(k\HH) + C/(k\HH)^2 =\alpha(k,\kcos,z)$ and $\hat F(k,\kcos,z) =\alpha(k,\kcos,z)\de_c(k)$ with Fourier transform $F(\bx,z)$ . Denoting
\be\label{e:Idef}
I(\chi(z)\bn,z)=\frac{2}{\chi(z)}\int_0^{\chi(z)}d\la\left[2-\frac{\chi(z)-\la}{\la}\De_\Om\right]\Phi\,,
\ee
we have
\bea
\xi_{\De\De}(\br, z) &=& \left\langle F(\chi_1\bn_1,z_1)F(\chi_2\bn_2,z_2)\right\rangle+
 \langle I(\chi_1\bn_1,z_1)F(\chi_2\bn_2,z_2)\rangle  
 \nonumber \\  &&  \label{e:corrfctn}
 +\langle F(\chi_1\bn_1,z_1)I(\chi_2\bn_2,z_2)\rangle + \langle  I(\chi_1\bn_1,z_1)I(\chi_2\bn_2,z_2)\rangle\,,
\eea
where $\chi_i=\chi(z_i)$ and $\br= \chi_2\bn_2-\chi_1\bn_1$, $ z=(z_1+z_2)/2$ and we assume both $\chi_i\gg r$ and the $z_i$ should not be very different. Using the relation between $\Phi$ and $\de_c$, the
 contribution of the cross term to the correlation function is then given by

\bea
\xi_{IF}(\br, z)&=& -\frac{3}{(2\pi)^3}\frac{\Om_mH_0^2(2-5s(z))}{2\chi_1}\int \frac{d^3k}{k^2}P_\de(k)e^{-i\bk\bn_2\chi_2}\al(k,\kcos,z_2)\times \nonumber\\
&& \qquad  \int_0^{\chi_1}d\la\left[\la(\chi_1-\la)k_\perp^{2}+2\right]D_1(z(\la))(1+z(\la))e^{i\bk\bn_1\la} \,.
\label{e:xiif1}
\eea
In the spirit of the flat sky approximation we now set $\bn_1=\bn_*+\De\bn/2$ and 
$\bn_2=\bn_*-\De\bn/2$ assuming that $\De\bn$ is very small. Splitting $\br=\br_\perp +\bn_*r_\pa$ with $ \br_\perp=\chi(z)\De\bn$ and $r_\pa=r\cos\al_2$, see Fig.~\ref{f:a-gammabeta}, we then perform the $k$-integral in the direction parallel to $\bn_*$, $dk_\pa \exp(-ik_\pa(\chi_2-\la))$. We neglect the slow dependence of the power spectrum on $k_\pa$ and only consider the rapidly oscillating exponential which gives $2\pi\de(\chi_2-\la)$. Hence the integral over $\la$ does not contribute if $\chi_2>\chi_1$, otherwise it reduces to the integrand at $\chi_2$, 
\bea
\xi_{IF}(\br,z) &=& -\frac{3}{(2\pi)^2}\frac{\Om_mH_0^2(2-5s(z))\Theta(\chi_1-\chi_2)}{2\chi_1}D_1(z_2)(1+z_2) \nonumber \\
&& \qquad  \int \frac{d^2k_\perp}{k_\perp^2} P_\de(k_\perp)e^{-i\bk_\perp\cdot\br_\perp}\al(k_\perp,0,z_2)\left[\chi_2(\chi_1\!-\!\chi_2)k_\perp^{2}\!+\!2\right] \,,
\label{e:xiif2}
\eea
where $\Theta$ is the Heaviside $\Theta$-function.

Using polar coordinates, $d^2\bk_\perp=dk_\perp k_\perp d\varphi$ we can perform the $\varphi$ integration which yields a Bessel function, $2\pi J_0(k_\perp r_\perp) =2\pi J_0(k_\perp r\sin\al_2)$. 
The term $\xi_{FI}(\br,\bar z)$ contributes in the same way with $z_1$ and $z_2$ exchanged. Setting $\chi_1-\chi_2=r_\pa =r\mu$ and neglecting the difference of $\chi_1$ and $\chi_2$ ($z_1$ and $z_2$) in all other places,  we find for the sum of both mixed terms
\bea
\xi_{IF+FI}(\br,z) &=& -\frac{3}{2\pi}\frac{\Om_mH_0^2(2-5s(z))}{2\chi}D_1(z)(1+z) \nonumber \\
&& \qquad  \int \frac{dk_\perp}{ k_\perp} P_\de(k_\perp)J_0(k_\perp r\sqrt{1-\mu^2})\al(k_\perp,0,z)\left[\chi |\mu| rk_\perp^{2}\!+\!2\right] \,.
\label{e:xiiffi}
\eea
Here we have also neglected the difference between $\cos\al_2$ and $\mu$. In the flat sky approximation all these angles are equal. (If we would want to be precise, actually in the case $z_1\equiv z_2$, hence $\mu=0$ the Shapiro time delay would obtain a factor $4$, not $2$, but we neglect this in the flat sky approximation.)

To obtain the Fourier transform of (\ref{e:xiiffi}) which is the contribution n.i.-I to the power spectrum we first multiply the equation with $\int dk_\pa\exp(-ik_\pa r_\pa)\de(k_\pa) = 1$. We then write the factor $|\chi_2-\chi_1|=|r_\pa|=|\mu| r$ inside the integral, 
$$ \int dk_\pa\exp(-ik_\pa r_\pa)|r_\pa|\de(k_\pa) = |r_\pa|$$
is the Fourier transform of
\be\label{e:deP}
\de^P(k_\pa) \equiv \frac{1}{2\pi}\int dr_\pa\exp(ik_\pa r_\pa)|r_\pa| \,.
\ee
Note that without the absolute value $\de^{P}$ would become $-i\de'$. This distribution is purely imaginary while $\de^P$ is real.
However, like $\de$ or $\de'$ its support is on $k_\pa=0$, i.e. for a function $f$ which vanishes in a small neighborhood around $k_\pa=0$ we have $\de^P\cdot f\equiv 0$.

Inserting \eqref{e:deP}, we can write the correlation function $\xi_{IF+FI}$ as the Fourier transform of
\bea
P_{\rm{n.i.}-I}(\bk,z) &=& -3\pi\frac{\Om_mH_0^2(2-5s(z))}{\chi}D_1(z)(1\!+\!z) 
P_\de(k_\perp)\al(k_\perp,0,z)\left[\de^P(k_\pa)+\frac{2}{k_\perp^{2}}\de(k_\pa)\right] \,.  \nonumber \\   &&
\label{e:Piffi}
\eea 
Note also that since $k_\pa=0$, in the flat sky limit, the integrated term is not correlated with redshift space distortions.

Let us finally compute the double integrated term,
\bea
\xi_{II}(\br,z) &=& \frac{(3\Om_mH_0^2(2-5s(z)))^2}{(2\pi)^34\chi^2}\int\frac{d^3k}{k^4}P_\de(k)
\int_0^{\chi_1} \hspace{-0.1cm}d\la\int_0^{\chi_2} \hspace{-0.1cm}d\la'\left[\la(\chi_1-\la)k_\perp^{2}+2\right]\times \hspace{2.7cm}   
\nonumber\\  &&  \hspace{-1.5cm}
\left[\la'(\chi_2-\la')k_\perp^{2}+2\right]D_1(z(\la))(1+z(\la))D_1(z(\la'))(1+z(\la'))e^{-i\bk(\bn_1\la-\bn_2\la')} \,.\nonumber \\
\eea
Via the same procedure as above, the integration over $k_\pa$ leads to $2\pi\de(\la-\la')$ and we find
 \bea
\xi_{II}(\br,z) &=& \frac{(3\Om_mH_0^2(2-5s(z)))^2}{(2\pi)^24\chi^2}\int\frac{d^2k_\perp}{k_\perp^4}P_\de(k_\perp)
\nonumber\\  &&
\int_0^{\chi}d\la\left[\la(\chi-\la)k_\perp^{2}+2\right]^2D^2(z(\la))(1+z(\la))^2e^{-i\bk_\perp\br_\perp(\la/\chi))} \,.
\eea
We now perform a change of variables, $\bk_\perp\mapsto (\la/\chi)\bk_\perp$. In terms of this new variable, the integral contribution to the correlation function becomes
\bea
\xi_{II}(\br,z) &=& \frac{(3\Om_mH_0^2(2-5s(z)))^2}{(2\pi)^24\chi^2}\int_0^{\chi}d\la\int\frac{d^2k_\perp}{k_\perp^4}P_\de(k_\perp\chi/\la)e^{-i\bk_\perp\br_\perp}\times
\nonumber\\  &&
\left(\frac{\la}{\chi}\right)^2\left[\frac{(\chi-\la)\chi^2}{\la}k_\perp^{2}+2\right]^2D^2(z(\la))(1+z(\la))^2 \,.
\eea
Again, performing the $\varphi$ integration we end up with
\bea
\xi_{II}(\br,z) &=& \frac{(3\Om_mH_0^2(2-5s(z)))^2}{8\pi\chi^2}\int_0^{\chi}d\la\int dk_\perp k_\perp P_\de(k_\perp\chi/\la)J_0(k_\perp r\sqrt{1-\mu^2})
\times \nonumber\\  &&
\left(\frac{\la}{\chi}\right)^2\left[\frac{(\chi-\la)\chi^2}{\la}+\frac{2}{k_\perp^2}\right]^2D^2(z(\la))(1+z(\la))^2 \,.
\eea
Inserting the same factor $1$ as for the mixed term above, we can read off the flat sky power spectrum of the integrated contribution,

\bea
P_{II}(\bk,z) &=& \frac{\pi(3\Om_mH_0^2(2-5s(z)))^2}{2\chi^2}\!\!\int_0^{\chi}\hspace{-1.5mm}d\la P_\de(k\chi/\la)\de(k_\pa)
\left(\frac{\la}{\chi}\right)^2\hspace{-1.mm}\left[\frac{(\chi-\la)\chi^2}{\la}+\frac{2}{k^2}\right]^2\hspace{-1.5mm}D^2(z(\la))(1\!+\!z(\la))^2 \,. \nonumber\\
&&  \label{e:PII}
\eea
Adding (\ref{e:Pni}, \ref{e:Piffi}, \ref{e:PII}) we obtain the result (\ref{e:Pflatint}).
For completeness, and since we use it for some of our results, we also write down the flat sky correlation function,
\bea
\xi_\De(\br,z) &=& \int dk_\perp  kk_\perp P_\de(k) J_0(k_\perp r\sqrt{1-\mu^2}) \int_{-1}^1d\kcos|\al(k,\kcos,z)|^2
e^{-ik\kcos \mu r} \nonumber\\
&& -\frac{3\Om_mH_0^2(2-5s(z))}{4\pi\chi}D_1(z)(1+z) \nonumber \\
&& \qquad  \int \frac{dk_\perp}{ k_\perp} P_\de(k_\perp)J_0(k_\perp r\sqrt{1-\mu^2})\al(k_\perp,0,z)\left[\chi |\mu| rk_\perp^{2}\!+\!2\right] +  \nonumber\\
&& \frac{(3\Om_mH_0^2(2-5s(z)))^2}{8\pi\chi^2}\int_0^{\chi}d\la\int\frac{dk_\perp}{k_\perp^3}P_\de(k_\perp\chi/\la)J_0(k_\perp r\sqrt{1-\mu^2})
\times \nonumber\\  &&
\left(\frac{\la}{\chi}\right)^2\left[\frac{(\chi-\la)\chi^2}{\la}k_\perp^{2}+2\right]^2D^2(z(\la))(1+z(\la))^2 \,. \label{e:xifinalflat}
\eea
Since $\al$ only contains terms which are constant, linear or quadratic in $\kcos$, the $\kcos$-integration is easily performed analytically.

\newpage
\bibliographystyle{utcaps}
\bibliography{pofk-refs}

\end{document}